\documentclass[fleqn,usenatbib]{mnras}

\usepackage{newtxtext,newtxmath}
\usepackage{gensymb}

\usepackage[T1]{fontenc}

\DeclareRobustCommand{\VAN}[3]{#2}
\let\VANthebibliography\thebibliography
\def\thebibliography{\DeclareRobustCommand{\VAN}[3]{##3}\VANthebibliography}

\usepackage{graphicx}

\usepackage[utf8]{inputenc}
\usepackage{amsmath}	

\usepackage{subfig}

\def\beq{\begin{equation}}
\def\eeq{\end{equation}}

\title[Dark Matter halo spin of UGC 5288]{Modelling Dark Matter Halo Spin using Observations and Simulations: application to UGC 5288}

\author[]{
Sioree Ansar,$^{1,2}$\thanks{E-mail: sioreeansar@gmail.com},
Sandeep Kumar Kataria,$^{3,4}$,
Mousumi Das$^{1}$
\\
$^{1}$Indian Institute of Astrophysics, Bangalore 560034, India \\
$^{2}$Pondicherry University, R.V. Nagar, Kalapet 605014, Puducherry, India 
\\
$^{3}$Department of Astronomy, School of Physics and Astronomy, Shanghai Jiao Tong University, 800 Dongchuan Road, Shanghai 200240, China\\
$^{4}$Key Laboratory for Particle Astrophysics and Cosmology (MOE) / Shanghai Key Laboratory for Particle Physics and Cosmology, Shanghai 200240, China\\
}

\date{Accepted XXX. Received YYY; in original form ZZZ}

\pubyear{2023}

\begin{document}
\label{firstpage}
\pagerange{\pageref{firstpage}--\pageref{lastpage}}
\maketitle

\begin{abstract}
Dark matter (DM) halo properties are extensively studied in cosmological simulations but are very challenging to estimate from observations. The DM halo density profile of observed galaxies is modelled using multiple probes that trace the dark matter potential. 
However, the angular momentum distribution of DM halos is still a subject of debate. In this study we investigate a method for estimating the halo spin and halo concentration of low surface brightness (LSB), gas-rich dwarf barred galaxy UGC 5288, by forward modelling disk properties derived from observations - stellar and gas surface densities, disk scale length, HI rotation curve, bar length and bar ellipticity. We combine semi-analytical techniques, N-body/SPH and cosmological simulations to model the DM halo of UGC 5288 with both a cuspy Hernquist profile and a flat-core pseudo-isothermal profile. We find that the best match with observations is a pseudo-isothermal halo model with a core radius of $r_{c} = 0.23$ kpc, and halo spin of $\lambda$= 0.08 at the virial radius. Although our findings are consistent with previous core radius estimates of the halo density profile of UGC 5288, as well as with the halo spin profiles of similar mass analogues of UGC5288 in the high-resolution cosmological-magneto-hydrodynamical simulation TNG50, there still remain some uncertainties as we are limited in our knowledge of the formation history of the galaxy. Additionally, we find that the inner halo spin ($ r< 10$ kpc) in barred galaxies is different from the unbarred ones, and the halo spin shows weak correlations with bar properties.
\end{abstract}


\section{Introduction}
The rotation curves of galaxies have been extensively used to model their dark matter (DM) halo mass distribution \citep{McGaugh.de.Blok.1998a,McGaugh.de.Blok.1998b,de.Blok.2001} as well as their density profiles \citep{Jimenez.et.al.2003, Kurapati.et.al.2020}. Based on observations and theories of galaxy formation and evolution, semi-analytical models have been developed to describe the structure and kinematics of the dark matter halos \citep{Mo.et.al.1998}. These studies, combined with modern N-body/hydrodynamical simulations \citep{Athanassoula.Misiriotis.2002, Springel.et.al.2005,Maccio.et.al.2007, Sharma.Steinmetz.2005, Yurin.Springel.2014, Schaye.et.al.2015} and Cosmological simulations \citep{Danovich.et.al.2015,Burkert.et.al.2016,Zjupa.Springel.2017,Jiang.et.al.2019,Bett.Carlos.2012,Bett.et.al.2007,Bett.et.al.2010, Hellwing.et.al.2020,2019TNG50_data_release}, have provided models by which many important questions regarding  the effect of DM halos on disks have been explored. For example, the effect of halo angular momentum on disk surface density \citep{Kim.Lee.2013}, disk scale height \citep{Klypin.et.al.2009} and the effect of halo spin on bar formation \citep{Saha.Naab.2013, Long.et.al.2014, Collier.et.al.2018, Collier.et.al.2019, Collier.Ann-Marie.2021,Kataria.Shen.2022}; as well as the importance of halo concentrations and shape on galaxy evolution \citep{Klypin.et.al.2016,Kumar.et.al.2022}. See \cite{Bullock.Boylan-Kolchin.2017} for a detailed review.

In simulations, the angular momentum of a DM halo is often studied using a dimensionless quantity called the halo spin $\lambda$. It was first introduced by \cite{Peebles1969} to quantify the fraction of total energy in a system that contributes to ordered rotational motion compared to the random motion,
\begin{equation} \label{Peebles_spin3}
    \lambda_{P}  = \frac{J |E|^{1/2}}{G M^{5/2}}
\end{equation}
where $J$ is the angular momentum of the halo, $E$ is the total energy and $M$ is the mass at a radius $r$. The distribution of halo spin over galaxies of different masses has been studied in various cosmological simulations \citep{Bullock.et.al.2001, Rodriguez-Puebla.et.al.2016}. The effect of baryonic processes on the spin or angular momentum of halos has also been studied \citep{ Sharma.Steinmetz.2005,Bett.et.al.2007,Bett.et.al.2010}. The effect of galaxy interactions on halos has been explored  \citep{Hetznecker.Burkert.2006}. The increase in halo angular momentum in massive galaxies at high redshifts has been explored using high resolution cosmological simulations \citep{Danovich.et.al.2015}. \cite{Bett.Carlos.2012} studied the changes in the direction of the halo spin vector due to merger events and it is found to play an important role in the morphological changes in disk galaxies. 

Another very important property of DM halos is the halo concentration parameter.
The concentration $c$ is defined for a cuspy centrally peaked NFW type halo as the ratio of the virial radius $r_{200}$ and the scale radius $r_{s}$.
\begin{equation}
    c=\frac{r_{200}}{r_{s}}
\end{equation}
It is found to be anti-correlated with the halo spin \citep{Maccio.et.al.2007,Jeeson-Daniel.et.al.2011}. Massive galaxies are found to have such cuspy profiles in cosmological simulations. But, in nature, there are also a large number of less massive galaxies that do not have a centrally peaked density profile but instead have a more flat core density profile. One such example is the pseudo-isothermal halo which has a flat-core density profile at its centre. In such profiles, the equivalent parameter of the scale radius is the core radius $r_{c}$ (more details in Section \ref{halo_models}). The structure and velocity distribution of DM in these two types of halo profiles are different. Hence, they may also show different spin distributions \citep{vandenBosch.et.al.2016}.   

Spin and halo concentration play a major role in the galaxy-halo scaling relations \citep{Kravtsov.2013,Somerville.Dave.2015,Huang.et.al.2017,Somerville.et.al.2018, Jiang.et.al.2019, Zanisi.et.al.2020}. The galaxy scaling relations are not only important for studying the galaxy-halo connection and galaxy evolution, but also very useful for understanding the large scale properties of halos in galaxies.

The DM halo spin or halo angular momentum cannot be estimated for galaxies directly as there is no tracer for the motion of the DM halo constituents. This is unlike the studies of galaxy disks that use the stellar mass density and gas disk rotation to estimate the specific disk angular momentum $j$ of galaxies \citep{fall.romanowsky.2018,Kurapati.et.al.2020}. Galaxy lensing studies can estimate the total angular momentum but are model dependent \citep{Bartelmann.and.Schneider.2001,Brimioulle.et.al.2013,Ardila.et.al.2021,Leauthaud.et.al.2020}. In a very recent model-dependent study \citep{Obreja.et.al.2021}, the Milky Way halo spin (using a contracted NFW profile) has been estimated to be $\lambda=0.061 \substack{+0.022 \\ -0.016}$. They use the existing relation between dark matter halo angular momentum to stellar disk angular momentum in the NIHAO suit of cosmological simulations to predict that the estimation of halo spin moves to the outer edge of a lognormal profile (peaking at 0.035) if NFW halo models are used.

To estimate the halo spin for a galaxy using equation (\ref{Peebles_spin3}), we need the halo mass, energy and total angular momentum.  We can approximate the halo mass up to the virial radius from the rotation curve and obtain an approximate value of the halo energy by assuming virial equilibrium and circular orbits for the halo particles \citep{Mo.et.al.1998}. But we are still unable to estimate the spin because the velocity structure of a dark matter halo and hence the angular momentum is unknown. In this study, we address this problem by first estimating the halo concentration and then determining  a probable halo spin profile using a model with the observed baryonic disk properties of a galaxy and N-body/SPH simulations. The technique that we present here has some limitations. (i)~It can be used to estimate the halo concentration and spin for isolated galaxies that have not undergone any recent interactions with satellite galaxies. (ii)~The galaxies should have low star formation rates as our model does not include star formation. A long-term goal would be to include star formation and different merger histories in our galaxy models so that they could be applied to a more general sample. This will be useful to build a phase space of halo parameters similar to distributions such as the log-normal distribution of halo spin $\lambda_{B}$  \citep{Bullock.et.al.2001} using observational data from galaxy surveys.

To verify our results we compare our halo spin profiles with one of the most recent high-resolution cosmological gravo-magnetohydrodynamical simulations, TNG50 \citep{2019TNG50_data_release, Nelson.et.al.2019,Pillepich.et.al.2019, Rosas-Guevara.et.al.2021} of the IllustrisTNG project. TNG50 has a reasonably high baryonic ($8.5\times 10^4$ M\textsubscript{\(\odot\)}) and dark matter particle mass resolution ($4.5\times 10^5$ M\textsubscript{\(\odot\)}) to study the inner disk and halo properties of galaxies with stellar mass $> 10^{10}$ M\textsubscript{\(\odot\)}. It has a box size of $\sim 51$ Mpc and it includes star formation, stellar feedback, AGN feedback, and galaxies undergo multiple interactions and flyby events with other galaxies throughout their evolution. These complex processes naturally arise in the cosmological simulations, and are very essential to study galaxy formation and evolution in a more realistic cosmological environment. These physical phenomena are not included in the codes we use. So, to check our assumptions and compare our results, we use samples of barred and unbarred galaxies from TNG50 suite to compare our findings. Furthermore, we search for analogues of UGC 5288 in the TNG50 sample.

In the following sections, we first describe our method and apply it to determine the halo concentration and probable halo spin of a gas-rich, barred dwarf galaxy UGC 5288 that lies in the lynx-Cancer void\citep{Makarova.1999,Van.zee.2000,van.Zee.2004, van.Zee.2006}. Since this is a pilot study, we modelled the halo using both Hernquist and pseudo-isothermal halo profiles. In the second part of the paper, we examine the halo profiles for TNG50 to see if our halo model for UGC5288 is similar to those produced by cosmological simulations.

We present the observed properties of UGC 5288 (Section \ref{galaxy_selection}), the galaxy model (including the assumptions from semi-analytical galaxy formation theories) in Section \ref{sec:galaxy_modeling}, the numerical methods and the simulated model results in Section \ref{sec:simulations}. We compare our results with cosmological simulations in Section \ref{sec:cosmological_simulations}. The discussion is presented in Section \ref{sec:discussion} and the summary in section \ref{sec:summary}. We adopt standard $\Lambda$CDM cosmology with ($\Omega_{m}=0.3$, $\Omega_{\Lambda}=0.7$, $h=0.7$).

\section{Galaxy Selection} \label{galaxy_selection}
We aim to estimate a probable halo spin profile of a galaxy. We select a galaxy which has the following properties. 1) Does not show signs of interactions with nearby galaxies and 2) has a low star formation rate. Isolated low surface brightness galaxies (LSBGs) in voids are thus ideal for our study as they have very little star formation but have extended gas disks which are good for measuring disk rotation velocities. Also, the larger spread of the disk in LSBGs is usually attributed to the high angular momentum of their dark matter halos \citep{Dalcanton.et.al.1997,Barnes.and.Efstathiou.1987} and the low star-formation rate could be due to larger dark matter to baryon mass ratios or lower disk surface densities \citep{Kim.Lee.2013}. 

\begin{figure}
\centering
	\includegraphics[width=0.8\columnwidth]{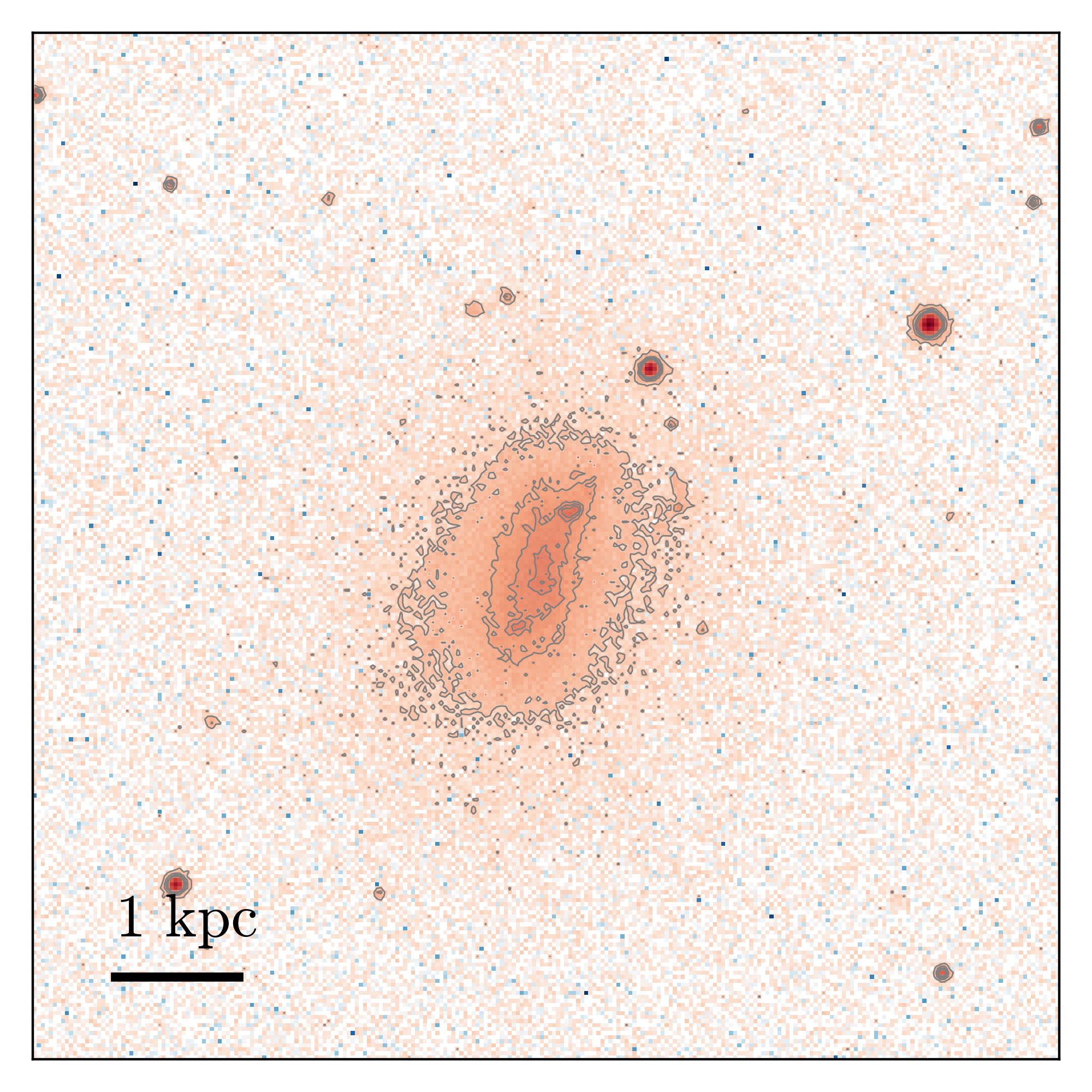}
    \caption{{\bf SDSS i-band image of UGC 5288}. The stellar disk is shown in light red. The stellar disk is mainly composed of the bar and is much less extended than the HI disk which has a radius of nearly 16.6 kpcs (see Figure  in \citep{Kurapati.et.al.2018,Kurapati.et.al.2020}) }
    \label{fig:example_figure}
\end{figure}

\begin{figure*}
\centering
    \includegraphics[width=0.8\textwidth]{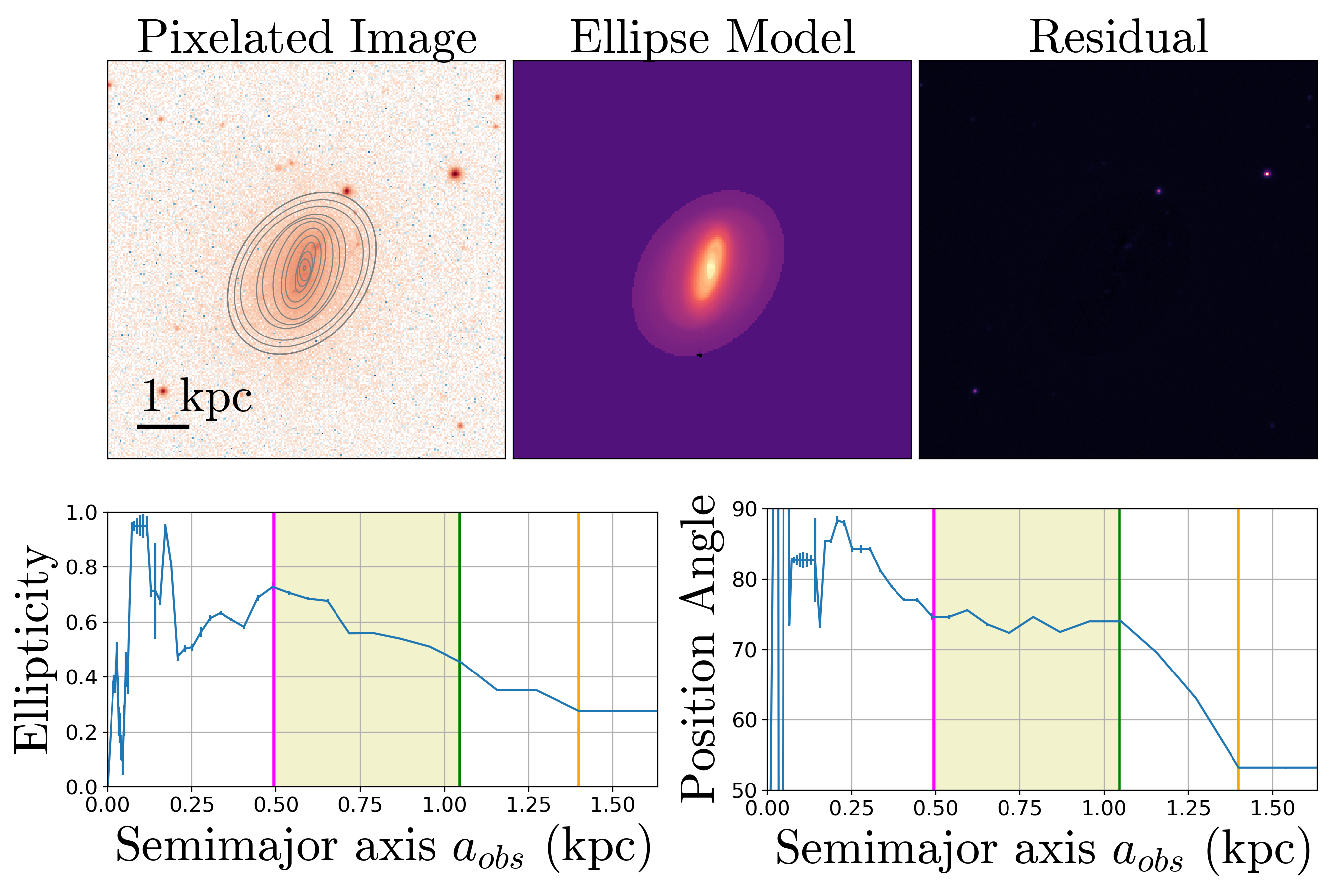}
    \caption{ {\bf The above figure illustrates the determination of the bar properties of UGC 5288 by fitting ellipses on the pixelated SDSS i-band  image of UGC 5288 where 1 pixel $ \approx 0.59012^{"}$}. Based on the fitting, an ellipse model is formed and the residual shows that the model represents the image very well. The peak in the ellipticity vs semi-major axis plot is marked with the magenta line; the orange line demarcates the beginning of the outer disk which has a position angle (PA) of $\sim 53\degree$ and is very different from the inner parts of the disk where PA$\sim 75\degree$. The green line marks the region where the PA of the bar changes sharply to the outer PA of the disk at $53\degree$. The yellow patch marks the region containing the peak ellipticity at 0.5 kpc (magenta line) and the beginning of the sharp decrease of position angle (green line) at 1.06 kpc.}
    \label{bar_fitting}
\end{figure*}

We searched through several studies of LSBGs in the literature and selected the dwarf LSBG UGC 5288, since it is an isolated galaxy and slightly inclined so that the measured rotation curve is reliable. It is a low luminosity galaxy in the Lynx Cancer void having a large HI disk \citep{Van.zee.2000,van.Zee.2004,van.Zee.2006, Kurapati.et.al.2018}, a small bulge (as seen from surface density distribution from WIRCAM data \citep{Fingerhut.2010}), and a low star formation rate (SFR) of $\sim 0.0063$ M\textsubscript{\(\odot\)} yr$^{-1}$ \citep{Werk.et.al.2011} which is confined to the central part of the stellar disk \citep{Van.zee.2000, van.Zee.2004}. 

The central surface brightness is $\sim 22$ magnitude/$arcsec^2$ in the B band \citep{Makarova.1999} i.e., $\sim 101$ L\textsubscript{\(\odot\)}/$\rm pc^{2}$ in physical units. So this galaxy is borderline LSB in nature and has been classified as an Sdm galaxy. The large HI disk extends out to $\sim 16.6$ kpc \citep{Kurapati.et.al.2020} and is completely devoid of any star formation activity \citep{van.Zee.2006}. The rotation curve has a flat rotation velocity of V$_{flat} \sim 72$ km s$^{-1}$ \citep{Kurapati.et.al.2018}. The galaxy also has a bar \citep{van.Zee.2004, Kurapati.et.al.2020}, which is very important for our study as it helps us
to identify the best fitting  stellar disk model as shown in Section 4.6. Our method is similar to previous studies that used bar potentials to determine the mass of the dark matter halo by modelling the non-circular streaming motion of gas flow in the bar \citep{weiner.etal.2001}. So we estimated the bar ellipticity ($\eta=1-b/a$, where $a$ and $b$ are the semi-major and semi-minor axes) and bar length of UGC 5288 (see next section).

\section{Modelling the disk and halo in UGC 5288} \label{sec:galaxy_modeling}
In this section, we model the isolated dwarf barred galaxy UGC 5288 with a multi-component disk containing stars and gas, and a dark matter halo. We aim to create galaxy models with all the observed properties of the disk and the HI rotation curve of UGC 5288.

\subsection{Determination of bar properties of UGC 5288} \label{obs_bar_prop}
To determine the bar ellipticity we use the ellipticity and position angle (PA) variation method previously used in multiple studies \citep{Zou.Shen.2014, Long.et.al.2014, Kataria.and.Das.2019, Fragkoudi.et.al.2021}. Figure (\ref{bar_fitting}) shows the ellipse fitting of the SDSS i-band image of UGC 5288 with the python module {\it photutils} \citep{larry_bradley_2020_4049061}. We derive the bar parameters by fitting ellipses and the residual plot in Figure (\ref{bar_fitting}) shows that the model matches very well with the image of the stellar disk. 
We note that the ellipticity change is more or less gradual up to the outer parts of the disk. The PA varies slowly in the inner disk, but changes sharply beyond $a_{obs}=1.06$ kpc, and attains a constant value of PA$\sim 53\degree$ in the outer disk. So, even if the bar ellipticity decreases, the PA does not show much variation until the point where $a_{obs}=1.06$ kpc, where there is a decrease in the peak ellipticity value by 37.6\%, from 0.72 to 0.45. So we take the range of possible values of ellipticity for the bar to be 0.72 to 0.45 (from the magenta line to the green line in Figure (\ref{bar_fitting})). The corresponding range for the semi-major axis of the bar is from $a_{obs}=0.5$ kpc to $1.06$ kpc. The intrinsic bar length and ellipticity can be derived from the observed values using the disk inclination $i$ and the difference in PA of the bar and the PA of the outer disk, $\alpha$. The relation between the observed bar length L$^{obs}_{b}$ and the intrinsic bar length L$_{b}$ can be expressed as \citep{Martin.1995,Gadotti.et.al.2007}, 
\begin{equation}
    L_{b}=L^{obs}_{b} \sqrt{\cos^{2}\alpha+ \sin^{2}\alpha \sec^{2}i}  \texttt{       .}
\end{equation}
The HI disk inclination is $\sim 38\degree$ \citep{Kurapati.et.al.2020}, (see Table (\ref{para5288})) but the stellar disk inclination is, $\cos^{-1}(b/a)=\cos^{-1}(1-\eta) = 43.28^{\degree}$, where the value of $\eta \sim  0.272$ is taken from the outer part of the disk from Figure (\ref{bar_fitting}).  There maybe two possibilities: (1) the disk is intrinsically asymmetric or, (2) there maybe an inclination between the stellar and gas disk. There have been multiple observations of asymmetric disks and they have also been found in cosmological simulations \citep{Lokas2021a,Lokas2021b}. Misalignment between gas and stellar disks has been observed for a large number of galaxies in the MaNGA survey and the corresponding misalignment in low redshift galaxies has been shown to be related to the halo spin at z=1 in the IllustrisTNG100 simulations \citep{Duckworth.et.al.2020}. Other possible reasons include the accretion of gas from the environment and flyby interactions with galaxies \citep{Starkenburg.et.al.2019}. 

UGC 5288 is an isolated void galaxy, so there are fewer chances of mass accretion from  neighbouring galaxies. Also, the gas disk is very regular and extended without any signs of flyby events in the recent past. The angle $\alpha=21\degree$, as seen in the PA-plot in the same Figure. So, considering the inclination of the stellar disk, the range of intrinsic semi-major axis lengths is 0.527 kpc $<a<$ 1.116 kpc. Again, it can be shown that the intrinsic ellipticity $\eta$ is related to the observed ellipticity $\eta_{obs}$ by the expression,
\begin{equation}
    \eta = 1- \left( 1-\eta_{obs} \right) \sqrt{\frac{\cos^{2}\alpha+ \sin^{2}\alpha \cos^{2}i}{\sin^{2}\alpha+ \cos^{2}\alpha \cos^{2}i}}    \texttt{       .}
\end{equation}
Hence, the observed range of ellipticity is intrinsically, $0.307<\eta<0.646$. Determining the bar ellipticity and length is very important as these parameters are used to narrow down to the best galaxy and halo model in UGC 5288, as described in Section 4.

\begin{table}
	\centering
	\caption{Parameters of UGC 5288}
	\label{para5288}
	\begin{tabular}{lccr} 
    \hline
 Parameter & Value \\
 \hline\hline
 RA, Dec & 09h51m17.00s, +07d49m39.0s \\ 
 Redshift & 0.00186 $\pm$0.00002 \\ 
 Distance (Mpc)  & 11.41 \\ 
 $\rm R_{25}$ (arcsecond) & 37 \\
 Apparent magnitude in K band & 11.49 $\pm$ 0.31 \\
 Apparent magnitude in SDSS g band & 20.08 $\pm$0.02 \\
 Apparent magnitude in SDSS r band & 19.76 $\pm$0.03 \\
 Scale length in I band (h$_{\rm I}$ in arcseconds) & 9.39 \\ 
Inclination (HI disk) & $38\degree$ \\
 \hline
	\end{tabular}
\end{table}

\subsection{Halo Model} \label{halo_models}
We model the dark matter halo of UGC 5288 with a spherically symmetric density profile. Following the classification of dwarf galaxies in \cite{Bullock.Boylan-Kolchin.2017}, UGC 5288 is a bright dwarf galaxy but can also be considered to be an LSBG. So we use both cuspy and flat core halo models, and then determine which one has a better match to observations of UGC 5288. The Hernquist \citep{Hernquist.1990} cuspy density profile is given by,
\beq \label{rhodm}
\rho_{dm}=\frac{M_{dm}}{2\pi} \frac{a}{r(r+a)^{3}} 	\texttt{  	,}
\eeq
where $a$ is the scale factor which can be related to the concentration $c$ of a corresponding NFW halo of mass $M(r_{200}) = M_{dm}$ as
\beq
a=\frac{r_{200}}{c}\sqrt{ 2\left[ \ln(1+c)-c/(1+c) \right]} \texttt{   .}
\eeq
Cosmological simulations indicate that Milky Way type galaxies and more massive galaxies have cuspy density profiles in their centres \citep{navarrow.frenk.white.1997}, whereas the rotation curves of LSBGs indicate that their dark matter profiles are more pseudo-isothermal in nature \citep{deblok.2010}. The pseudo-isothermal profile is given by,
\beq
\rho_{iso}=\frac{\rho_{0}}{1+\left( \frac{r}{r_{c}}\right)^{2}} 	\texttt{  	,}
\eeq
where, $r_{c}$ is the core radius and $\rho_{0}= M_{dm}/(4\pi r^{3}_{c}\left[ c_{iso} -\tan^{-1}(c_{iso}) \right])$ and $c_{iso}=r_{200}/r_{c}$. For simplicity, we use the density profile of an isothermal sphere $\rho(r)= v_{c}^{2}/4\pi G r^{2}$ and integrate it up to the virial radius $r_{200}$ to determine the virial mass $M_{200}$:
\beq \label{masstot1}
 M(r_{200})=\frac{ v^{2}_{c}  r_{200}}{ G} 	\texttt{  	,}
\eeq
where $v_{c}$ is the rotation velocity at the flat part of the rotation curve and $ r_{200}= v_{c}/10H(z)$.
Using the flat rotation velocity of the galaxy UGC 5288 the total mass $M(r_{200})$ is $1.23\times 10^{11}$ M\textsubscript{\(\odot\)}. In the subsequent sections, we show the rotation curve of UGC 5288, where the maximum velocity from the rotation curve, $V_{max}\sim V_{flat}$, the flat velocity of the rotation curve. For this particular galaxy, we can apply Equation \ref{masstot1} to calculate the total DM mass, but a more general method can be implemented for a general case. For example, \cite{Kurapati.et.al.2020} implement mass modelling of HI rotation curves and fit the circular velocity component of the dark matter halo with analytical expressions of circular velocity for different dark matter profiles.

\subsection{Disk Model}
The baryonic disk consists of stars and gas. We know from observations that the density profile of a stellar disk can be modelled with an exponential profile having scale length $R_{d}$ and with a $sech^{2}$ profile in the z-direction with scale length $z_{0}$; i.e.,
\begin{equation}\label{rhostars}
\rho_{\star}(r, z)=\frac{M_{\star}}{4\pi z_{0} R^{2}_{d}}  sech^{2}\left( \frac{z}{z_{0}} \right) \exp(-r/R_{d})   \texttt{		.}
\end{equation} 
So the stellar disk has a high surface density in the centre which falls exponentially with the radius. However, the gas disk does not usually follow such a profile and contributes significantly to the disk surface density only in the outer disk where the stellar surface density is low.  Since UGC 5288 contains a large HI disk \citep{Van.zee.2000,van.Zee.2004}, we model the gas disk in the radial direction with the function,
\begin{equation}\label{func_gas}
 \frac{1}{\left(1+\exp((r-r_{max})/r_{1})\right)} \texttt{       .}
\end{equation}
Here $r_{1}$ is a parameter that controls the rate of decrease of surface density at the outer edge of the HI disk; $r_{max}$ controls the extent of the HI disk. In the z-direction, the gas disk is modelled with a $sech^{2}$ function with a scale length z$_{g0}$. Thus the gas density profile can be expressed as
\begin{equation} \label{rho_g}
\rho_{g}(r, z)= \frac{M_{g}}{2\pi \bar{I}(\infty , r_{1}, r_{max})} \frac{sech^{2}\left(\frac{z}{z_{g0}}\right)}{2z_{g0}}\frac{1}{\left(1+\exp((r-r_{max})/r_{1})\right)}
\end{equation}
where, $\bar{I}(\infty, r_{1}, r_{max})$ is the normalization constant (for brief derivation, see Appendix \ref{normalisation}). We estimate the values of the scale lengths $r_{1}$ and $r_{max}$ from the HI rotation curve and find them to be: $r_{1}=1.8$ kpc and $r_{max}=11.0$ kpc. The detailed method will be explained in the next section once we have the mass of the gas disk. The gas disk scale length $z_{g0}$ is fixed using the condition that the FWHM of the $sech^{2}$ profile is 2 kpc, i.e., $2 z_{1/2}=1.76 z_{g0}= 2$ kpc. Hence, we obtain $z_{g0}= 1.13$ kpc.

\subsection{Disk Mass and Halo Mass}
We estimate the total stellar mass of the galaxy using the K-band luminosity. The mass-to-light ratio in K-band, $\frac{M}{L}|_{K}$ is used to calculate the total mass of the old stars, which constitutes the most massive stellar component in a galaxy. Using the $(B - V)$ colour, we calculate $\frac{M}{L}|_{K}$ using the empirical relation from \cite{Bell.deJong.2001}, $\log_{10}\left(\frac{M}{L}\right)_{K}=a_{K}+b_{K}\times(B-V)$; where, $a_{K}$ and $b_{K}$ are empirical parameters. Again, colour $(B-V)$ is related to the $(g-r)$ colour of the object as described in \cite{Jester.2005}, $B-V=0.98\times(g-r)+0.22$. Using SDSS g and r band magnitudes for this galaxy from the NASA/IPAC Extragalactic Database (NED), $\frac{M}{L} \rvert_{ K}$ comes out to be $0.54$. From the basic relations between apparent magnitude $m$, absolute magnitude $M$, distance and luminosity of an object, we have the total luminosity of UGC 5288 in K-band as $L_{ K}=L\textsubscript{\(\odot\)}_{ K} \times 6.70766 \times 10^{8}$. Finally, using the $\frac{M}{L} \big\rvert_{ K}$ ratio we get the total stellar mass in K-band to be $\sim 3.64 \times 10^{8}$ M\textsubscript{\(\odot\)}. 

The mass of the gas disk in our models is taken from \citet{Kurapati.et.al.2018,Kurapati.et.al.2020} where the HI mass $M_{g}\approx 1.2\times 10^{9}$ M\textsubscript{\(\odot\)}. We do not consider the molecular gas mass because it is not detected in UGC 5288 \citep{boker.etal.2003}. Thus, the baryonic mass fraction of the disk is $m_{d}=( M_{\star}+M_{g})/M(r_{200})= 0.01267$ and the halo dark matter mass is $M_{dm}=( M(r_{200})- M_{\star} - M_{g})=1.21\times 10^{11}$ M\textsubscript{\(\odot\)}.

\subsection{Estimation of stellar disk scale length}
We estimate the disk scale length $R_{d}$ for an exponential disk from the total luminosity of the disk, $L_{\lambda}={\int_{0}}^{\infty} 2\pi r I_{\lambda}(r) dr=2\pi {I_{0}}_{\lambda}{\int_{0}}^{\infty} r  exp(-r/R_{d}) dr=2\pi {I_{0}}_{\lambda}R^{2}_{d}$. Hence, the scale length $R_{d}=\sqrt{ L_{\lambda}/ 2\pi  {I_{0}}_{\lambda}}$. Using the central surface brightness of UGC 5288 in the $\rm K_{s}$ band to be 18.4 \citep{Fingerhut.2010} and M\textsubscript{\(\odot\)}$_{K}=3.27$ \citep{Willmer.2018}, the total luminosity in $\rm K_{s}$ band is L\textsubscript{\(\odot\)}$_{ \rm K_{s}}\Delta\lambda$= $6.71\times 10^{8}$ L\textsubscript{\(\odot\)}$_{ \rm K_{s}}$. Consequently, using the above relation we get $R_{d}=0.53$ kpc. This value is in agreement with the values reported by \cite{Makarova.1999} where they estimate the exponential scale length in I-band to be $9.39^{"} \approx 0.52$ kpc, with the distance to the galaxy at 11.41 Mpc \citep{Kurapati.et.al.2018}.

\subsection{Estimation of gas disk scale lengths}
The gas disk scale lengths $r_{1}$ and $r_{max}$ that we introduce in our modelling in Equation (\ref{func_gas}) are constrained by the circular velocity of the gas disk as previously derived and shown in Figure 2 in \cite{Kurapati.et.al.2020}. We use a qualitative comparison between our model and the derived circular velocity of the gas disk to determine the scale lengths $r_{1}$ and $r_{max}$ for our model. The circular velocity for the gas disk can, in principle, be derived from the surface density using the concept of estimating disk potentials from homoeoids with the thin disk approximation as shown in Section (2.6.1) in \cite{Binney.Tremaine.2008}.
\begin{equation}
    v^{2}_{c, gas}(r)= r\frac{\partial \Phi}{\partial r}= -4 G {\int^{r}_{0}} \frac{a da}{\sqrt{r^{2}-a^{2}}} \frac{d}{da} {\int^{\infty}_{a}} \frac{\Sigma_{gas} (r^{'}) r^{'} dr^{'}}{\sqrt{{r^{'}}^{2}-a^{2}}}
\end{equation}
where $\Phi$ is the gas disk potential and the gas surface density profile $\Sigma_{gas}(r)= {\int^{\infty}_{-\infty}} \rho_{g}( r, z) dz$ is evaluated from Equation (\ref{rho_g}). The evaluation of the above integral is presented in Appendix (\ref{circ_vel_gas_disk}). We compare the above analytical expression for circular velocity and the corresponding circular velocity curve (in Figure (2) in \cite{Kurapati.et.al.2020}) and found a good match for r$\geqslant 5$ kpc for the values of $r_{1}=1.8$ kpc and $r_{max}=11.0$ kpc ( see in Figure(\ref{vc_gas_comp}) in Appendix \ref{circ_vel_gas_disk}). To match with the inner part ($\leqslant 5$ kpc) of the curve a more complex gas disk modelling is needed, and the dark matter halo should be modelled similarly to that in \cite{Kurapati.et.al.2020}. But we do not aim for such detailed modelling as the initial profile is also subject to change as we evolve the system for a few Gyrs.

\begin{table} 
	\centering
	\caption{Structural and dynamical parameters of UGC 5288 used in the model}
		\label{table:ugc5288-model-parameters}
	\begin{tabular}{lccr} 
		 \hline
  Model parameters & observed value \\
 \hline\hline
  flat rotation velocity $V_{flat}$ (km s$^{-1}$) & 72 \\ 
  Halo mass $M_{dm}$ ($10^{11}$ M\textsubscript{\(\odot\)}) & 1.22 \\
  Stellar mass $M_{\star}$ ($10^{8}$ M\textsubscript{\(\odot\)}) & 3.64 \\
  Gas disk mass $M_{g}$ ($10^{9}$ M\textsubscript{\(\odot\)}) & 1.2 \\
  Stellar disk scale length $R_{d}$ (kpc) & 0.53 \\
  Gas disk scale length $r_{1}$ (kpc) & 1.8 \\
  Gas disk scale length $r_{max}$ (kpc) & 11.0 \\
  Gas disk scale height $z_{g0}$ (kpc) & 1.13 \\
 \hline
\end{tabular}
\end{table}
\subsection{Velocity structure of different components} \label{vel_structure}
We determine the initial velocity structure of the halo, stellar disk and gas disk following the formalism in the code GalIC \citep{Yurin.Springel.2014}. We consider the most general velocity structure for the stellar and gas disk with the third integral of motion in the distribution function $f( E, J_{z}, I_{3})$. We take the combined density of the stellar and gas disk $\rho_{\star}(r, z)+ \rho_{g}(r, z)$ in the Jeans equation. We aim to match the observed rotation curve of UGC 5288 with the rotation curves from our simulated galaxy models. Observations show the flat rotation velocity is $\sim 72$ km s$^{-1}$ for radii $>4$ kpc (see Figure (2) in \cite{Kurapati.et.al.2020}) and the velocity dispersion $\sigma \sim 9$ km s$^{-1}$ (see Table (1) in \cite{Kurapati.et.al.2020}).  Additionally, we know from observations of Milky Way-type galaxies, the dispersion in the plane of the galaxy disk is twice the dispersion perpendicular to the plane, i.e., $\sigma_{\phi}=\sigma_{R} =2\sigma_{z}$ \citep{Binney.Tremaine.2008}. The same relation may not hold for dwarf galaxies in general. But for simplicity, we assume this relation to hold for the stellar disk in our models. 
Thus, we fix $\sigma_{\phi}=\sigma_{R} =2\sigma_{z}$ for the disk in the initial models.  We also test our models with an initial isotropic velocity dispersion distribution and find it to converge with the results of the simulations with anisotropic dispersion within $\sim 1$ Gyr. Our final results remain unchanged and independent of the choice of stellar disk dispersion (see Appendix \ref{appendix:isotropic-vel-dispersion}).

For the halo, we consider the general velocity structure with the distribution function to be $f(E, J_{z}, I_{3})$ \citep{Yurin.Springel.2014}. The velocity structure of the halo determines the angular momentum at different radii and hence the spin of the halo. One way to vary the angular momentum is to change the streaming velocity $\langle v_{\phi}\rangle$ in Equation (37) in \cite{Yurin.Springel.2014} by changing the 'k-parameter'. The maximum value of the 'k-parameter', $k_{max}$ is fixed and depends on the dispersion $\sigma_{R}$ and $\langle v^{2}_{\phi}\rangle$. As $\langle v^{2}_{\phi}\rangle$ is fixed for a given background potential, density distribution and $\sigma_{R}$, we can increase $k_{max}$ only by varying the $\sigma_{R}/\sigma_{z}$ ratio. Hence, we are able to generate dark matter halos with high angular momentum and higher spin.

\subsection{Angular momentum and spin parameter $\lambda$} \label{diskang}
The total angular momentum of the disk ($J_{d}$), which mainly consists of the z-component angular momentum of the stellar disk ($J_{d,\star}$) and gas disk ($J_{d, gas}$), is related to the total angular momentum of the dark matter halo ($J$) by, 
\begin{equation} \label{Jdh}
 J_{d}= J_{d,\star}+ J_{d, gas}= j_{d} \times J \texttt{		,}
\end{equation}
where $j_{d}$ is the fraction of the angular momentum of the halo contained in the disk. If we consider the conservation of specific angular momentum of the gas during disk formation, $j_{d}= (M_{\star} + M_{g})/M(r_{200})$ \citep{Mo.et.al.1998}. The ratio of disk and halo masses can give estimates of $j_{d}$ only during the initial period of formation of the halo-disk system. As the galaxy evolves, it may undergo merger events, accretion of gas, mass in-flow and out-flow  during stellar feedback.  The $j_{d}$ of a galaxy encapsulates all the physical phenomena that can potentially change the angular momentum of the galaxy throughout its evolution. As a first approximation, from the above ratio of masses, $j_{d} \approx 0.012$ for the following theoretical calculations. $J_{d}$ can be expressed in terms of the overall properties of the disk. For the two disk components, the angular momentum is expressed as,
\begin{equation} \label{angularmom}
 J_{d, \star (gas)}= {\int^{\infty}_{0}} \Sigma_{\star (gas)}(r) 2\pi r^{2} v_{c}(r) dr 
\end{equation}
where, $v_{c}(r)$ is the circular velocity of the system at radius $r$ on the plane of the disk. It is expressed as,
\begin{equation}
\begin{aligned}
 V^{2}_{c}(r)=  \frac{ G M_{dm}}{r(1+a/r)^{2}} + \frac{2 G M_{\star}}{R_{d}} y^{2} \left[ I_{0}(y)K_{0}(y) - I_{1}(y)K_{1}(y) \right] \\
 -4 G {\int^{r}_{0}} \frac{a da}{\sqrt{r^{2}-a^{2}}} \frac{d}{da} {\int^{\infty}_{a}} \frac{\Sigma_{gas} (r^{'}) r^{'} dr^{'}}{\sqrt{{r^{'}}^{2}-a^{2}}} \texttt{        ,}
\end{aligned}
\end{equation}
where the first term is due to the dark matter halo, the second term is due to the stellar disk and the third term is for the gas disk with surface density profile as explained in the previous section (also see \cite{Binney.Tremaine.2008}); $I_{0}(y)$ \& $K_{0}(y)$ are the Bessel functions and $y=r/2R_{d}$. 

The angular momentum of the halo $J$ is represented in terms of a dimensionless parameter $\lambda$ \citep{Peebles1969} having an explicit dependence on halo mass $ M$ and energy $ E$ \citep{Fall.Efstathiou.1980} as introduced previously in Equation (\ref{Peebles_spin3}). 
It is difficult to estimate the total energy of a halo, but considering a virialized halo with the halo particles following circular orbits, we can derive an expression of the total energy at the virial radius $r_{200}$. \cite{Mo.et.al.1998} and \cite{Springel.White.1999} showed that for a truncated isothermal halo the total energy is given by $-G M(r_{200})/2r_{200}$ and for an NFW halo it is $-\left(G M(r_{200})/2r_{200} \right) f_{c}(c)$, where $c$ is the NFW concentration parameter and
\begin{equation*}
    f_{c}(c)=\frac{c \left[ 1-1/(1+c)^{2}-2 \ln(1+c)/(1+c) \right]  }{  2\left[ \ln(1+c) - c/(1+c) \right]^{2}} \texttt{      .}
\end{equation*}
Similarly, using the same assumptions we find the total energy for a Hernquist halo, 
\begin{equation}
E=-E_{KE}=-\frac{G M^{2}_{dm}}{2r_{200}} \tilde{f}_{\tilde{c}}(\tilde{c})
\end{equation}
where $\tilde{c}= r_{200}/a = c/\sqrt{2 \left( \ln(1+c)- c/(1+c)\right)} $ and
\begin{equation*}
    \tilde{f}_{\tilde{c}}(\tilde{c}) = \tilde{c} \left( \left( 1+ \tilde{c} \right)^{4} -6\left( 1+ \tilde{c} \right)^{2}+ 8\left( 1+ \tilde{c} \right)-3 \right)/6\left( 1+ \tilde{c} \right)^{4}   \texttt{       .}
\end{equation*}
Once the total energy is known we can relate the halo angular momentum and disk angular momentum through Equation (\ref{Jdh}), and we can express the halo spin for the Hernquist halo as
\begin{equation}\label{LambdaH}
    \lambda_{H} = \frac{J_{d} M^{-3/2}_{dm} }{j_{d}\sqrt{G}} \left( \frac{1+\tilde{c}}{\tilde{c}} \right)^{5} \sqrt{ \frac{ \tilde{f}_{\tilde{c}}(\tilde{c}) }{2 r_{200}}} \texttt{      .}
\end{equation}
Similarly for NFW halo the spin at r$_{200}$ is given as
\begin{equation*}
   \lambda_{NFW} = \frac{J_{d} M^{-3/2}(r_{200})}{j_{d}\sqrt{G}} \sqrt{\frac{f_{c}(c)}{2r_{200}}}
\end{equation*}
\citep{Mo.et.al.1998, Springel.White.1999, Yurin.Springel.2014} and for the truncated isothermal profile the spin is the \cite{Bullock.et.al.2001} spin parameter
\begin{equation*}
    \lambda_{B}= \frac{J}{\sqrt{2} M v_{c} r_{200}}      \texttt{       .}
\end{equation*}
 
\begin{table}
	\centering
	\caption{Spin values for different dark matter profiles using the physical parameters of UGC 5288}
	\label{spin_values}
	\begin{tabular}{lccr} 
		 \hline
 c & $\lambda_{\rm NFW}$ & $\lambda_{\rm H}$ & $\lambda_{B}$ \\
 \hline\hline
 6 & 0.0308 & 0.08152 & 0.03625 \\ 
 8 & 0.0362 & 0.08752 & 0.041139 \\ 
 10 & 0.0412 & 0.0937 & 0.04539 \\ 
 \hline
 \end{tabular}
\end{table}
The estimates of the above three initial spins for UGC 5288 are shown in Table (\ref{spin_values}). Also, the analytical expression for the spin parameter of the pseudo-isothermal halo is provided in Appendix (\ref{appendix_Iso_halo}).  All these expressions show the connection between halo properties (i.e., halo mass $M_{dm}$, spin $\lambda_{H}$(or $\lambda_{iso}$) and concentration $\tilde{c}$ (or core radius $r_{c}$)) and disk properties ($R_{d}$, $r_{max}$, $r_{1}$, $M_{\star}$ and $M_{g}$). 
\begin{figure*}
\centering
	\includegraphics[width=\linewidth]{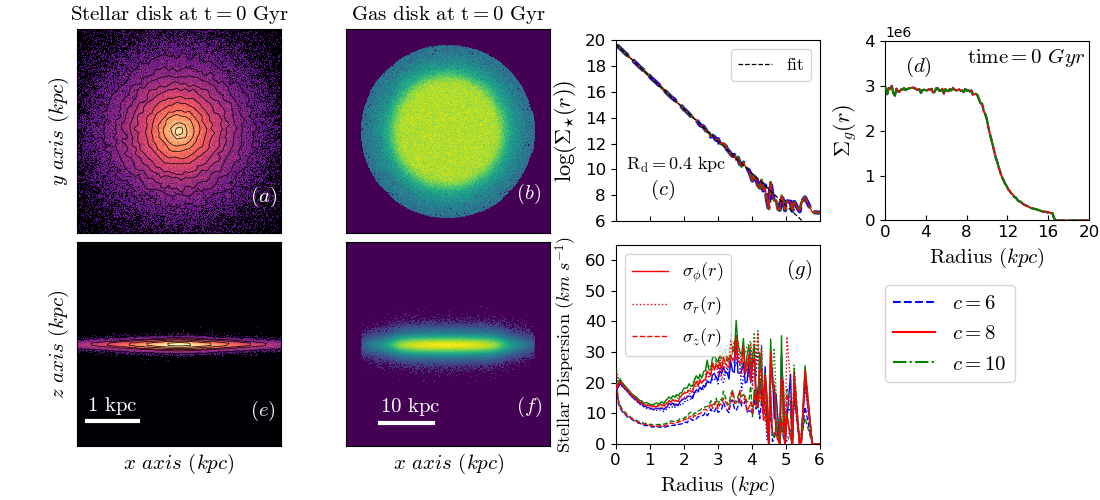}
    \caption{ {\bf Initial configuration of the stellar disk and gas disk}. In the left half of the figure, one of the initial models with $c=8$ and $z_{0}/R_{d}=0.2$ is shown; (a), (b), (e) and (f) are the log-histogram distributions of the stellar disk and gas disk in the face-on and edge-on orientations respectively. Models with $c=6$ and $c=10$ have similar disk structures as evident from the panels (c), (d) and (g). It shows the stellar disk and gas disk properties for models with three different concentrations: $c=6$ (Blue), 8 (Red) and 10 (Green) at $t= 0$ Gyr; (c) shows the initial stellar surface density $\Sigma_{\star}(r)$ for the three models and the fitted curve with disk scale length $R_{d}=0.4$ kpc, (d) shows the gas surface density $\Sigma_{g}(r)$, (g) shows the stellar velocity dispersion with $\sigma_{\phi}$ highlighted by the solid line, $\sigma_{r}$ by the dotted line and $\sigma_{z}$ by the dashed lines. For the stellar disk, $\sigma_{r}=2 \sigma_{z}$.}
    \label{initial_disk}
\end{figure*}
Equation (\ref{LambdaH}) (Equation (\ref{lambda_iso}))  gives the theoretical relation between concentration c (core radius $r_{c}$) and spin $\lambda$. Other than these two parameters our disk galaxy models are dependent on another quantity, the stellar disk thickness $z_{0}$, which determines the stellar velocity dispersion and hence controls the formation of a bar in the stellar disk. Larger values of stellar disk scale height $z_{0}$ increase dispersion and suppresses bar formation, but decrease the central density in the disk and hence the phase space density and lead to the formation of large bars that cover the whole stellar disk \citep{Klypin.et.al.2009}. All three quantities ($c$, $\lambda$ and $z_{0}$) depend on the formation history of the galaxy and determine disk-halo interactions. So we use these parameters to construct our model galaxies, and then use the disk thickness and bar properties to zero in on the best-fitting model.

\section{Numerical methods and Simulations} \label{sec:simulations}
In this section, we present the steps for constructing stable models of UGC 5288 using the structural and dynamical parameters from Table \ref{table:ugc5288-model-parameters}. Since this is a pilot study we explored models with both the Hernquist halo profile and the pseudo-isothermal halo profile.

\subsection{Initial conditions of model galaxies}
We use the N-body code GalIC \citep{Yurin.Springel.2014} to generate the initial galaxy. The GalIC code uses techniques similar to the orbit-based method introduced by Schwarzchild (1979). Spherical and axisymmetric density profiles for the galaxy halo and the stellar disk are fixed according to Equation (\ref{rhodm}) and (\ref{rhostars}) respectively. We modified the GalIC code in two major ways. First, we included gas particles, so we modified the code to include a gas disk profile following Equation (\ref{rho_g}) and estimated the probability distribution function that generates position coordinates following the gas profile. This is explained in detail in  Appendix (\ref{gas_pdf}). The N-body particles that follow the gas profile structure are treated as star particles as there is no provision for treating them as SPH particles in GalIC. But they are finally evolved as SPH-particles using the galaxy evolution code GADGET2 \citep{Springel.Volker.2005}. Thus, we constructed two separate disks - a stellar and a gas disk by tagging the gas particles; this required further changes in the code. 

Secondly, we changed the code to include the pseudo-isothermal halo profile. A point to note is that GalIC does not generate a dark matter halo with the same halo spin as the input halo spin parameter. So we had to impart angular momentum and hence spin by changing the streaming velocity $\langle v_{\phi}\rangle$ of the DM halo to ensure that our initial halo models have a specified initial halo spin at the radius of $r_{200}$. Once the initial conditions are generated by GalIC, we provide only streaming velocities to the gas particles and evolve the system with GADGET2. 

In our models, we fixed the number of disk particles to be 10$^{6}$. The Hernquist halo has 10$^{6}$ DM particles and the pseudo-isothermal halo has $3\times 10^{6}$ particles. We fix the number of stellar and gas particles according to the stellar mass and gas mass ratios in the disk, i.e., $M_{\star}/( M_{\star}+M_{g}) \times 10^{6} \approx 232737$ stellar disk particle and the rest 767263 number of gas particles. The mass resolution for the disk particles is 1099 h$^{-1}$ M\textsubscript{\(\odot\)}, for dark matter particles of Hernquist halo it is 85664 h$^{-1}$ M\textsubscript{\(\odot\)} and for the pseudo-isothermal halo, it is 90515 h$^{-1}$ M\textsubscript{\(\odot\)}. We also ran some of our models at higher resolution, with $6\times 10^{6}$ DM particles having particle mass 14277 h$^{-1}$ M\textsubscript{\(\odot\)}. The softening for disk particles and dark matter particles is 30 h$^{-1}$ pc and 50 h$^{-1}$ pc respectively, and the force accuracy f$_{\rm acc}$ or $\rm ErrTolForceAcc$ in GADGET2 is fixed at 0.005.

\subsection{Introducing the parameter space for all models}
We fix the model parameters of UGC 5288 from the observations as shown in Table (\ref{table:ugc5288-model-parameters}). The three quantities in our model that are not determined from observations are the halo spin $\lambda$, concentration $c$ for the Hernquist profile (or, core radius $r_{c}$ for the pseudo-isothermal halo) and disk scale height $z_{0}$. These three form our initial parameter space. We estimate the halo properties in two stages. First, we constrain the halo concentration by comparing the gas rotation curves from simulated models to the observed HI rotation curve of UGC 5288. Once the halo concentration is fixed, in the second stage we explore the parameter space of halo spin and disk scale height ($\lambda - z_{0}$ parameter space) and match the barred disk property namely bar length and ellipticity of our models with the estimates from observations of UGC 5288. What we have at the end is a probable halo spin profile for the galaxy UGC 5288.

One may wonder if we could as well estimate the halo spin theoretically using the spin expressions for the Hernquist halo (Equation (\ref{LambdaH})) and the pseudo-isothermal halo (Appendix (\ref{appendix_Iso_halo})). But these expressions are derived assuming some very special criteria like circular orbits of all the halo particles and virial equilibrium. These conditions may not hold true for all cases and at all radii. So we vary the halo spin within a range of values and check which of the models leads to the creation of a stellar bar with the same properties (see Section (\ref{obs_bar_prop})) as the bar in UGC 5288. 

In the first step, we set the initial conditions of the baryonic disk as follows. Figure (\ref{initial_disk}) shows the initial stellar and gas disk, and their properties before evolution. We want the evolved disk properties to match the observed disk properties of UGC 5288. So, the initial stellar disk scale length is fixed at $R_{d}=0.4$ kpc, smaller than the observed disk scale length $\sim 0.53$ kpc of UGC 5288, as the disk spreads during evolution. Comparatively, the gas disk does not spread much. The flat rotation velocity $V_{flat}$ of the gas disk varies between 60 km s$^{-1}$ to 78 km s$^{-1}$ for $c=6$ to $c=10$. Figure (\ref{initial_disk}) shows the stellar and the gas disk for one of the models with $c=8$, $\lambda_{\rm H}=0.0875$ and $z_{0}/R_{d}=0.2$.
\begin{figure}
\centering
	\includegraphics[width=\columnwidth]{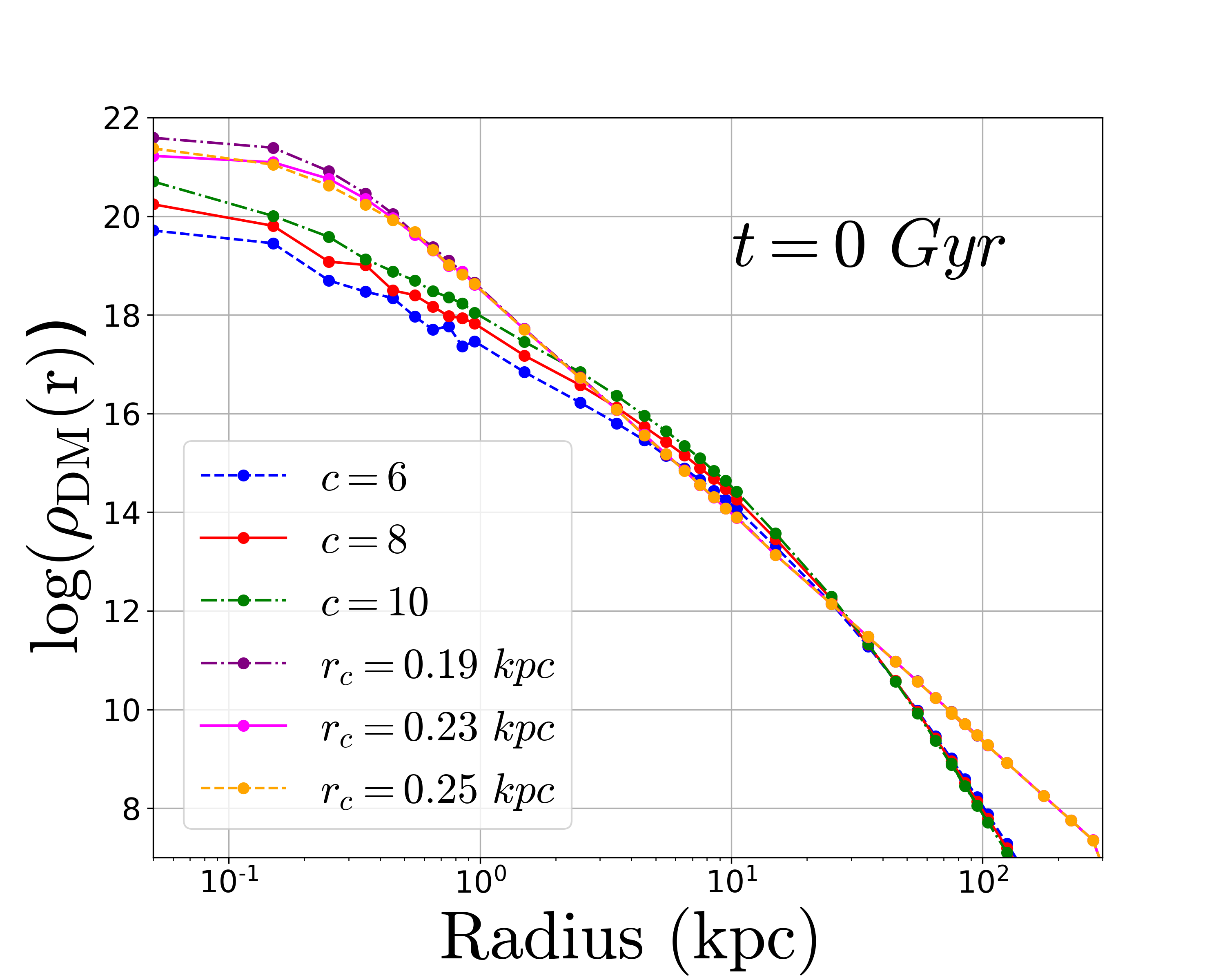}
	\includegraphics[width=\columnwidth]{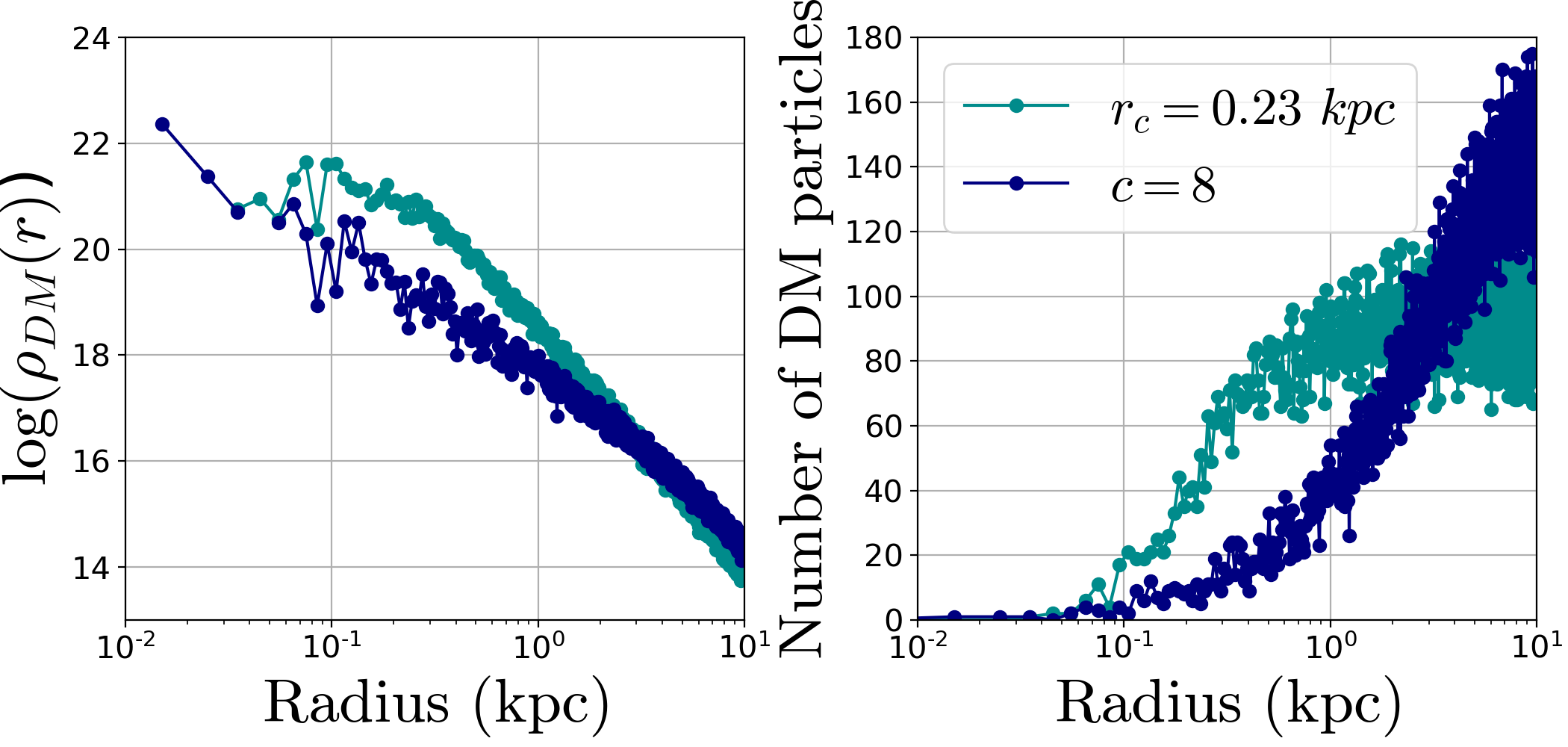}
    \caption{ {\bf The initial dark matter halo profiles for the Hernquist halo and the pseudo-isothermal halo}. The three different models for the Hernquist profile have halo concentration, $c=6$ (Blue), 8 (Red) and 10 (Green) at $t= 0$ Gyr. The corresponding halo core radius $r_{c}=0.19$ kpc (Purple), 0.23 kpc (Magenta) and 0.25 kpc (Orange) (corresponding halo concentrations are $c_{iso}=541$ (Purple), 447 (Magenta) and 411 (Orange)). The same colour scheme is used throughout the article except for the second figure here, which shows the difference between the central DM density distribution and particle number in bins of size $\Delta r=10$ pc for the two halo profiles. }
    \label{halo_profiles}
\end{figure}

\begin{figure}
\centering
\includegraphics[width=\columnwidth]{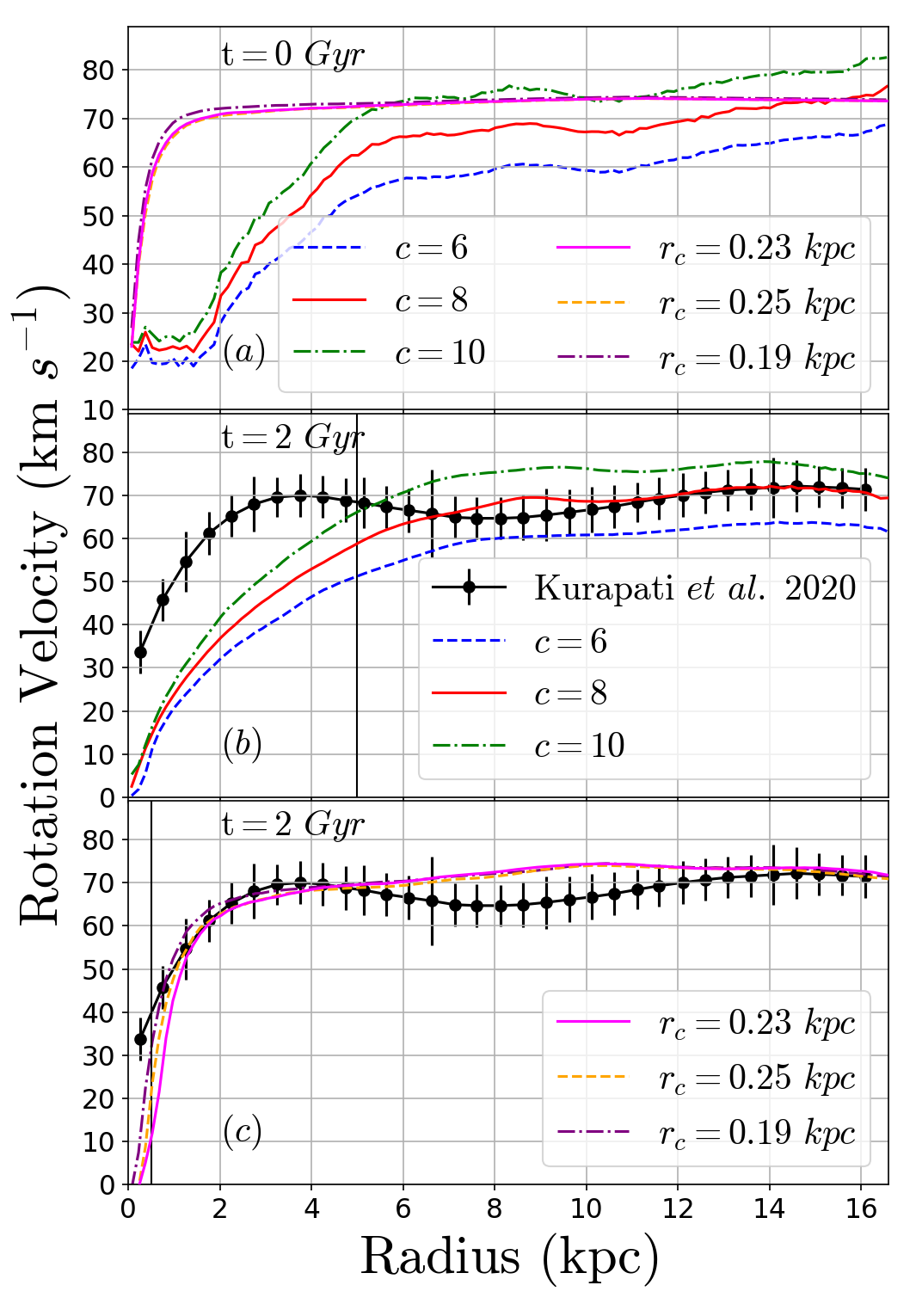}
\caption{{\bf The observed HI rotation curve is used to constrain the DM halo density profile}. This figure shows the initial and the evolved rotation curves for Hernquist and pseudo-isothermal halo profiles and compares them with the observed rotation curve of UGC 5288 (black curve with error bars) \citep{Kurapati.et.al.2020}. Figure (a) shows the initial rotation curve for models with Hernquist profile and with pseudo-isothermal profile, for three different halo concentrations ($c=6$, 8 and 10) and core radii ($r_{c}=0.23$ kpc, $r_{c}=0.25$ kpc and $r_{c}=0.19$ kpc). Figures (b) and (c) compare the final rotation curves after 2 Gyrs evolution for both halo models with observed rotation curves. The black vertical line indicates the radius (for Hernquist, $r=5$ kpc and for pseudo-isothermal $r=0.5$ kpc) above which $\chi^{2}$ is estimated. To constrain the halo concentration we did a $\chi^{2}$ fitting of the observed rotation curve with our models as shown in table (\ref{chi_square}) and (\ref{chi_square_iso}). }
\label{rotation_curve}
\end{figure}

The initial halo properties of the Hernquist halo and pseudo-isothermal halo for three halo concentrations $c$ and core radii $r_{c}$ are shown in Figure (\ref{halo_profiles}). The HI rotation curve of a galaxy changes with different halo concentrations or core radii. The initial gas rotation curves for the two halo profiles are shown in panel (a) of Figure (\ref{rotation_curve}). For the Hernquist halo profile, we consider the velocity structure of the gas disk as generated by GalIC. The mean $v_{\phi}$ matches closely with the circular velocity $v_{c}$, but for the pseudo-isothermal halo, we observe that the velocity structure generated by GalIC does not match with the expected circular velocity curve. So for the pseudo-isothermal halo, we fix the circular velocity as the initial mean $v_{\phi}$. Once we evolve the initial models for 2 Gyrs the final rotation curves of the stable models are seen in panels (b) and (c) of the same Figure. 

\begin{figure*}
\centering
\includegraphics[width=\textwidth]{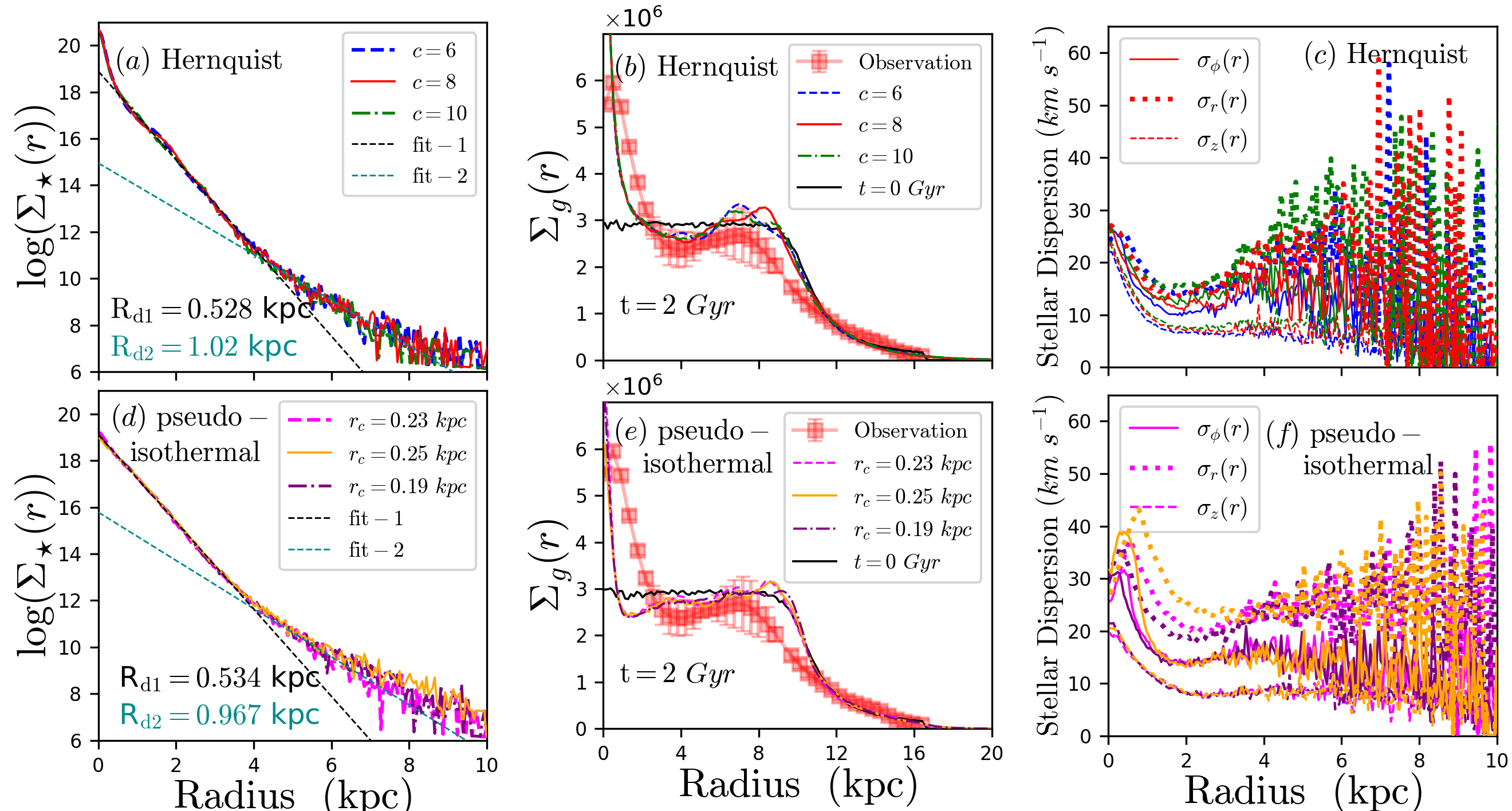}
    \caption{ {\bf Evolution of the properties of the galaxy disk over 2 Gyrs}, for models with three halo concentration parameters ($c=6$, 8 and 10 for Hernquist halo in (a), (b) and (c)), and three core radii ($r_{c}=0.23$ kpc, 0.25 kpc, 0.19 kpc for the pseudo-isothermal halo in (d), (e) and (f)). Panel (a) shows the surface density of the stellar disk with two disk scale lengths, one at the centre with $\rm R_{d1}=0.528$ kpc and the other at the outer edge of the disk with $\rm R_{d2}=1.02$ kpc.  $\rm  R_{d1}$ matches with observed disk scale length (see Table \ref{table:ugc5288-model-parameters} ). Panel (b) shows the evolved gas disk surface density with gas accumulation in the central region, along with the observed surface density \citep{Kurapati.et.al.2020}. Panel (c) shows the stellar dispersion. Panels (d), (e) and (f) are the corresponding figures for the pseudo-isothermal halo.  All the evolved disk properties match (roughly) the observed structural and dynamical properties of UGC 5288. This shows that our models have similar disk properties even though the underlying dark matter properties (concentration or core radius) are different. The degeneracy is broken with the gas rotation curve in section \ref{c_constrain}. }
   \label{evolution_model_properties}
\end{figure*}

In Table (\ref{spin_values}) we present the initial halo spin for a Hernquist halo ($\lambda_{H}$) and for a pseudo-isothermal halo ($\lambda_{iso}$) at $r_{200}$. To understand the structural and dynamic properties of the halo we have to study the halo spin at different radii. To fix the initial spin of the dark matter halos we changed the streaming motion of halo particles as explained previously in Section (\ref{vel_structure}). We explored a range of parameter values of $f_{R}$ (i.e., $\sigma_{R}/\sigma_{z}$) and the 'k-parameter', and found that different combinations can give the same spin value at $r_{200}$. We chose two values of $f_{R}=1.0$ \& $0.75$ as initial parameters in GalIC and for different choices of k-parameter, we generated different spin values. For more details on the range of choices of k-parameter for different halo spin $\lambda$ at virial radius $r_{200}$ see Table (\ref{kparameter_fr_spin}) in Appendix \ref{appendix:kparameter}).

One important point to note here is that it is not possible to attain a very high value of spin ($\sim 0.08$) in our models using isotropic velocity distribution in GalIC. This is because by keeping $f_{R}=1$ as the maximum value of the k-parameter, we obtain an upper limit to the spin value that is much less than 0.08. Thus, in our models, it is necessary to consider an initial anisotropy in $\sigma_{R}/\sigma_{z}$ ratio to attain a larger spin at $r_{200}$.

\subsection{Galaxy evolution} \label{gal_evo}
Once we fix all the initial conditions according to the observed properties of the baryonic disk and construct the models using GalIC, we check their stability by further evolving them in isolation for 2 Gyrs, which is about twice the dynamical time scale for the system.  We expect the models to become stable within 2 Gyr and attain disk properties similar to that of UGC 5288. 

\begin{figure}
\centering
	\includegraphics[width=\columnwidth]{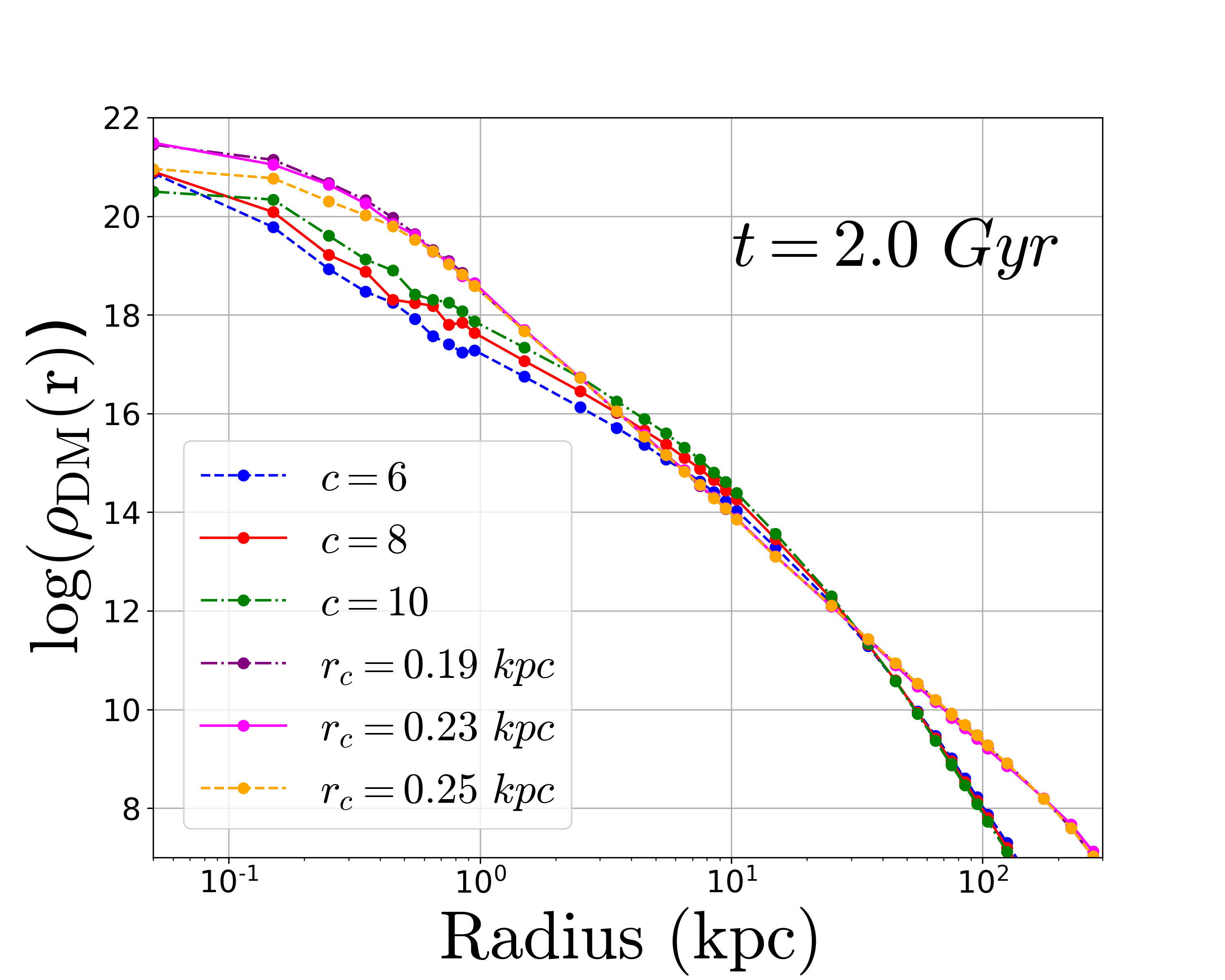}
    \caption{ {\bf The final dark matter halo profiles for the Hernquist halo and the pseudo-isothermal halo}. The three different models for the Hernquist profile have initial halo concentration, $c=6$, $8$ \& $10$ at $t= 2.0$ Gyr. The corresponding halo concentrations (core radius) for the pseudo-isothermal halo are $c_{iso}=447$ ($r_{c}=0.23$ kpc), $411$ ($r_{c}=0.19$ kpc) \& $541$ ($r_{c}=0.19$ kpc). The DM halo density changes only slightly at a radius $r < 1$ kpc. So the halo concentrations are maintained throughout evolution. }
    \label{halo_profiles_final}
\end{figure}

Figure (\ref{evolution_model_properties}) shows the evolved stellar disk and gas disk surface densities and stellar dispersion for the models with Hernquist and pseudo-isothermal halos. All structural properties are maintained within a few \% for 2 Gyrs of evolution and the galaxy models attain equilibrium by the end of evolution. Figure (\ref{rotation_curve}) shows the evolved rotation curves for the two kinds of halo profiles shown in Figure (\ref{halo_profiles_final}) for different halo concentrations and core radii. 
For the Hernquist halo, the stellar disk spreads from an initial disk scale length of $R_{d}\sim$ 0.4 kpc to 0.528 kpc in the inner regions of the disk, which is close to the observed scale length of 0.53 kpc of UGC 5288. For the pseudo-isothermal halo the inner disk hardly spreads and the final disk scale length is 0.534, which is very similar to the initial length. The outer part of the stellar disk spreads for both profiles with final disk scale length $\sim 1.02$ kpc and $\sim 0.96$ kpc respectively as shown in panels $(a)$ and $(d)$ in the same Figure. The inner part of stellar disk develops a bar in some of our models with a small bulge component that is seen in the stellar and gas surface density panel (a) in Figure (\ref{evolution_model_properties}). The stellar disk dispersion is maintained. The fluctuations in the outer regions of the stellar disk are due to the very low number of star particles at the outskirts. The dark matter density profile does not vary much with evolution as seen from Figure (\ref{halo_profiles_final}), and hence we can take the concentration to be nearly constant throughout the evolution. 

All the above structural and dynamical properties of the baryonic disk match with the observed surface densities, dispersion, and disk scale lengths of UGC 5288 \citep{Van.zee.2000, van.Zee.2004, van.Zee.2006, Werk.et.al.2011, Fingerhut.2010, Kurapati.et.al.2018, Kurapati.et.al.2020}, except the rotation curves of our models which differ for different concentrations of the dark matter halo. The value of $V_{flat}$ and the nature of the rotation curve depends on the dark matter halo density profile. So we constrain the halo concentration and core radius of the Hernquist and pseudo-isothermal halos by comparing the simulated and observed HI rotation curve.

\subsection{Constraining halo concentration}\label{c_constrain}
The simulated gas rotation curve depends on the halo concentration or core radius of the Hernquist and pseudo-isothermal halos. In panels $(b)$ and $(c)$ of Figure (\ref{rotation_curve}), we present the rotation curves of our models at $t=2$ Gyr, for three different concentrations, $c = 6$, 8 and 10 and core radii $r_{c}=0.23$ kpc, $0.25$ kpc and $0.19$ kpc, and compare with the observed rotation curve from \cite{Kurapati.et.al.2020}. Halos with low concentration values have lower flat rotation velocities compared to halos with higher concentrations.
This difference is not prominent for haloes with different core radii as seen in the initial ($t=0$ Gyr) rotation curves in panel (a). 
For the models with Hernquist halo, the rotation curve gradually increases and reaches a constant value at $\sim 6$ kpc. While in the observed rotation curve, there is a steep rise in rotation velocity in the inner regions and the $V_{flat}$ is reached by the radius of 4 kpcs. We observe in panel (c) in Figure (\ref{rotation_curve}) that the rotation curve of the pseudo-isothermal halo profile is a better match with the observed rotation curve than the models with the Hernquist DM profile.

Additionally, we compared the flat portion of the model rotation curves for the Hernquist halo beyond a 5 kpc radius for different halo concentrations, with the flat portion of the observed rotation curve. Along with the models $c=6$, 8 \& 10 we checked the $\chi^{2}$ for two other models $c=7.5$ and $c=8.5$. We find that $\chi^{2}$ is a minimum for the model $c=8$ (see Table \ref{chi_square}). Similarly, we found the minimum $\chi^{2}$ for the pseudo-isothermal halo with $r_{c}=0.23$ kpc (see Table (\ref{chi_square_iso})). This value is within the uncertainty of the measured value $r_{c}=0.25 \pm 0.05$ kpc in \cite{Kurapati.et.al.2020}. So, henceforth we fixed the concentration parameter at $c=8$ for the Hernquist halo and the core radius at $r_{c}=0.23$ for the pseudo-isothermal halo respectively.

\subsection{Halo spin and disk scale height}
In this section, we match the stellar bar properties in our models to the bar properties of UGC 5288. Once the halo concentration or core radius is constrained using the HI rotation curve, we need to form a disk in our models where the bar properties match with that of the bar in UGC 5288. We explore a range of values for the initial free parameters-- halo spin $\lambda$ and disk scale height $z_{0}$, in our simulations. We examine the bars formed in the different models and compare the resulting bars' properties with the observed bar in UGC 5288. For the Hernquist halo, we used halo spin values starting from 0.0362 (similar to $\lambda_{NFW}$ for $c=8$ in Table \ref{spin_values}), intermediate values of 0.0463 (similar to $\lambda_{B}$ values in Table \ref{spin_values}), and higher values $\sim 0.0874$ (similar to $\lambda_{H}$ for $c=8$; see Table \ref{spin_values} for different initial spin values), all at the virial radius $r_{200}$. Within the range of values, the theoretically expected value of halo spin is the Hernquist spin value of $\lambda_{H}=0.087$.

Table (\ref{models_c8}) shows the initial parameter space for simulated galaxy models for the Hernquist halo and the pseudo-isothermal halo, within a range of initial halo spin values at r$_{200}$ and for a few values of disk scale heights. Many of the models form a bar in the simulations but as we show in the following sections, only in few of them have bar properties similar to UGC 5288. We find that for the pseudo-isothermal halo models bars do not form as easily as the cuspy Hernquist halo models, and those that do form are weaker, similar to the weak bar of UGC 5288.
\begin{table*}
	\centering
	\caption{Hernquist halo models with $c=8$ and Pseudo-isothermal halo models with $r_{c}=0.23$ kpc}
	\label{models_c8}
	\begin{tabular}{lcc||ccr} 
		 \hline
Hernquist halo Model No. & Halo spin $\lambda$ (at $r_{200}$) & $z_{0}/R_{d}$ & Pseudo-isothermal halo Model No. & Halo spin $\lambda$ (at $r_{200}$) & $z_{0}/R_{d}$ \\
 \hline\hline
 1 &  0.0874 & 0.125 &  1 &  0.157 & 0.07\\ 
 2 &  0.0463 & 0.125 & 2 &  0.126 & 0.07 \\ 
 3 &  0.0363 & 0.125 & 3 &  0.157 & 0.08 \\  
 4 &  0.0874 & 0.2 & 4 &  0.142 & 0.08\\   
 5 &  0.0463 & 0.2 & 5 &  0.157 & 0.09 \\   
 6 &  0.0356 & 0.2 & 6 &  0.157 & 0.1 \\  
 \hline
\end{tabular}
\end{table*}
For the Hernquist profile, the bar is relatively stronger for the models with lower initial disk scale height $z_{0}=0.125 R_{d}$ compared to the ones with $z_{0}=0.2 R_{d}$. If we keep increasing the disk scale height, the stellar density decreases and the self-gravity of the stellar disk is not high enough to support bar formation. Hence, for higher disk scale heights of $z_{0}\geq 0.3 R_{d}$, the bar does not form. For the pseudo-isothermal halo models, there are different sets of initial disk scale heights for which the bar appears. One point to note is that the final disk scale heights are very similar for all the models (see the last column of Figure (\ref{evolution_models_c8}) and Figure (\ref{evolution_models_ciso}) ). Unlike the stellar disk, the gas disk does not show bar formation, although there is an accumulation of gas particles at the centre as seen in the panels $(b)$ \& $(e)$ of Figure (\ref{evolution_model_properties}). In the following section, we determine the bar length and ellipticity for all the models and compare it with the observed bar length and ellipticity from Section (\ref{obs_bar_prop}) to determine the best matching model for UGC 5288.

\subsection{Determining bar properties of the models} \label{bar_prop}
\begin{figure*}
\centering
	\includegraphics[width=0.8\linewidth]{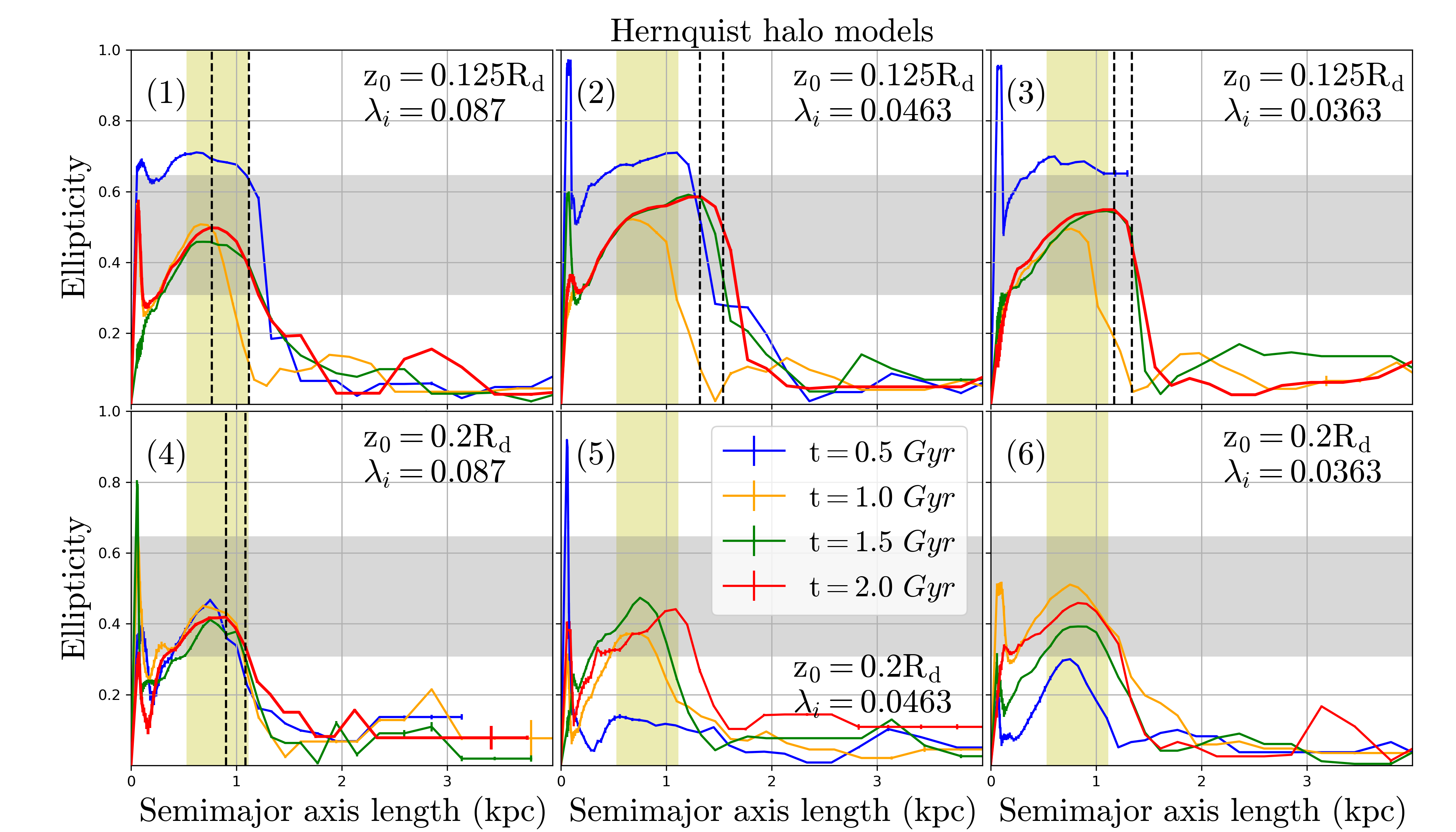}
    \caption{ {\bf This Figure shows the variation of ellipticity with semi-major axis length} for the different ellipses fitted to the stellar disk in the Hernquist halo models at 0.5 Gyr (blue), 1.0 Gyr (orange), 1.5 Gyr (green) and 2 Gyrs (red) of evolution. Model $(1)$, $(2)$ \& $(3)$ have initial disk scale height $z_{0}=0.125R_{d}$ and models $(4)$, $(5)$ \& $(6)$ have $z_{0}=0.2R_{d}$.  The grey patch is used to compare the bar ellipticity of our simulated models to that of UGC 5288. It is bounded by the peak ellipticity (at $\eta=0.646$) of the bar of UGC 5288 (see Section \ref{obs_bar_prop}) and the beginning of the outer stellar disk in UGC 5288 (at $\eta=0.307$). Similarly, the yellow patch is bounded by the corresponding semi-major axis lengths (0.527 kpc $<a_{obs}<$ 1.116 kpc). The black dashed lines in model $(1)$, $(2)$, $(3)$ and $(4)$ marks the semi-major axis length at maximum ellipticity and the 80\% of peak ellipticity. For models (1) and (4), the black dashed lines lie inside the yellow patch indicating their similarity with the observed bar length and ellipticity of UGC 5288.}
    \label{fitted_bar_simulations_c8}
\end{figure*}
\begin{figure*}
\centering
	\includegraphics[width=0.8\textwidth]{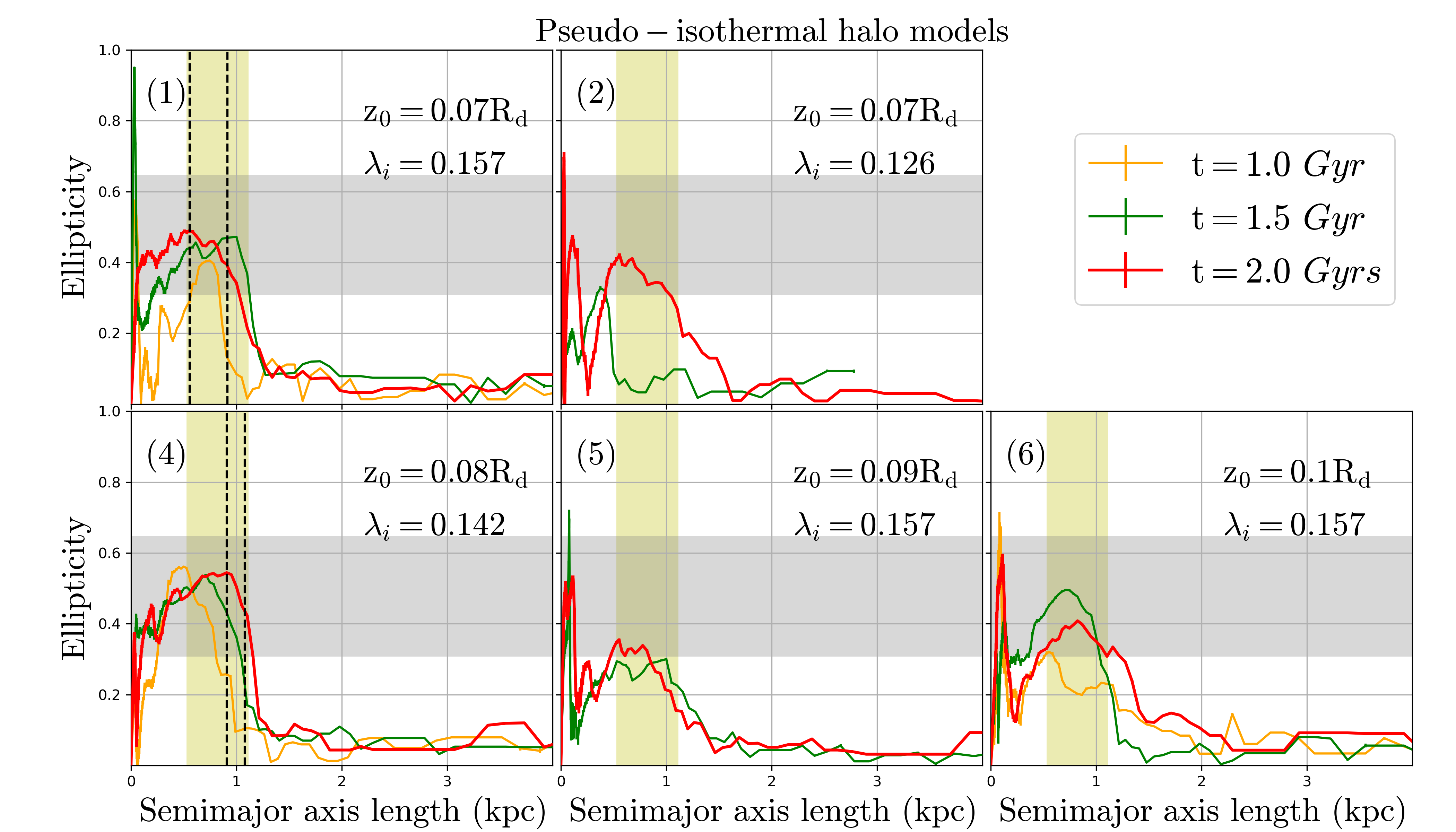}
    \caption{{\bf This Figure shows the variation of ellipticity with semi-major axis length} for the different ellipses fitted to the stellar disk in the Pseudo-isothermal halo models at 1.0 Gyr (orange), 1.5 Gyr (green) \& 2 Gyrs (red) of evolution. The initial disk scale height and spin of the Models are from Table (\ref{models_c8}). Similar to Figure (\ref{fitted_bar_simulations_c8}) the yellow patch and the grey patch denote the observed range of semi-major axis and ellipticity values that is used to match with our models. All models except $(1)$ \& $(4)$ either show a very weak bar or the instability in the disk has not settled yet. The black dashed lines in model $(1)$ \& $(4)$ mark the maximum ellipticity and the 80\% peak ellipticity and in both cases they lie inside the yellow patch. Among the two models $(4)$ is more stable for a longer time and has the peak ellipticity closer to the observed peak ellipticity (i.e., the upper boundary to the grey patch). Model $(3)$ only forms a deformed instability that has very low bar strength; so do not present it here. }
    \label{fitted_bar_iso_rc23}
\end{figure*}
In this section, we estimate the bar ellipticity and bar length in our simulated models and compare them with the observed values in Section (\ref{obs_bar_prop}). When fitting ellipses to the bar in the SDSS i-band image of Figure (\ref{bar_fitting}), the observed ellipticity value peaks at a semi-major axis length of $\sim 0.5$ kpc and gradually decreases by 37\% at the semi-major axis length of 1.06 kpc, after which the PA of the inner disk at $\sim 74^{\degree}$ sharply changes to $\sim 53^{\degree}$ in the outer disk. So our best fit should have similar bar properties i.e., the ellipticity should lie between $0.307<\eta_{obs}<0.646$ and the corresponding semi-major axis length is between 0.527 kpc $<a_{obs}<$ 1.116 kpc. We estimate the ellipticity of the simulated models using the same method as described in Section (\ref{obs_bar_prop}). Figure (\ref{fitted_bar_simulations_c8}) and Figure (\ref{fitted_bar_iso_rc23}) show the ellipticity versus semi-major axis plots for the Hernquist and pseudo-isothermal halo models from Table (\ref{models_c8}). Note that not all models form a strong bar and it is difficult to choose the time at which we should compare the bar properties of our evolved models with the observed bar in UGC 5288. Hence, we set certain criteria to compare the bar properties of the simulated models with the bar in UGC 5288:
\begin{itemize}
    \item {The evolved bar properties of the simulated models should be stable for at least two consecutive time intervals of 0.5 Gyr. We note that the Hernquist models (5) and (6) have significantly varying ellipticity within 0.5 Gyr intervals and are not favourable models according to this criteria. On the other hand, models (1), (2) and (3) are more favourable and model (4) is the most favourable in terms of stability. We chose the ellipticity distribution at 2 Gyrs of evolution (red curve) as it matches with the ellipticity distribution at 1.5 Gyrs (green curve), indicating that the disk is stable for more than 0.5 Gyrs. Most of the disks in the pseudo-isothermal models do not host strong, stable bars except models (1) and (4) with model (4) being the more stable one for about a Gyr.}
    \item {The maximum value of bar ellipticity of a model should lie in the range $0.307<\eta_{obs}<0.646$ as marked by the grey patch, and the corresponding semi-major axis length should be in the range 0.527 kpc $<a_{obs}<$ 1.116 kpc marked by the yellow patch in Figure ($\ref{fitted_bar_simulations_c8}$) and Figure (\ref{fitted_bar_iso_rc23}).}
    \item {The semi-major axis length of the bars in our simulated models is defined as the radius of 20\% decrease in ellipticity. So we set the criteria for a model to be favourable, that the peak of the ellipticity of our models and 80\% of peak value should lie within the range 0.527 kpc $<a_{obs}<$ 1.116 kpc ( i.e., within the yellow patch). So considering the first criteria, we marked the peak ellipticity and the 20\% decreased value with black dashed lines. In models (1) and (4) both the peak ellipticity and the 20\% decrease in peak ellipticity lie inside the yellow patch for both types of halo models  (see Figure (\ref{fitted_bar_simulations_c8}) and (\ref{fitted_bar_iso_rc23})).}
\end{itemize}
In the Hernquist halo models, all the models with initial disk scale length $z_{0}=0.125 R_{d}$, have highly elliptic bars within 0.5 Gyr of evolution while models with $z_{0}=0.2 R_{d}$ never reach such high bar ellipticity values during the evolution. So based on the above criteria, models (1) and (4) are the most favoured models for the Hernquist profile. But model (1) reaches the observed peak ellipticity at some point during evolution while model (4) does not reach the peak ellipticity within 2 Gyrs. For the evolution of model $(1)$ between 0.5 Gyr and 1 Gyr, there is one phase of evolution where the bar ellipticity matches very well with the observed peak ellipticity of UGC 5288. So among all the Hernquist models, model (1) is the most favourable model for UGC5288 with a cuspy halo profile. Among the favourable pseudo-isothermal halo models (1) and (4), model (4) forms a stable bar earlier while the bar in model (1) is not very strong and is not stable even until the end of evolution. So model (4) is the best match for the models with pseudo-isothermal halo profiles. 

Now that we have the best models with the Hernquist halo and Pseudo-isothermal halo profiles, in the following section, we study the halo spin profile of these models.
\begin{figure*}
\centering
	\includegraphics[width=0.9\textwidth]{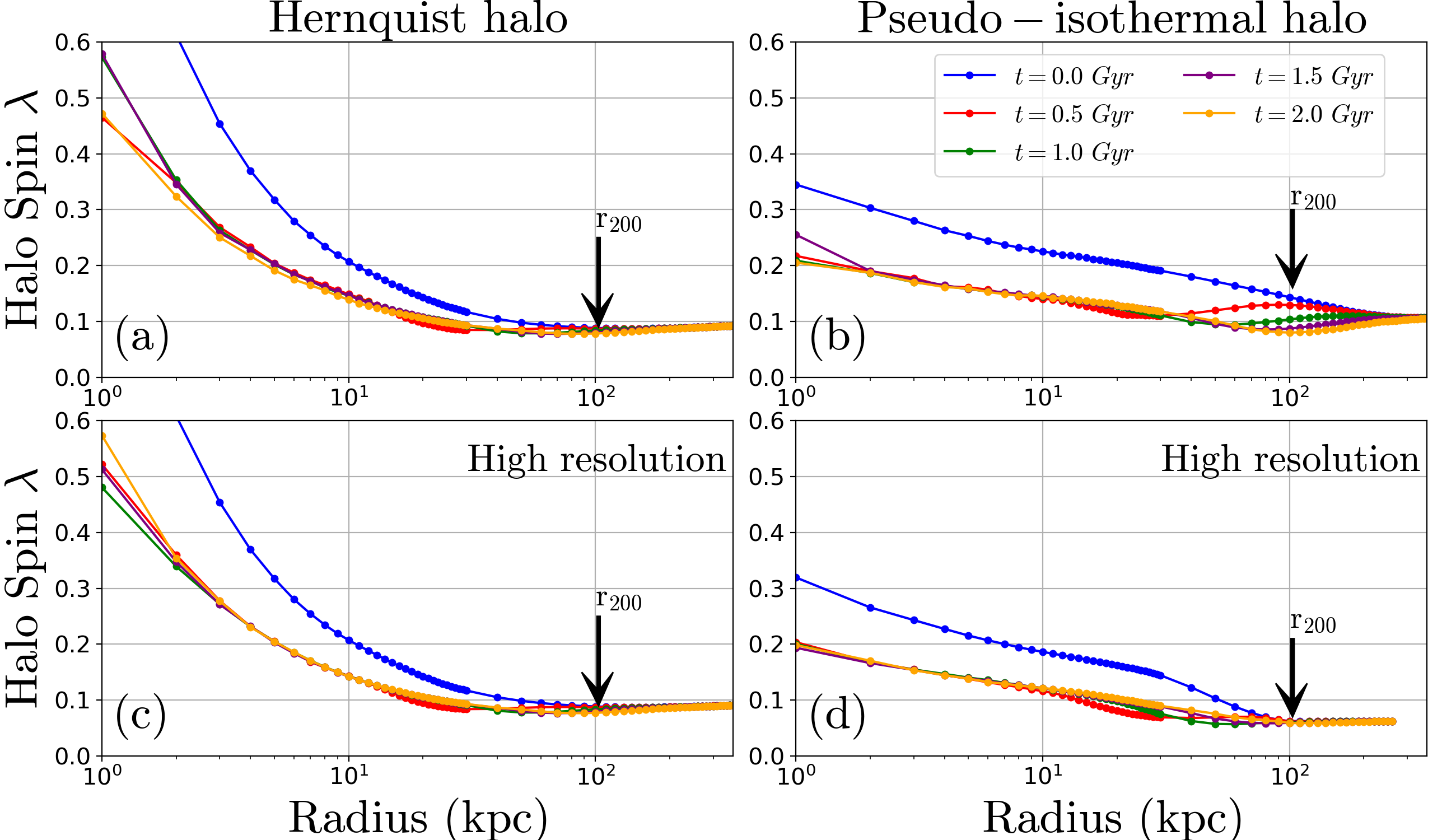}
    \caption{{\bf The evolution of halo spin $\lambda$ in the Hernquist halo and the pseudo-isothermal halo of our best-fit models with a different resolution of DM particles}. The halo spin of the inner part of the DM halo undergoes some changes but converges to constant values for both the Hernquist and pseudo-isothermal halo models; the halo spin does not change much outside the virial radius $r_{200}\sim 103$ kpc (marked by the black arrow). Though the spin values are very similar for both halo profiles at virial radius  $r_{200}$, the nature of the distribution is different for the two halos. The values of halo spin match closely at the central region for low resolution and high resolution simulations. They only differ slightly in the outer parts of the halo for the pseudo-isothermal halo model. The central values are very different in both profiles. This may be one of the reasons why the bars in these potentials look slightly different (see Figure (\ref{best_bars})).}
    \label{evolution_spin}
\end{figure*}
\begin{figure}
\centering
	\includegraphics[width=\columnwidth]{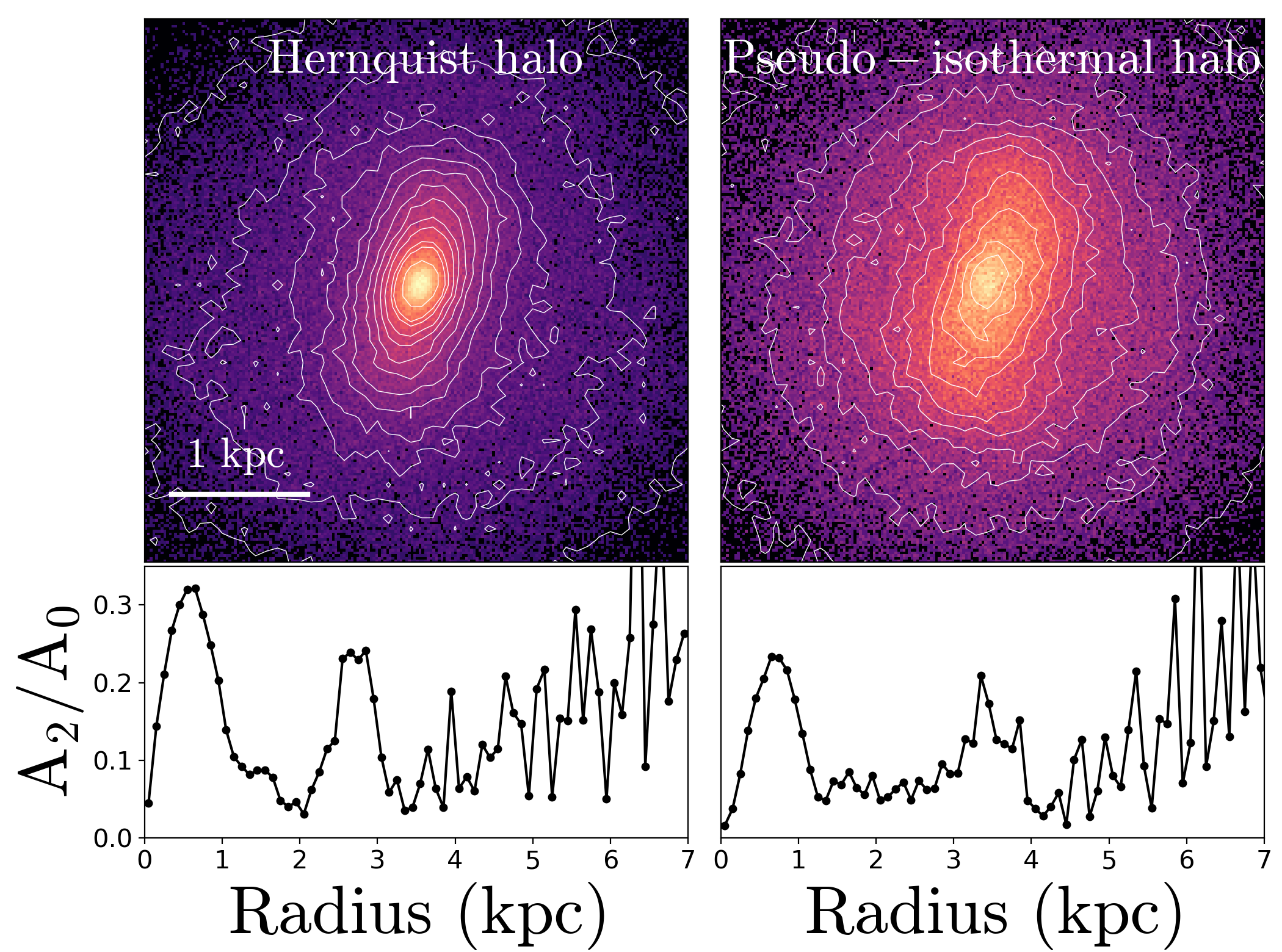}
    \caption{{\bf The two simulated models of UGC 5288 where the bar properties have the closest match with the observed bar properties of the galaxy}. Both models (one with Hernquist halo and another with Pseudo-isothermal halo) have similar peak ellipticity ($\sim 0.5$) and bar length of (2.3 kpcs and 2.1 kpcs considering semi-major axis at 80\% decrease in peak ellipticity), but still, the dissimilarity in thickness and surface density distribution can be clearly noticed in the Figure. The bar strength $\left(A_{2}/A_{0}\right)_{max}$ for the Hernquist halo is greater by $\sim 0.1$ compared to the Pseudo-isothermal halo (see Appendix (\ref{bar_strength}) for bar strength description). }
    \label{best_bars}
\end{figure}

\begin{table}
	\centering
	\caption{ Parameters of Best Models}
	\label{models_best}
	\begin{tabular}{lccr} 
		 \hline
 Pseudo-isothermal halo &  Hernquist halo \\
 \hline\hline
 $r_{c}=0.23$ kpc & $c=8$  \\
 $\lambda_{i}|_{r_{200}}=0.142$ & $\lambda_{i}|_{r_{200}}=0.0874$ \\ 
 $z_{0}/R_{d}|_{i}=0.125$  & $z_{0}/R_{d}|_{i}=0.08$ \\
  $\lambda_{f}|_{r_{200}}=0.08$ & $\lambda_{f}|_{r_{200}}=0.08$ \\ 
 $\lambda_{f}|_{10 kpc}=0.12$ & $\lambda_{f}|_{10 kpc}=0.15$ \\ 
 \hline
\end{tabular}
\end{table}
\subsection{Determining halo spin evolution for the best-fit models}
In this section, we investigate the evolution of the halo spin profile for our best models with the Hernquist halo profile and the pseudo-isothermal halo profile.
Figure (\ref{evolution_spin}) shows the halo spin profile after evolution of 2 Gyrs for the most favourable models with $c=8$ and $\lambda \sim 0.5$ at 1 kpc ($\lambda=0.08$ at $r_{200}$) for the Hernquist profile and $r_{c}=0.23$ kpc and  $\lambda=0.2$ at 1 kpc ($\lambda=0.08$ at $r_{200}$) for the pseudo-isothermal profile. The halo spin profile tends to converge within 2 Gyrs of evolution. The spin at the outer halo does not change much but the inner halo spin profile changes quite a lot for both profiles. This is because during evolution the halo particles undergo a net rearrangement process that redistributes angular momentum, energy and mass in the inner regions of the halo. The signature of the redistribution of angular momentum is seen in the evolution of the halo spin in the Figure. Overall we find that the  pseudo-isothermal halo model gives a better match to the observed properties of UGC 5288, mainly because it fits the rotation curve much better than the Hernquist model.

In this section, we have constructed a multi-component simulated model of the observed galaxy UGC 5288. We have considered a relatively isolated dwarf void galaxy UGC 5288 with very negligible star formation and a weak stellar bar. The bar in the Hernquist halo appears more compact than that in the pseudo-isothermal halo (see the bars for both models in Figure (\ref{best_bars})) and we find that the pseudo-isothermal halo model gives a better match with the observed HI rotation curve of the galaxy. In the following section, we investigate the cosmological simulations to test our isolated galaxy simulation models.

\begin{figure*}
\centering
\includegraphics[width=\textwidth]{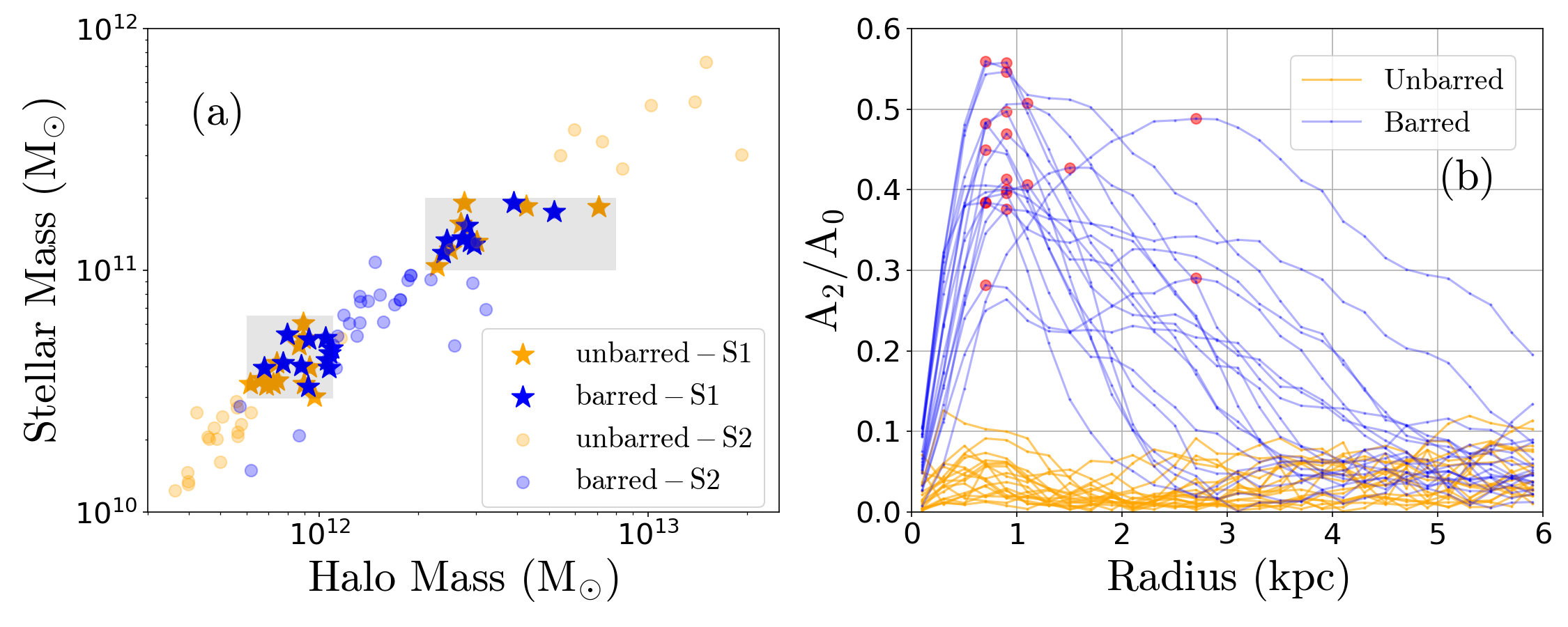}
    \caption{{\bf The sample of barred and unbarred galaxies from TNG50}. In panel (a) there are two samples - S1 and S2. S1 is marked with the stars and is within the grey rectangular patches, while S2 is a bigger sample marked with lighter colour circles (see Table (\ref{table:appendix_bar_properties_tng})). Different shades of blue represent the barred galaxies and orange represents the unbarred galaxies. Panel (b) shows the bar strength distribution of barred and unbarred galaxies in the S1 sample. The red circles indicate the maximum bar strength in the barred galaxies. } 
    \label{fig:tng_galaxy_sample_barstrength}
\end{figure*}

\section{Comparison with Cosmological Simulations} \label{sec:cosmological_simulations}
In this section, we investigate how our model halo spin profile compares with the halo spin from the cosmological simulations.
In previous sections, we have modelled the dark matter halo of UGC 5288 and estimated an approximate halo spin by using N-body/SPH simulations with all the available observed properties as inputs. In earlier studies, dark matter halos had been modelled using parameters such as halo concentration or core radius which were derived from the HI or optical rotation curves. Our method of modelling the halo in UGC 5288 is completely different. We have replaced the analytical models with simulated ones and included a live halo composed of particles. One point that deserves mention is that we are not modelling the evolutionary history of the galaxy. For example the star formation history, merger history etc. that the galaxy has undergone. We are modelling the present state of the galaxy by recreating all the observable properties seen in different wavelength bands. 

One of the ways to test our method is to compare the spin profiles of our models with those of cosmological simulations. We compare the halo spin profile and spin estimates of our simulated models of UGC 5288 with that of the DM halos in TNG50 \citep{2019TNG50_data_release, Nelson.et.al.2019,Pillepich.et.al.2019, Rosas-Guevara.et.al.2021}, which is one of the most recent cosmological magneto-hydrodynamical simulations and has publicly available data. The main steps are the following. \\
(i)~We expect that there should be some correlation between bar properties and the inner regions of the halo. This is based on several studies that show that bars transfer angular momentum to the dark matter halo \citep{Athanassoula.2002, Saha.Naab.2013,Long.et.al.2014, Kurapati.et.al.2018,Collier.et.al.2018,kataria.das.2018,Collier.et.al.2019}. Previously \cite{Rosas-Guevara.et.al.2021} had shown that barred galaxies in the TNG50 simulations have relatively lower halo spin than unbarred galaxies. So first, we construct a sample of galaxies to compare the halo spin profiles of barred and unbarred galaxies in TNG50. \\
(ii)~With the sample of barred galaxies, we check if there is any correlation between the bar properties and the halo spin in the inner regions of the galaxy. \\
(iii)~Finally, we search for analogues of UGC 5288 in the TNG50 data to compare the halo spin profile of our model with the UGC 5288 analogues in TNG50. We are not aiming for an exact comparison of a galaxy in TNG50 with UGC 5288, because such low-mass dwarfs galaxies with bars are difficult to find in good resolution in cosmological simulation. Instead, we want to see if our model halo follows similar trends as in TNG50. \\
To investigate the above points, we need to find the barred and unbarred galaxies in TNG50 and estimate the halo spin profiles of these galaxies and conduct multiple tests as described in the following sections.

\subsection{Sample Selection}
We selected barred and unbarred galaxies at redshift $z=0$ in the stellar mass range of $10^{10} - 10^{12}$ M\textsubscript{\(\odot\)} from the TNG50 suit of cosmological simulations \citep{2019TNG50_data_release, Nelson.et.al.2019,Pillepich.et.al.2019, Rosas-Guevara.et.al.2021}. The advantage of choosing this mass range is that the galaxy disks and dark matter halos are highly resolved. So, more reliable results can be expected. Out of the total number of 253 barred galaxies with bar strength $(A_{2}/A_{0})_{max}> 0.25$ and 270 unbarred galaxies having $(A_{2}/A_{0})_{max}< 0.2$, we selected a sample of 19 strongly barred galaxies with $(A_{2}/A_{0})_{max} \gtrsim 0.3$ and an equal number of unbarred galaxies with $(A_{2}/A_{0})_{max}< 0.1$ having very similar stellar masses. This is sample-1 (S1) as shown in panel (a) in Figure (\ref{fig:tng_galaxy_sample_barstrength}). A second larger sample of 90 barred and unbarred galaxies S2 (also in panel (a) in Figure (\ref{fig:tng_galaxy_sample_barstrength})) is also used. The difference between the two samples is that S2 is not uniformly distributed in mass over the whole stellar mass-halo mass range. 

To search for galaxies having high bar strength we first centred the stellar disk about the minimum of the potential. Next, we estimated the angular momentum vector of the stellar disk inside a sphere of radius 5 kpc centred at the potential minimum of the stellar disk. We aligned the axis of maximum angular momentum towards the z-axis of the coordinate system. We rechecked the final increase in angular momentum in the z-direction and visually verified the face-on and edge-on orientation of the disks in the present configuration. We then estimated the bar strengths in concentric annular regions of the stellar disk having $\Delta r=0.2$ kpc and $|z|<1.0$ kpc. Panel (b) in Figure (\ref{fig:tng_galaxy_sample_barstrength}) shows the bar strength distribution of galaxies in sample S1.
\begin{figure}
\centering
	\includegraphics[width=\columnwidth]{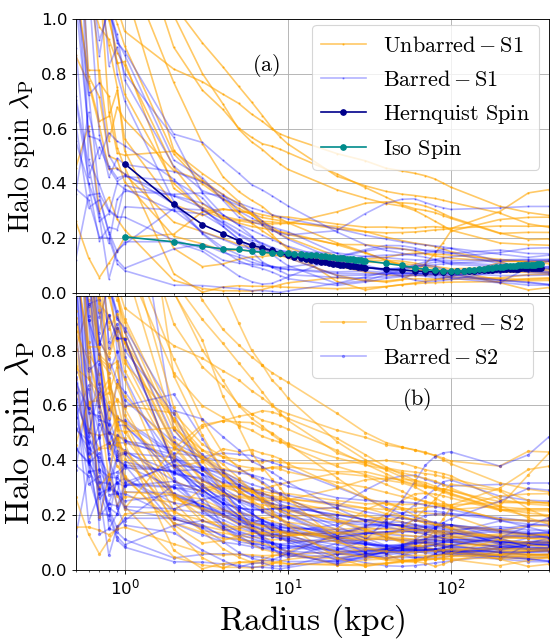}
    \caption{{\bf The halo spin profile of the barred and unbarred sample of galaxies from TNG50 occupy different regions in the above Figures}. The halo spin profile has less spread for the barred sample at all radii compared to the unbarred sample. Panel (a) shows sample-1 (S1) with 38 galaxies and panel (b) shows the sample-2 (S2) with a total of 90 barred and unbarred galaxies as shown in panel (a) of Figure (\ref{fig:tng_galaxy_sample_barstrength}). Irrespective of stellar/halo mass the spin distribution is similar for barred and unbarred galaxies in samples S1 and S2. We compare the model halo spin profiles for the Hernquist (Hernquist Spin) and Pseudo-isothermal halo (Iso Spin) models of UGC 5288 with the spin profiles from TNG50.}
    \label{tng_bars_spin}
\end{figure}
\subsection{Comparison of halo properties of barred and unbarred galaxies}
\begin{figure}
\centering
	\includegraphics[width=\columnwidth]{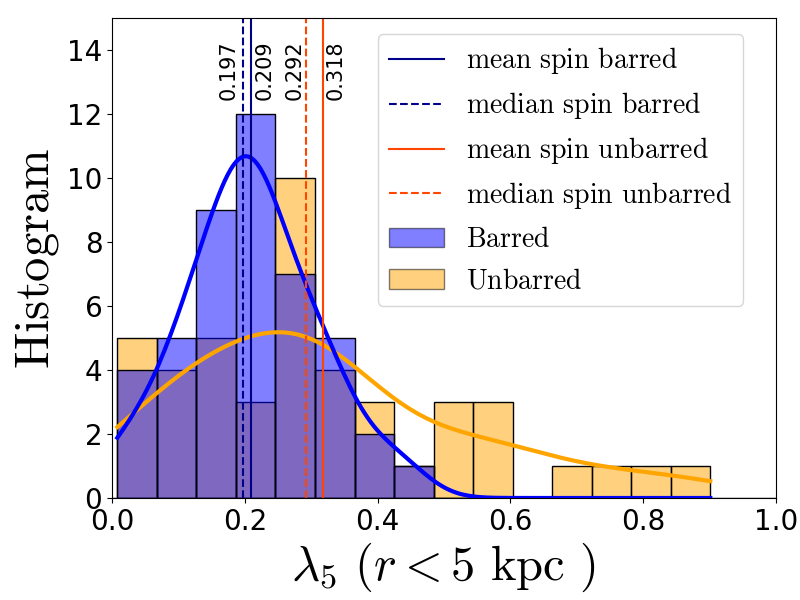}
    \caption{{\bf Distribution of halo spin at inner regions of $\sim 5$ kpc is different for barred and unbarred galaxies}. We plot the normalized histograms and the corresponding smoothed probability density function estimated with a KDE, as shown by the solid blue and orange curves on top of the histograms. The histograms are for the S2 sample of 90 galaxies with a bin size of 1/15. The halo spin profile and the median value of halo spin for the barred and unbarred galaxies have a significant difference.}
    \label{tng_bars_spin5kpc}
\end{figure}
Figure (\ref{tng_bars_spin}) shows the difference in the halo spin profile for the sample of strongly barred and unbarred galaxies. At a fixed radius, the distribution of spin values for the barred galaxies follows a more narrow distribution than the unbarred galaxies (more clearly evident in Figure (\ref{tng_bars_spin5kpc})). There is a drop in halo spin for the barred galaxies within the inner disk radius, which is not seen for the unbarred galaxies. Furthermore, the spin values at the central disk region are higher than the values in the outer region of the halo. The galaxies in S1 and S2 samples have very similar values of halo spin, although the stellar and halo mass are not uniformly distributed in S2. Hence, the general trend that halo spin is independent of the stellar mass or halo mass is reaffirmed in this sample, even though we can see a clear difference between the spin profiles for barred and unbarred galaxies. The spin distribution from our simulated model of UGC 5288 (Figure (\ref{evolution_spin})) is similar in nature to the spin distribution of halos of the barred galaxies in the S1 and S2 samples, where the galaxies and their halos are of cosmological origin. This reassures us that our halo model and the model of spin distribution of UGC 5288 are similar to a cosmologically evolved halo model.

Figure (\ref{tng_bars_spin5kpc}) shows the distribution of halo spin values estimated at a fixed radius of 5 kpcs. It shows quantitatively how halo spin is different for barred and unbarred galaxies. The mean values of halo spin in barred and unbarred galaxies are significantly different, and both are much higher than the average halo spin at virial radius $\sim 0.035$, for dark matter halos in cosmological simulations \citep{Bullock.et.al.2001}. We notice that the mean halo spin is lower for the barred galaxy distribution. A similar trend is also seen for halo spin estimated at the virial radius in the recent study of TNG50 bars by \cite{Rosas-Guevara.et.al.2021}. The possible reason may be that the stellar bar interacts with the dark matter halo as it evolves and exchanges angular momentum at different resonances.

From the halo spin models of the pseudo-isothermal halo and the Hernquist halo in Figure \ref{evolution_spin}, we note that the halo spin is sensitive to the dark matter halo density profile. To check if halo spin is also affected by the presence of a bar in the disk and not only the halo density profile, we do some tests to probe the effect of halo spin in barred and unbarred galaxies having very similar dark matter circular velocity curves. Here also we find that the median halo spin of the distribution of barred galaxies is lower than the unbarred galaxies. See Appendix \ref{appendix:test_CV_bars} for more details of the tests. \\ 
In the following section, we explore how the bar properties are correlated with the halo spin. 

\subsection{Bar properties and their connection with halo spin}
We have seen that the halo spin profile and the central halo spin values in barred galaxies are different from unbarred galaxies. These differences maybe due to bar-halo interaction and thus some of the bar properties maybe correlated with the halo properties also. So we estimated some of the properties of the stellar bar -- peak bar strength, bar length, and ellipticity at bar region, in S1 and S2 samples. We show an example of how we estimate bar length and bar ellipticity using {\it photutils} by ellipse fitting one of the bars in TNG50, in Figure (\ref{fig:appendix-tng_bar_length}) in Appendix \ref{appendix:bar_properties}. The bar length, bar strength and ellipticity are estimated for the galaxies in S1 and S2 and presented in Table (\ref{table:appendix_bar_properties_tng}), along with their subhalo ID. 

\begin{figure}
\centering
	\includegraphics[width=\columnwidth]{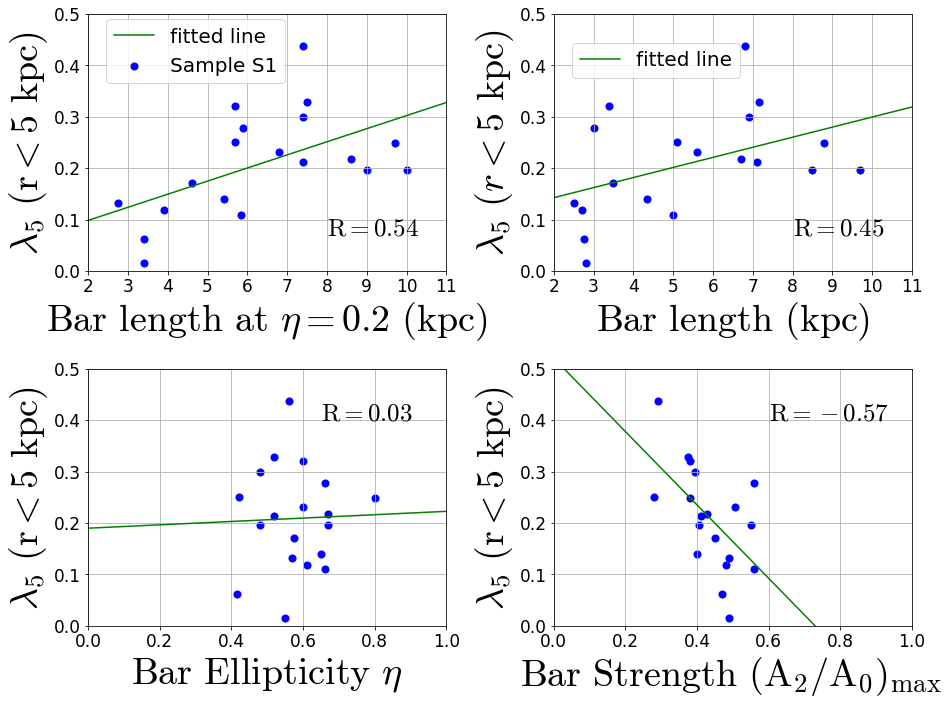}
    \caption{{\bf Correlation between bar properties and halo spin is weak}. Sample S1 is used to estimate the correlation coefficients. The Pearson correlation coefficient $R$ for all the cases is low. With a larger sample S2 the correlations even decrease.}
    \label{tng_bar_property_correlation}
\end{figure}

The bar length, as estimated through different methods, has a slight correlation with the inner halo spin $\lambda_{5}$, estimated at a radius of 5 kpcs, while bar ellipticity has a very weak correlation and bar strength is anti-correlated with the spin as seen in Figure (\ref{tng_bar_property_correlation}). The anti-correlation is consistent with the trend observed for the mean halo spin for barred and unbarred galaxies, discussed in the previous section. Although we find that the spin distribution for barred and unbarred galaxies are different (seen in Figure (\ref{tng_bars_spin}) and Figure (\ref{tng_bars_spin5kpc})) there is a very weak correlation with some of the bar properties. We do not show it here, but we have checked that as expected, the halo spin at larger radii is less correlated to the bar length. We have also checked with the larger sample S2 and found that the correlation decreases slightly. This may be due to several reasons. The initial angular momentum of a galaxy can change during the course of evolution depending on the environment of the galaxy. Galaxies undergo interactions and mergers with nearby galaxies and exchange angular momentum with the disk through disk asymmetries like bars or spiral arms. Or, it may even be that some dark matter halos with relatively lower halo spin in the central region are more favourable to host a stellar bar. To answer these questions we intend to conduct a more in-depth study in future.
\subsection{UGC 5288 analogues in TNG50}
\begin{figure*}
\centering
\includegraphics[width=0.85\textwidth]{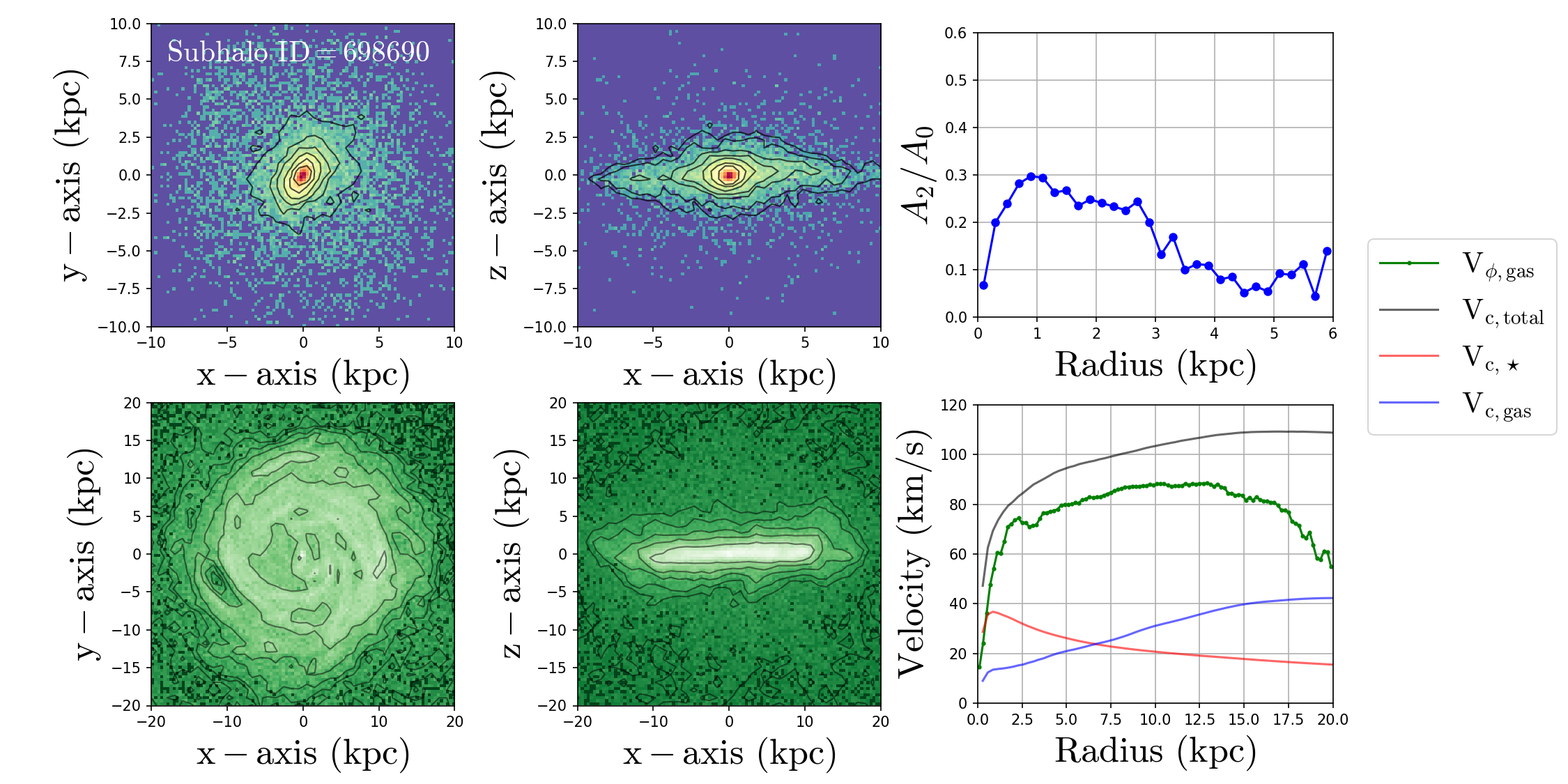}
\caption{ {\bf The stellar disk and gas disk of one of the analogues of UGC 5288 (Subhalo ID-698690) with its bar strength $A_{2}/A_{0}$ distribution over time and the velocity curves at redshift $z=0$ }. This galaxy has a small stellar disk (top panel) that has a bar-like structure at its centre with low bar strength and an extended gas disk (bottom panel). The gas rotation curve 
$\rm V_{\phi, gas}$ (green), the stellar circular velocity curve $\rm V_{c, \star}$ (red) and the gas circular velocity curve $\rm V_{c, gas}$ is lower than the total circular velocity curve $\rm V_{c, total}$ (black). }
    \label{fig:tng50-ugc5288-analogues}
\end{figure*}

\begin{figure*}
\centering
\includegraphics[width=0.45\textwidth]{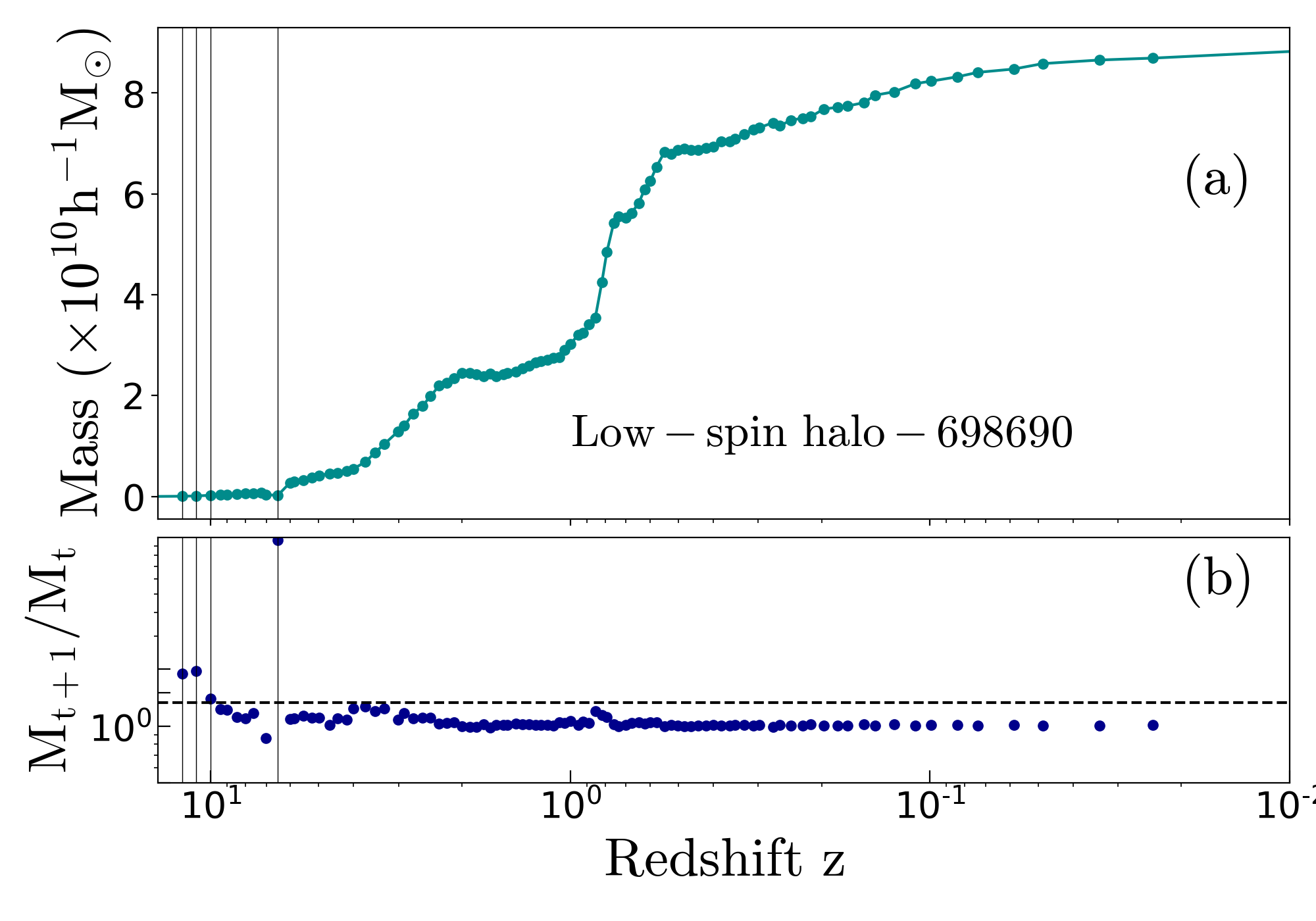}
\includegraphics[width=0.45\textwidth]{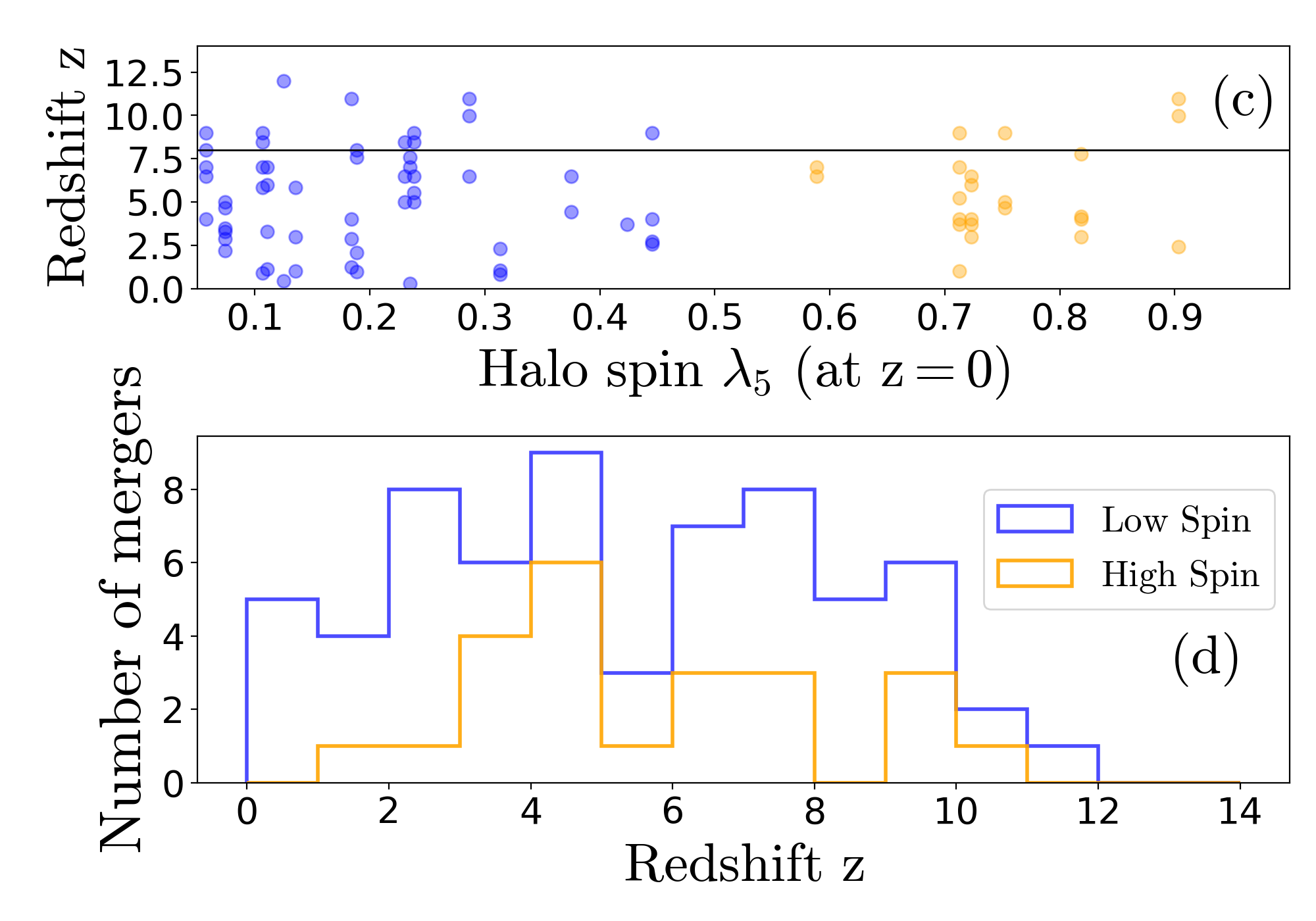}
\caption{{\bf Merger history of the UGC5288 analogue galaxies}. Panel (a) shows the evolution of the total subhalo mass with redshift for one of the low spin halos (Subhalo ID-698690 at $z=0$), which is used in panel (b) to obtain the mass ratio of the main halo in consecutive time ($t, t+1$). The dashed horizontal line marks the upper bound (4/3) above which a major merger is counted. The black vertical lines denote the instances when there is a major merger with $\rm M_{t+1}/M_{t}> 4/3$. Panel (c) shows all the major mergers at different redshifts in the 24 galaxies with their halo spin (at z=0) on the x-axis. The low spin lows are marked with blue and the high spin halos are marked with orange.}
\label{fig:tng50-ugc5288-analogues-merger-history}
\end{figure*}

\begin{figure*}
\centering
\includegraphics[width=0.45\textwidth]{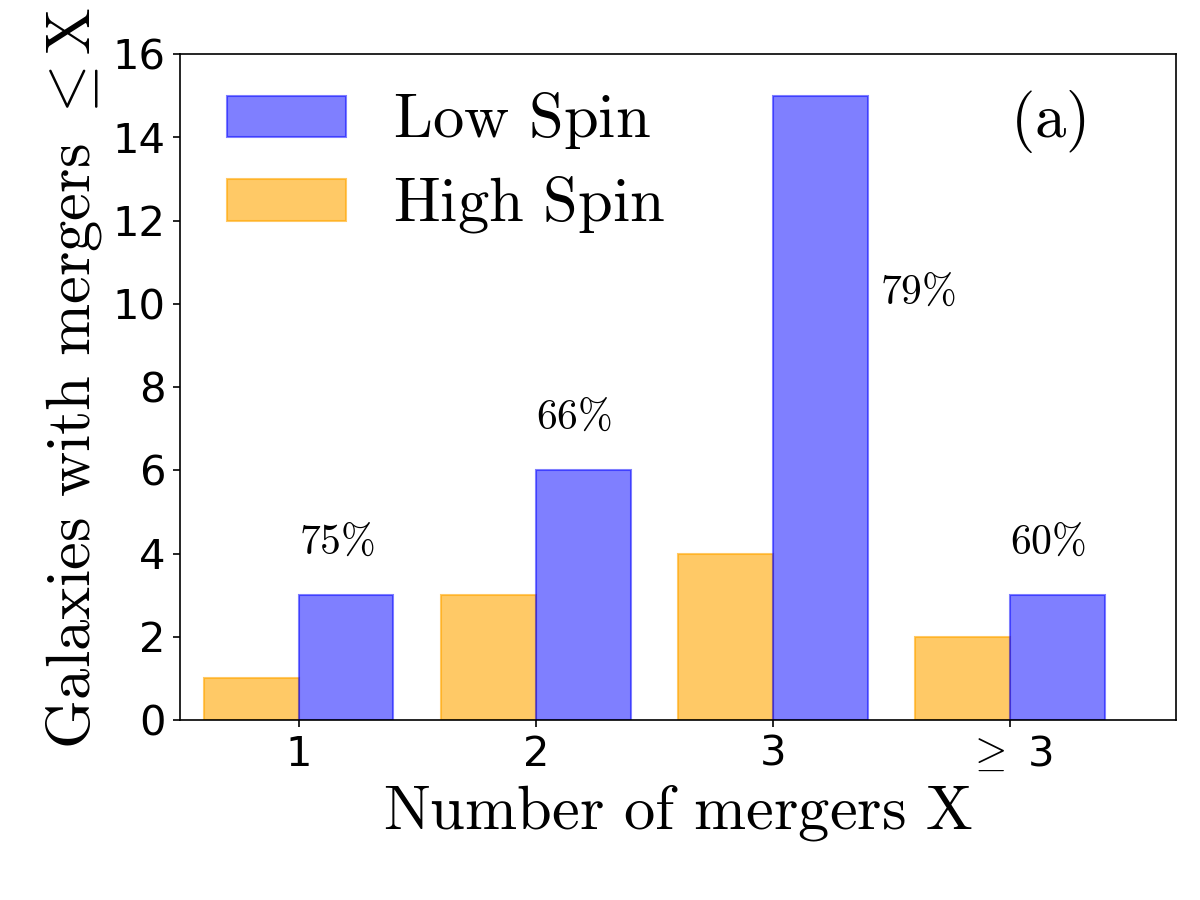}
\includegraphics[width=0.45\textwidth]{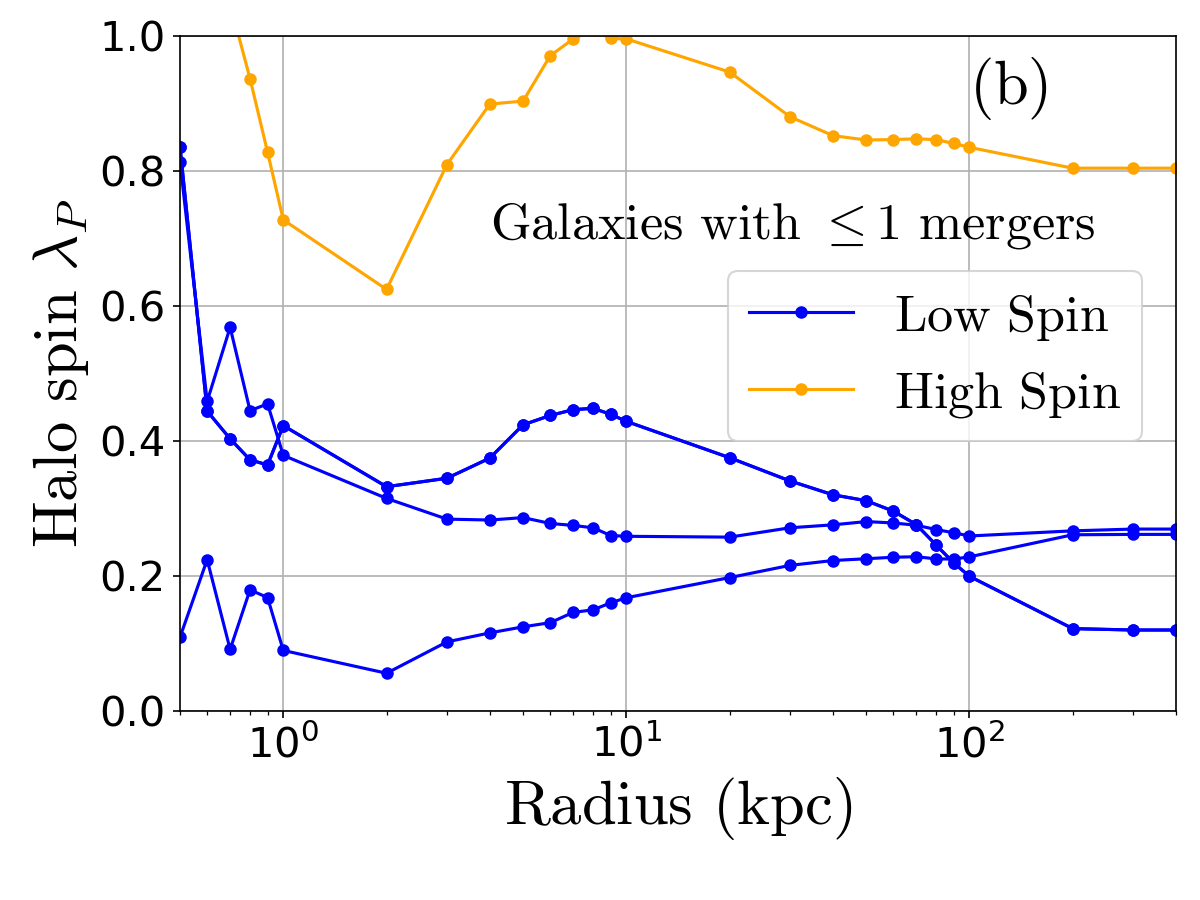}
\caption{ {\bf Halo spin and major mergers of UGC 5288 analogues}. Panel (a) shows the number of galaxies with low halo spin (blue) and high halo spin (orange) undergoing $\leq X$ number of total major mergers during their evolution. The first pair of histograms at $\rm X=1$ consists of 3 galaxies with low halo spin and 1 galaxy with high halo spin, also shown in panel (b) that undergo $\leq X$ number of mergers. Similarly the spin distribution of galaxies with $X\leq 2, 3$ (2$^{nd}$ and 3$^{rd}$ pair of histograms) is shown in panel (a) and (c) in Figure \ref{fig:appendix-halospin_all_galaxies_energy}. The fluctuations in the spin value within a radius of 1 kpc are due to very few DM particles ($< 1000$) and are less reliable. }
    \label{fig:tng50-ugc5288-analogues-halospin}
\end{figure*}
In this section, we search for analogues of UGC 5288 among the TNG50 dwarf galaxies. To find UGC 5288 analogues we search for galaxies with similar morphology: a weakly barred galaxy with an extended gas disk that appears undisturbed, i.e., has undergone very few major mergers in the recent past. 

We first searched for galaxies that have similar stellar and gas mass as UGC 5288 (see Table \ref{table:ugc5288-model-parameters}); the mass ranges are $2\times 10^8 <\rm M_{\star}/M_{\odot} < 4\times 10^8 $ and $1.1\times 10^9 <\rm M_{gas}/M_{\odot} < 1.3\times 10^9$. However, we found only 18 galaxies all of which have irregular stellar and gas disks, and are unbarred. So we relaxed our selection criteria to a broader mass range: $1.5\times10^8 <\rm M_{\star}/M_{\odot} < 9\times 10^8 $ and found $\sim 2073$ galaxies. We searched for barred galaxies by setting the criteria: $A_{2}/A_{0}> 0.2$. We selected 24 barred galaxies that have extended gas disks and small stellar disks, and visually appear to have a bar in the face-on stellar surface density distribution (for example, see Figure \ref{fig:tng50-ugc5288-analogues}). We call the 24 galaxies candidate analogues of UGC 5288. 

As discussed in section \ref{galaxy_selection}, UGC 5288 is a void galaxy and such galaxies tend to undergo less number of major mergers throughout their evolution compared to galaxies in denser environments, such as those in filaments and sheets in the cosmic large scale structure. Since UGC 5288 has a very regular, undisturbed, extended HI disk, there is a very high probability that it has not undergone any recent major merger. To find out possible analogues of UGC 5288, we examined the merger histories of the 24 candidate galaxies and checked which of them has undergone a less number of mergers. We examined the evolution of the total dark matter mass of the subhalos from $\rm z=0$ to $\rm z=15$ as provided in the 100 snapshot files in the TNG50 data set.

We present the evolution history of a typical subhalo in Figure \ref{fig:tng50-ugc5288-analogues-merger-history}. We searched each candidate galaxy for an increase in mass at all redshifts, and classify an event to be a major merger if the ratio of the host mass at redshift $\rm z_{n}$ and the next redshift $\rm z_{n+1}$, i.e., $\rm M_{t_{n+1}}/M_{t_{n}}> 4/3$. Panel (a) of Figure \ref{fig:tng50-ugc5288-analogues-merger-history} shows one example of the total subhalo mass evolution of an example low halo spin galaxy (subhalo ID= 698690) where the black vertical lines indicate the major mergers and in panel (b) we show the mass ratio $\rm M_{t+1}/M_{t}$ for consecutive redshift. In this example, we have multiple major merger events in the early redshift, all of which may not be important for spin distribution at present. At very early redshifts the galaxies would have formed from the hot turbulent gaseous medium and settled into a disk. Hence, we consider galaxy major mergers within a redshift range of $0< z < 8$. Furthermore, we examine each of the galaxy mass evolution plots for false merger cases termed "subhalo switching problem" which are sudden unrealistic jumps in the mass evolution. See \citet{Rodriguez-Gomez.et.al.2015} for more details. After discarding these false merger cases we determined the total number of mergers each galaxy undergoes and their corresponding redshifts.

We also estimated the halo spin profile for each of the 24 candidate galaxies and we present the halo spin $\lambda_{5}$ (at $ r=5 $ kpc) at $\rm z=0$ along with the redshifts of major mergers for all the 24 candidate galaxies in panel (c) of Figure \ref{fig:tng50-ugc5288-analogues-merger-history}. We observe that there is a similar distribution in the number of major mergers among the low halo spin ($\lambda_{5}< 0.5$) and high halo spin ($\lambda_{5}> 0.5$) galaxies over redshift. There are a larger number of low-spin halos (18, coloured in blue) as compared to the low number of high-spin halo (6, coloured in orange) in our sample as seen from panel (c). Also in panel (d) of Figure \ref{fig:tng50-ugc5288-analogues-merger-history} we show that the total number of major mergers for the low spin and the high spin halos in our sample of candidate galaxies are distributed over a larger redshift range and we do not observe any correlation between the halo spin and the number of major mergers. 

Now with the frequency of major mergers in the galaxies from our sample, we group the galaxies according to the number of major mergers they undergo throughout their evolution. 4 out of 24 galaxies undergo $\leq 1$ merger event throughout their evolution. The very rare number of major interactions in these 4 galaxies has a close equivalence to an isolated non-interacting void-like environment. Even though we have a very less number of galaxies in the sub-sample of the 24 galaxies, $75\%$ of the galaxies have low halo spin with their halo spin profile in panel (b) of Figure \ref{fig:tng50-ugc5288-analogues-halospin}. If we consider the total number of major mergers to be $\leq 2$ then $66\%$ of galaxies have a low spin and if the number of mergers $\leq 3$ it is $79\%$. The rest of the galaxies with the number of mergers $\geq 3$ are fewer in number, with $60\%$ showing low halo spin (see Figure \ref{fig:appendix-halospin_all_galaxies_energy} for the spin distribution of these galaxies).

One of the 4 galaxies that undergo $\leq 1$ merger has a high halo spin. This indicates that other than the frequency and time of major mergers, several factors can play important role in influencing the halo spin. Mergers can happen with different orbital properties, including multiple pericenter passages that can have different impacts over the DM halo. The orbital inclination and direction are important, for example, the satellites can have planar or polar orbits, and the direction can be retrograde or prograde with respect to the disk and halo. These parameters can influence the halo spin profile. Furthermore, we observe that the total energy is more negative for the high-spin halos as compared to the low-spin ones (see panel (d) in Figure \ref{fig:appendix-halospin_all_galaxies_energy}). The highly negative value of energy suggests that the high spin halos are in relatively denser environments that have deeper potential well, a good example being the filaments and sheets of the large-scale structure. While galaxies with more positive values of total energy maybe in less dense, void-like environments. Currently, it is beyond the scope of this article to investigate in more detail the properties and environments of the mergers but it is clear that there are many factors playing a role.

Thus, we have investigated a sample of dwarf galaxies in the TNG50 data set and found 3 analogues of UGC 5288 (Subhalo ID- 698690, 721346 and 712591) that have a small stellar disk and an extended gas disk, as well as low spin values. One issue with the TNG50 dwarf galaxy data is that the galaxies and their dark matter halos have a very less number of star, gas and dark matter particles in the dwarf galaxy mass range ($\rm M_{\star} < 10^9 M_{\odot}$). For example, the galaxy in Figure (\ref{fig:tng50-ugc5288-analogues}) has $\sim 13,000$ star particles, $\sim 69,000$ gas particles and about $\sim 68,000$ dark matter particles within a radius of 20 kpc from the centre and $\sim 10,000$ star particles, $\sim 7,000$ gas particles and $13,000$ dark matter particles within a radius of 5 kpc. With the low mass resolution of these galaxies, it is difficult to get a smooth halo spin profile for an individual dark matter halo. We can only match our simulated halo models with the median halo spin profile for a bunch of galaxies as shown in Figure \ref{fig:appendix-halospin_all_galaxies_energy}. The TNG50 DM halos match with the halo spin  profile of our pseudo-isothermal halo model of UGC 5288  within an order of magnitude (see Figure \ref{fig:appendix-halospin_all_galaxies_energy}). If we match with individual dark matter halos it shows significant uncertainties.

\section{Discussion} \label{sec:discussion}
We investigate a new method for modelling barred galaxies from observations. This method is hybrid in nature, combining a detailed multi-component disk and dark matter halo model by N-body/SPH simulations and forward modelling of galaxy properties from observations. Numerical forward modelling is required because 1) We need to self-consistently generate a bar in the galaxy that is difficult to consider in an analytical approach; 2) prediction of halo spin profile using an analytical approach consists of a few assumptions (for example, virial equilibrium, circular orbits of DM particles and the lack of non-linear dynamics) that may not hold true at every radius. We test this method for a barred galaxy as the asymmetric structure in the disk will interact with the spherical DM halo by exchanging angular momentum. The presence of the bar will affect the final spin distribution and the spin will be different from that of an unbarred disk galaxy. 
Our simulated models of UGC 5288 give us an estimate of the probable internal spin profile as shown in Figure (\ref{evolution_spin}). Our spin distribution is similar to {the median halo spin profile } of barred galaxies in TNG50, while it shows significant uncertainties with individual halo spin profiles of the galaxies. Since we used both cuspy (Hernquist) and flat-core (pseudo-isothermal) halo profiles we 
can compare the effect of different dark matter density profiles on the spin distribution and its evolution. We find that although the halo spin at virial radius $\lambda\vert_{r_{200}}$ is not very different for the cuspy and flat-core halo profiles, the spin in the inner 10 kpc rises sharply in the Hernquist profile but nearly constant in the pseudo-isothermal profiles. This affects the bar formation as discussed earlier in Section \ref{bar_prop}. 

An important question is whether the internal spin distribution uniquely determines the halo-bar interaction while all other disk properties are held unchanged. Different ways to generate a bar instability have been used in the literature by modifying the velocity distribution and hence the spin distribution of DM haloes. In previous studies, \citep{Saha.Naab.2013, Collier.et.al.2019,Kataria.Shen.2022} showed that a combination of prograde/retrograde orbits of halo particles results in bar formation. They used the property that one can in principle generate multiple solutions of the Jeans equation by changing the sign of the azimuthal velocities. This technique has been used to control the motion of DM particles such that the disk-halo interactions can be tuned to generate early or late bars in simulations. We use a different technique to generate the DM haloes of different spin by multiplying the $\langle v_{\phi}\rangle$ with a scalar, k$_{parameter}$. This gives another solution to the Jeans equation used to generate the initial conditions of the DM halo particles in simulations. The velocity distribution and mass distribution at the central region determine the nature of bar formation and this may not be correlated to the spin at very large distances like $r_{200}$.  We see that the bar properties among our two most favourable models are a little different and this is also reflected in the halo spin profile at the very central regions ($r < 10$ kpc). 
Thus, different spin distributions may in principle host bars that are similar in some properties, like in this case, bar length and ellipticity but maybe dissimilar in other properties like surface density distribution around the bar region. Our analysis of the barred galaxies in the TNG50 simulations shows that there might be a weak correlation between the halo spin in the inner regions of the galaxy and the presence of the bar. Additionally, the correlations between the bar properties and halo spin are not strongly reflected with the small sample of barred galaxy data that we have analysed in this work. We are working with a larger data set for a thorough analysis (Ansar et al. in prep).

Galaxy mergers can also influence the mass and velocity distribution of dark matter halos \citep{Bett.Carlos.2012,Hetznecker.Burkert.2006}. Here we have presented one possible spin distribution for the galaxy UGC 5288, which is located in a void and has hence probably not undergone any major mergers in the recent past. So this method may not be applicable to galaxies that have undergone mergers in the recent past. Another drawback of this method is that we have not included star formation. In the case of UGC 5288 star formation may not be important because of two major reasons -- (1) the estimated star formation rate is very low ($\sim 0.0063$) \citep{Werk.et.al.2011} (as mentioned in Section \ref{galaxy_selection}), (2) the stellar disk is very small compared to the large HI disk and the star formation is limited to the central regions. The galaxy may have had a star formation history, however here we are not modelling the star formation history or the merger history of the galaxy. We are modelling the present observation of the galaxy just as it has been done previously with analytical models of halos, disks and bulges. The new approach in our method is that we are modelling the galaxy with a live simulation. Currently, our simulations do not include star formation, but we intend to include it in future applications. Additionally, AGN activity has not been reported in UGC5288 and so we do aim to model it here.

We also ran high resolution simulations for the two best-fit models in Figure (\ref{best_bars}) with $6\times 10^{6}$ DM particles and $10^{6}$ disk particles. Here the DM particle to disk particle mass is $\sim 13$. For both halos, the DM density profiles, and the stellar and gas surface density profiles remain intact for more than 2 Gyrs of evolution. Hence, the halo concentration/core radius is unchanged during evolution. The bar formation time scale is similar to the low resolution simulations for the Hernquist case but for the pseudo-isothermal halo, it forms later than 2.25 Gyr. However, the overall bar properties, bar length and bar ellipticity are similar to the previous simulations. The halo spin values and the internal halo spin profile in high resolution are very similar to that of the low resolution simulations (see Figure (\ref{evolution_spin})). Even though there is a slight difference in spin values at $r>r_{200}$ for the pseudo-isothermal halo, the central values remain the same in the high resolution models. After 2 Gyrs of evolution, the spin at $r_{200}$ is slightly different from the theoretically expected spin of $0.087$ for Hernquist halo and $0.04$ for the Pseudo-isothermal halo. For both halo profiles, the halo spins at $r_{200}$, estimated using the Peebles (1969) spin (Equation 1) is $\sim 0.08$, which is higher than the mean halo spin of $\lambda_{B}\sim 0.035$ (from \cite{Bullock.et.al.2001} definition of spin) estimated for a large mass range of dark matter halos from cosmological simulations. This is expected because dwarf LSBGs such as UGC5288 are more dark matter dominated and are supposed to have higher halo spin than the cosmological average, which aids in the increase of the disk scale length over time, and makes the disk less dense and less star-forming.

Finally, we test our assumption of the correlation of bar properties with halo spin by comparing our results of halo spin profile with the corresponding estimates from a publicly available cosmological magneto-hydrodynamical simulation suite TNG50 \citep{2019TNG50_data_release,Nelson.et.al.2019,Pillepich.et.al.2019,Rosas-Guevara.et.al.2021}. We find that indeed there is a difference in the halo spin profiles between the barred and unbarred galaxies irrespective of the stellar and halo mass range we are looking at. We further probe if the difference in the distributions is related to any of the bar properties. We find a weak correlation between bar length and the halo spin in the central 5 kpc region. Although we have not shown it here with a plot, we have also checked that this correlation fades as we estimate halo spin at larger radii. Hence, there is surely a difference in the halo properties in the presence of a stellar bar in the disk. We have further found a few galaxies analogues to that of our modelled dwarf galaxy UGC5288 in TNG50 simulations. We can clearly see that a major fraction of these analogues systems (3 out of 4) have a lower value of halo spin (comparable to the predicted value of halo spin from our N-body model of UGC5288). Note that we use the terms "low" and "high" spin for the analogue galaxies in TNG50 relative to each other and not with respect to the global mean of halo spin at virial radius, $\lambda_{r_{200}}\sim 0.035-0.04$ from cosmological simulations \citep{Bullock.et.al.2001}, which is smaller than the values we obtain at large radii. For the TNG50 UGC 5288-analogue galaxies do not expect the halo spin at outer radii to be exactly comparable to the mean halo spin ($\sim 0.035-0.4$) from cosmological simulations, as the dwarf analogues form a very small data set compared to the large dynamical range of dark matter halo mass that is considered for the estimation of a global mean halo spin (for the first time in \cite{Bullock.et.al.2001}).

We note that the method we present here has its limitations. It can only be applied to non-interacting isolated galaxies having very low star formation and no AGN activity, and most importantly having a stellar bar. We also note the fact that there is still significant uncertainty in the estimate of the halo spin parameter, even after we have (1) selected an isolated, relatively unperturbed system to analyze, and (2) performed a detailed multi-component halo model and numerical forward model with SPH simulation. The model is sophisticated enough, yet we still have significant uncertainties, partly due to our lack of knowledge about the formation history and the present velocity structure of the constituents of the dark matter halo of a real galaxy like UGC 5288. We plan to test and validate this method with a larger sample of virtual galaxies from the TNG50 simulations for which the velocity structure and galaxy formation history are known (Ansar et al. in prep).

\section{Summary} \label{sec:summary}
\noindent
(i)~In this study we investigate a new method to estimate the halo spin of a dwarf-isolated barred galaxy UGC5288 and constrain its halo concentration using N-body/SPH galaxy simulations and by forward modelling the galaxy properties. Our model includes the -- stellar mass, gas mass, stellar disk and gas disk scale lengths, gas rotation curve and the bar properties. We have adopted the typical stellar velocity dispersion observed in LSB galaxies and varied the disk scale height and halo spin to obtain stable galaxy disk models.\\
(ii)~We model the dark matter halo with two types of halo profiles -- the Hernquist model and the pseudo-isothermal model, with the most general distribution function $f(\rm E, L_{z}, I_{3})$. On comparing the rotation curve of our simulated models to the HI rotation curve of UGC 5288, we find that the pseudo-isothermal halo model with $r_{c}=0.23$ kpc has the least $\chi^{2}$ and is the best match to the observed rotation curve.\\
(iii) After fixing the halo concentration and core radius from the dynamical modelling of the HI rotation curve, we explore a range of initial spin values for the two halo potentials. First, we fix the initial disk scale height and vary the halo spin to find models that have bars at the end of 2 Gyrs of evolution. Then we repeat the process for disks with different initial disk scale heights. We present the 6 most favourable models which show bar instabilities for both types of halo profiles. We find two best models, one with the Hernquist profile and another with the pseudo-isothermal halo profile, that best fits the observed bar properties of UGC 5288. \\ 
(iv)~The bar length and ellipticity for the above models are similar, but the bar morphology is different. The central stellar surface density and central halo spin are higher for the bar in the Hernquist model compared to the pseudo-isothermal model. The central mass distribution of dark matter halo affects the bar formation and bar morphology \citep{Athanassoula.Misiriotis.2002} and this is reflected in the difference in bar morphology  in the cuspy Hernquist halo and flat-core pseudo-isothermal halo model. \\
(v)~The halo spin profile is sensitive to the dark matter density profile, as expected from the expression of halo spin in Equation (\ref{Peebles_spin3}) and also clearly seen from the different halo spin profiles of Hernquist and pseudo-isothermal halo profiles in Figure (\ref{evolution_spin}). The halo spin is also sensitive to the presence of the bar as evident from the analysis of the TNG50 simulations (see Section \ref{sec:cosmological_simulations} and Appendix \ref{appendix:test_CV_bars}). If we only use the bar properties and not the HI rotation curve to constrain the dark matter distribution then we cannot narrow down the halo spin profile. For example, the bar properties are similar for the Hernquist and pseudo-isothermal halo models, although the spin distributions are quite different. Hence, first, we need to constrain the halo density distribution and then the bar properties in the disk. For modelling unbarred galaxies, we need to constrain the dark matter halo density first, and then explore the parameter space for models that do not form a bar instability. \\
(vi)~The halo spin at the central region (1 kpc $ <r< 10$ kpc), which is most relevant for bar characteristics and evolution, is different for the two cases. It varies between $0.6-0.15$ for the Hernquist halo and between $0.2-0.1$ for the pseudo-isothermal halo. The Hernquist halo has a larger spin at the central regions compared to the pseudo-isothermal halo. Hence, as seen in previous studies \citep{Saha.Naab.2013, Collier.et.al.2018, Collier.et.al.2019,Kataria.Shen.2022}, we observe the bar to form later ($> 0.5$ Gyr) in the case of the pseudo-isothermal halo model.  In a subsequent paper, we are studying this aspect more quantitatively (Ansar et. al. 2023). \\
(vii)~We compare the halo spin profiles from our best-fit models to spin profiles of barred galaxies derived from the publicly available cosmological magneto-hydrodynamical simulation TNG50 \citep{2019TNG50_data_release, Nelson.et.al.2019,Pillepich.et.al.2019, Rosas-Guevara.et.al.2021}. We find that the central halo spin ($r< 10$ kpc) of barred galaxies is lower than the unbarred galaxies in TNG50 (also observed for the outer disk in \cite{Rosas-Guevara.et.al.2021}). We find that our model halo spin profile is similar to that of the median halo spin profile of the barred galaxies in TNG50 but has significant uncertainties when compared to the individual galaxy halo spin profiles. \\
(viii)~We also find 3 analogue galaxies of UGC 5288 in the TNG50 data set that have similar disk mass and that undergo $\leq 1$ major merger during their evolution and have a low value of halo spin which is comparable to the spin profile values of our pseudo-isothermal halo spin models. However, we note that the low mass resolution for the dwarf galaxies in the TNG50 data can significantly impact the estimation of halo spin in the central regions ($\rm r<10$ kpc) and also affect the identification of poorly resolved bars in the disk. 

\section*{Acknowledgements}
We are thankful to use the high-performance computing facility ‘NOVA’ at the Indian Institute of Astrophysics, Bengaluru, India where all the simulations were run. We thank Denis Yurin and Volker Springel for providing the codes GalIC and GADGET2 publicly. This research is heavily based on these two codes and their modifications of it. We thank the Illustris and IllustrisTNG teams for making the data publicly available for use by the community. We thank the anonymous referee for very insightful suggestions that have improved the article. This research made use of Photutils, an Astropy package for the detection and photometry of astronomical sources (\cite{larry_bradley_2020_4049061}). We have also used Numpy \citep{harris2020array}, Matplotlib \citep{Hunter:2007} and pynbody \citep{pynbody} in this work. SA is grateful for the very constructive discussions with Girish Kulkarni and Yogesh Wadadekar at the Astronomical Society of India Meeting 2020. SA grateful for the very useful discussions with Shy Genel and the important suggestions from Ariyeh Maller. SA is grateful to the CCA, Flatiron Institute, Simons Foundation for hosting her from February 2021 - June 2021 while attending the Pre-Doctoral Program 2020-21, during which the above discussions happened.  MD acknowledges the support of the Science and Engineering Research Board (SERB) MATRICS grant MTR/2020/000266 for this research.

\section*{Data availability}
The data underlying this article will be shared on reasonable request to the corresponding author.



\bibliographystyle{mnras}
\bibliography{article}



\appendix

\section{Normalization of density profile} \label{normalisation}
The normalization constant in the gas density profile Equation(\ref{rho_g}) is given by $\bar{ I} (\infty, r_{1}, r_{max})$ where $\bar{ I}$ is 
\beq
\bar{ I} (r,r_{1}, r_{max})=r^{2}_{1} \bar{I}_{1}(r,r_{1}, r_{max}) +  r_{1} r_{max} \bar{I}_{2}(r,r_{1}, r_{max}) 
\eeq
where 
\begin{equation*}
    \bar{I}_{1}(r, r_{1},r_{max})={\int^{\frac{r-r_{max}}{r_{1}}}_{-\frac{r_{max}}{r_{1}}}} \frac{x dx}{1+\exp(x)}
\end{equation*}
and 
\begin{equation*}
    \bar{I}_{2}(r, r_{1},r_{max})={\int^{\frac{r-r_{max}}{r_{1}}}_{-\frac{r_{max}}{r_{1}}}} \frac{ dx}{1+\exp(x)} \texttt{        .}
\end{equation*}
The normalization constant $\bar{ I} (\infty, r_{1}, r_{max})$ can be evaluated with $ \bar{I}_{1}(\infty, r_{1},r_{max})=-1.3605 \times 10^{2}$ and $ \bar{I}_{2}(\infty, r_{1},r_{max})=1.65951 \times 10^{1}$. The normalization also appears in the total disk mass estimation within radius $r$. It is estimated as
\beq
M_{disk}(r)= M_{\star}\left[ 1-\left( 1+\frac{r}{R_{d}}\right)exp(-\frac{r}{R_{d}})\right]+\frac{M_{gas} \bar{I}(r,r_{1}, r_{max})}{\bar{I}(\infty ,r_{1}, r_{max})} \texttt{          .}
\eeq

\section{Circular velocity for gas disk} \label{circ_vel_gas_disk}
The total circular velocity for all the components of the galaxy is,
\begin{equation}
\begin{aligned}
V^{2}_{c}(r)=  \frac{G M_{dm}}{r(1+a/r)^{2}} + \frac{2 G M_{\star}}{R_{d}} y^{2} \left[ I_{0}(y)K_{0}(y) - I_{1}(y)K_{1}(y) \right] \\
-\frac{4 G M_{g}}{2 \pi \bar{I}} {\int^{r}_{0}} \frac{a da}{\sqrt{r^{2}-a^{2}}} \frac{d}{da} {\int^{\infty}_{a}}   \frac{r^{'} dr^{'}}{\left(1+\exp((r^{'}-r_{max})/r_{1})\right) \sqrt{{r^{'}}^{2}-a^{2}}}
\end{aligned}
\end{equation}
where $ \bar{I}= \bar{I}(\infty, r_{1}, r_{max})$.
The integral in the above expression is the contribution of the gas disk to the circular velocity and can be expressed as,
\beq
\begin{aligned}
 V^{2}_{c, g}(r)=-\frac{4 G M_{g}}{2 \pi \bar{I}(\infty, r_{1}, r_{max})} {\int^{r}_{0}} \frac{a da}{\sqrt{r^{2}-a^{2}}} \times \\
  \frac{d}{da} {\int^{\infty}_{0}}   \frac{1}{\left(1+\exp((z^{2}+a-r_{max})/r_{1})\right)} \frac{2(z^{2}+a) dz}{\sqrt{z^{2}+2a}}
\end{aligned}
\eeq
The above integral is an improper integral and can be evaluated using the extended midpoint rule of integration.
\beq
\int^{x_{N}}_{x_{0}} f(x) dx =h \left[ f_{1/2} +f_{3/2} + ... + f_{N-1/2} \right] + O\left(\frac{1}{(N+1)^{2}}\right)
\eeq

\begin{figure}
\centering
\includegraphics[width=\columnwidth]{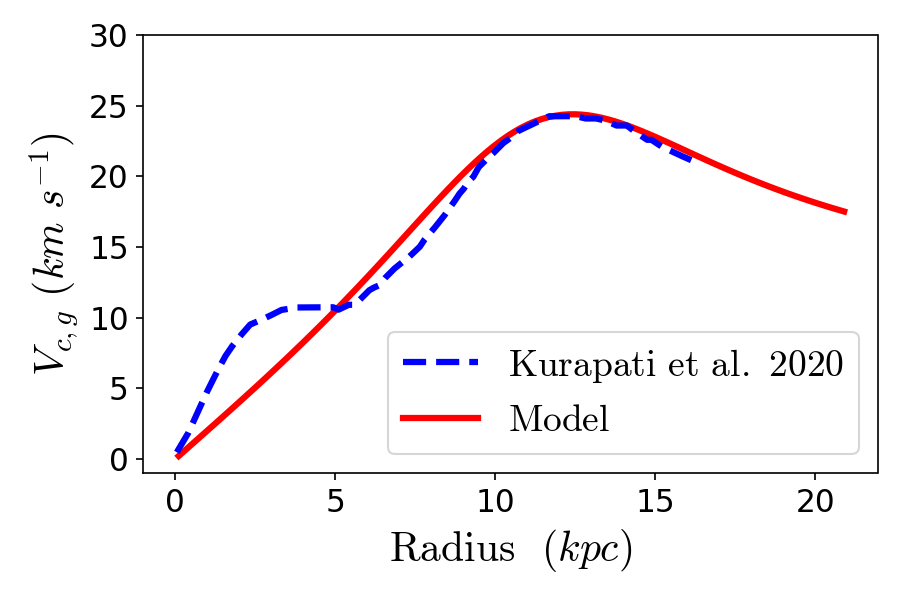}
\caption{In this figure we compare the circular velocity due to gas disk (different from gas rotation curve) obtained from the observations of \citep{Kurapati.et.al.2020} (blue dashed curve) with that estimated from modelling the gas disk in this study (red curve), where $r_{1}=1.8$ kpc and $r_{max}=11.0$ kpc.}
    \label{vc_gas_comp}
\end{figure}

\section{Method to include gas disk profile in GalIC} \label{gas_pdf}
\begin{figure}
\centering
	\includegraphics[width=\columnwidth]{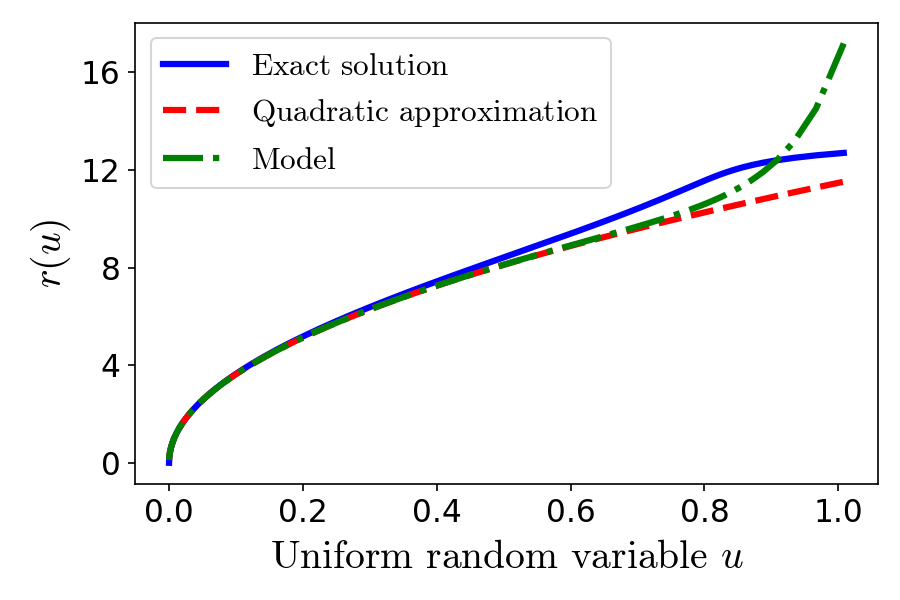}
    \caption{This figure shows the profile for $r(u)$ versus $u$. The blue line is the exact solution of Equation (\ref{bigeq}); the red dashed line is the quadratic approximation from Equation (\ref{qua_approx}) and the green dot-dash line is our approximated model for the exact solution from Equation (\ref{fitted}).}
    \label{ru_profile}
\end{figure}

The gas profile function Equation(\ref{rho_g}) is included in the density of the whole disk as $\rho (r, z)= \rho_{\star}(r, z) + \rho_{g}(r, z)$ while estimating the potential of the system, the dispersion measure for the disk, etc. But including this alone is not sufficient to generate particle positions with GalIC. To generate random position coordinates, we use the same technique as used in GalIC. For each particle, we have to randomly determine ($r$, $\phi$, $z$) where they have the usual meaning in cylindrical coordinates. \\
We note that all the particles for the disk are of the same mass. In that case, the probability of finding a particle in the elemental volume $d\tau$, $P(r, z)\tau$ should be proportional to $\rho_{g}(r, z) d\tau$. From Equation(\ref{rho_g}) we can write  $\rho_{g}(r, z)\propto \rho_{1g}(z) \times \rho_{2g}(r)$. So the probability can be decomposed into $P(r, z) d\tau=P_{1}(z) dz \times P_{2}(r) 2\pi r dr$. \\
First, we consider the z coordinate of the particles. Before specifically explaining the method we briefly introduce the concept of random numbers with uniform distribution. For a random variable $u$ that follows a uniform distribution with $0 \leqslant u \leqslant 1$, the probability distribution function of $u$, $P(u)= 1$. So the expectation value of any function of $u$, $f(u)$ is given as
\beq \label{PDFu}
\left\langle f(u) \right\rangle = {\int^{1}_{0}} f(u) P(u) du = {\int^{1}_{0}} f(u) du 
\eeq
We know that the z coordinates of the particles follow a distribution $P_{1g}(z)=sech^{2}(z/z_{g0})/2 z_{g0}$. The z coordinate can be considered as a random variable. So the expectation of value for all the z coordinates can be expressed as
\beq
\left\langle z \right\rangle = {\int^{\infty}_{-\infty}} z P_{1g}(z) dz
\eeq
The above expression can further be written in terms of the uniform random variable $u$ as
\beq
\left\langle z \right\rangle = {\int^{1}_{0}} z(u) P_{1g}(z(u)) \frac{dz}{du} du = {\int^{1}_{0}} z(u) \times 1 du
\eeq
where we have used Equation(\ref{PDFu}) and we have,
\beq
P_{1g}(z(u)) \frac{dz}{du}=1 \texttt{				.}
\eeq
Solving the above with proper limits, we express the z coordinate in terms of the uniform random number $u$.
\beq \label{zu}
z=\frac{z_{0}}{2} \ln \left( \frac{u}{1-u} \right)
\eeq
In a similar way, the radial coordinates $r$ of all the gas particles can be considered to be a random variable and we have 
\beq
\left\langle r \right\rangle = {\int^{\infty}_{0}} r P_{2g}(r) 2 \pi r dr = {\int^{1}_{0}} r(u) \times 1 du
\eeq
where we using Equation(\ref{PDFu}) we have, $2 \pi r P_{2g}(r)\frac{dr}{du}=1$. With 
\begin{equation*}
    P_{2g}(r)=  \frac{1}{2\pi \bar{I}(\infty, r_{1}, r_{max}) \left(1+\exp((r^{'}-r_{max})/r_{1})\right) }
\end{equation*}
we solve the following equation to obtain $r=r(u)$. 
\beq
\frac{1}{\bar{I}(\infty, r_{1}, r_{max})} {\int^{}_{}} \frac{r dr}{ \left(1+\exp((r^{'}-r_{max})/r_{1})\right)} = {\int^{}_{}} du
\eeq
We obtain,
\begin{equation}
\begin{aligned} \label{bigeq}
r^{2}_{1} \left\lbrace  \frac{x}{2} \left(x- 2\ln \left(e^{x}+1\right) \right) - Li_{2}(-e^{x}) \right\rbrace +r_{1}r_{max} \left\lbrace x - \ln\left(e^{x}+1 \right) \right\rbrace \\
= \bar{I}(\infty, r_{1}, r_{max}) \left(u + c^{'} \right)
\end{aligned}
\end{equation}
where $x=(r-r_{max})/r_{1}$ and $ Li_{2}(z)= \sum_{n= 1}^{\infty} \frac{z^{n}}{n^{2}}$ is the poly-logarithm function.  With proper limits and by taking $e^{-r_{max}/r_{1}}= e^{-16.6}\approx 0$, we get the integration constant $c^{'}$ to be,
\beq
c^{'} = \frac{r^{2}_{1}-r^{2}_{max}/2}{\bar{I}(\infty, r_{1}, r_{max})} \texttt{    .}
\eeq
The above equation can be approximated with the three most dominant terms as:
\begin{equation} \label{qua_approx}
  \frac{r^{2}_{1}}{2} x^{2}+ r_{1} r_{max} x + \frac{r^{2}_{max}}{2} \approx \bar{I}(\infty, r_{1}, r_{max}) u
\end{equation}
The solution of the above equation is closely represented by the following form of $r(u)\approx \sqrt{2 \bar{I}(\infty, r_{1}, r_{max}) u} $ but to match with the solution of Equation (\ref{bigeq}) and to have a more realistic disk which shows a gradual decrease in density at the edges, the quadratic solution has to be multiplied with a function such as
\beq \label{fitted}
r(u)=\sqrt{2 \bar{I}(\infty, r_{1}, r_{max}) u} \left( 1+ 10^{-\alpha}\exp(\beta u) \right)
\eeq
where, the values $\alpha=6.0$ and $\beta=13.0$ depends on the choice of r$_{1}$ and r$_{max}$. See Figure(\ref{ru_profile}) for r comparison of Equation (\ref{bigeq}) \& (\ref{fitted}). Finally, we get three random coordinates for each particle with the $z$-coordinate given by Equation(\ref{zu}) and
\beq
x= r(u) cos\phi  \quad \texttt{       \it   and    }  \quad y= r(u) sin\phi
\eeq
where $\phi=2\pi q$, and $u$ \& $q$ are the uniform random variable varying within [0, 1].

\section{Initial dispersion in gas disk}
\begin{figure}
\centering
	\includegraphics[width=\columnwidth]{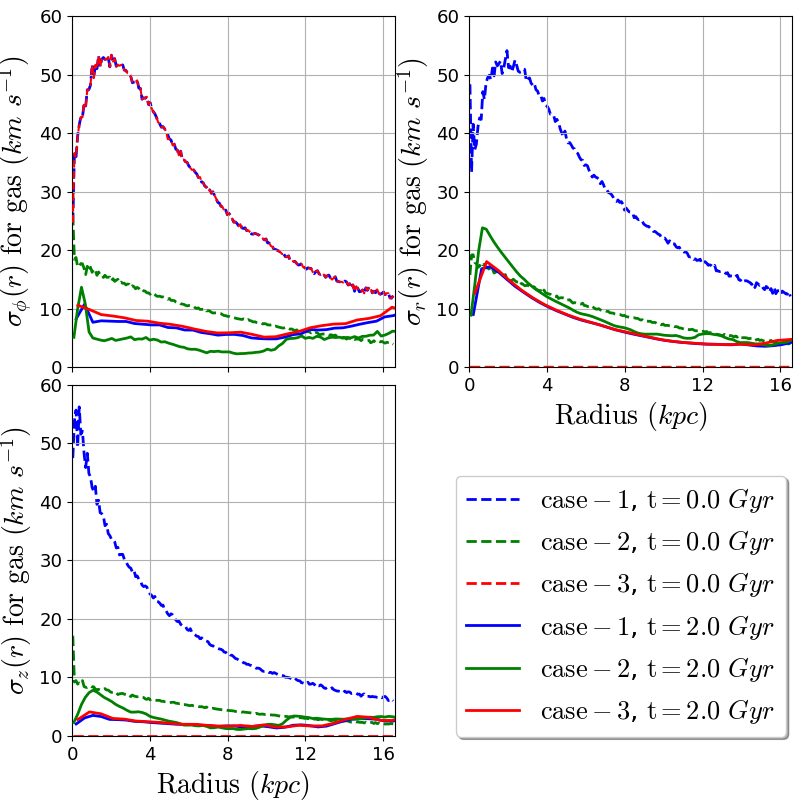}
    \caption{This figure shows the three different choices of initial gas dispersion represented by three colours and the corresponding evolved gas dispersion for the three different cases.}
    \label{evolution_dispersion}
\end{figure}
The mean dispersion of gas in UGC 5288 is $\sigma \sim 9.1$ km s$^{-1}$ \cite{Kurapati.et.al.2020} but the radial distribution of gas dispersion is in general different from this. We test our simulations with three different initial gas dispersions. We compare the resultant dispersion for the three different cases after 2 Gyr of evolution. In case-1 we keep the gas dispersion similar to the stellar particle dispersion as fixed by GalIC, i.e., when we tag the star particles as gas particles we do not change the velocities of the gas particles. Though the gas dispersion is very different from the stellar dispersion as seen from observations, we still want to check whether a detailed knowledge of initial gas dispersion is required at all. So in Case-2, we consider a model dispersion at different radii. We estimate a measure of the dispersion in the z-direction of HI, $\sigma_{HI,z}= 6.2$ km s$^{-1}$ using the expression in Equation (2) in \cite{Das.et.al.2019}. We take the central z-dispersion to be this value and we take the dispersion at the outer edges (16.6 kpcs) of the HI disk to be $\sim 1$ km s$^{-1}$. Hence we construct an analytic function to represent the initial gas dispersion, $\sigma_{HIz}=6.2\times \exp(-0.11r)$ and we assume $\sigma_{\phi}=\sigma_{R}=2\sigma_{z}$. In case-3 we provide the $\phi-$dispersion along with the streaming motion to the gas particles and we set the initial $\sigma_{R}=\sigma_{z}=0$. All three cases of initial dispersion and the final evolved state are shown in Figure ($\ref{evolution_dispersion}$).

\section{Velocity dispersion of stellar disk} 
\label{appendix:isotropic-vel-dispersion}
\begin{figure*}
\centering
	\includegraphics[width=0.45\textwidth]{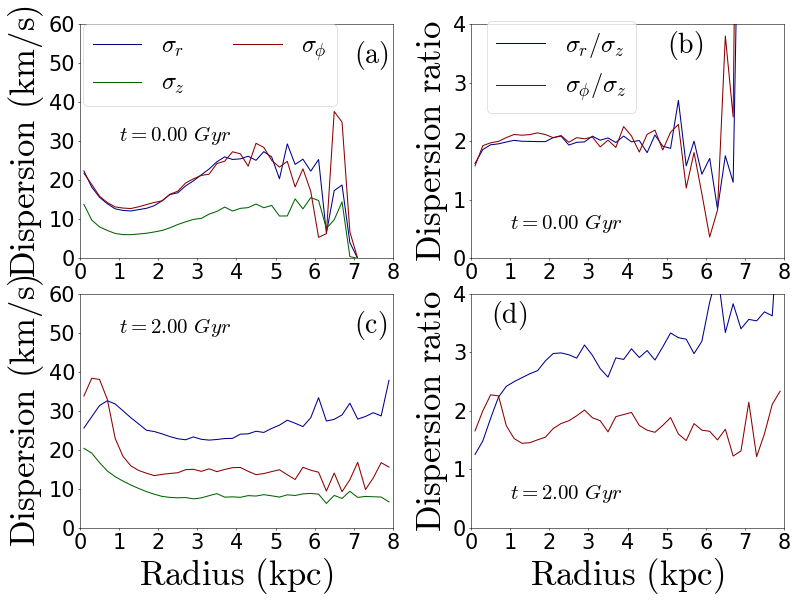}
	\includegraphics[width=0.45\textwidth]{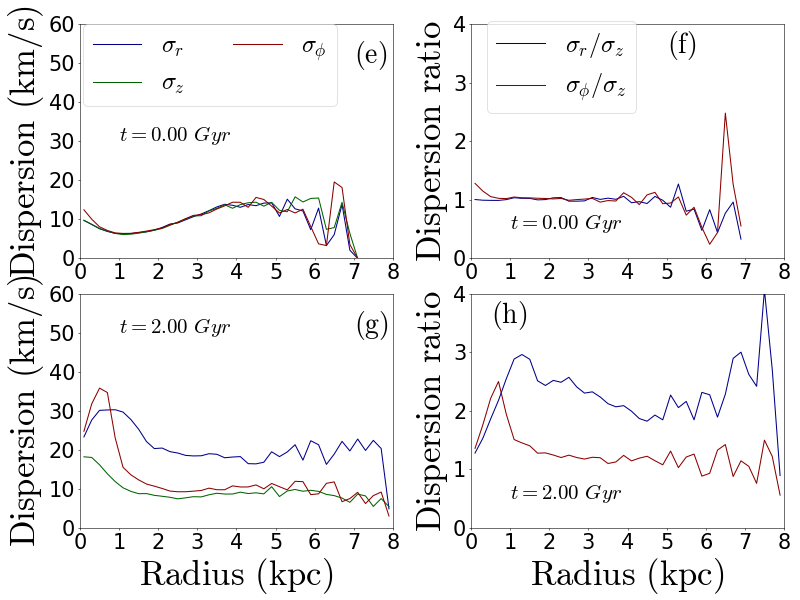}
\caption{ Comparison of stellar velocity dispersion evolution in disks with MW-like dispersion ratio (panel (a)-(d)) and isotropic velocity dispersion (panel (e)-(f)). The velocity dispersion distribution evolves to very similar values within 2 Gyrs of evolution ((c) and (g)). }
    \label{fig:appendix-stellar-vel-dist}
\end{figure*}

We investigate the effect of initial stellar velocity dispersion on the evolution of our models. Figure (\ref{fig:appendix-stellar-vel-dist}) shows the initial and final velocity dispersion ($\sigma_{R}$, $\sigma_{\phi}$ and $\sigma_{z}$) and dispersion ratios ($\sigma_{R}/\sigma_{z}$ and $\sigma_{\phi}/\sigma_{z}$) for two models: (1) Milky Way (MW) like velocity dispersion ratios, i.e., with $\sigma_{R}=\sigma_{\phi}=2\sigma_{z}$ and (2) isotropic velocity distribution, i.e., with $\sigma_{R}=\sigma_{\phi}=\sigma_{z}$. We observe that after an evolution of 2 Gyrs the velocity dispersion in both models are very similar irrespective of the initial values of the stellar velocity dispersion. Additionally, the disk forms a weak bar during the later part of evolution at around $\sim 3.3$ Gyrs.

\section{Values of the k-paramater}\label{appendix:kparameter}
\begin{table}
	\centering
	\caption{Model spin parameters}
	\label{kparameter_fr_spin}
	\begin{tabular}{lcccr} 
		 \hline
 $f_{R}$ & $k-parameter$ & Spin $\lambda$ at r$_{200}$ \\
 \hline\hline
 1.0 & 0.7 & 0.031 \\ 
 1.0 & 0.768 & 0.0365 \\ 
 1.0 & 0.837 & 0.0413 \\ 
 0.75 & 0.95 & 0.0813 \\
 0.75 & 1.00 & 0.08748 \\
 0.75 & 1.06 & 0.0959 \\
 \hline
	\end{tabular}
\end{table}

\section{$\chi^{2}$ of Hernquist and pseudo-isothermal halo models}
\begin{table}
\centering
\caption{$\chi^{2}$ fitting for Rotation Curve of Hernquist halo}
\label{chi_square}
\begin{tabular}{lr} 
\hline
 Halo concentration (c) & $\chi^{2}$ ($r\geqslant 5$ kpc) \\
 \hline\hline
 6 & 61.7  \\
 7.5 & 7.9 \\
 8 & 6.3 \\ 
 8.5 & 12.3 \\
 10 & 46.0  \\ 
 \hline
\end{tabular}
\end{table}

\begin{table}
\centering
\caption{$\chi^{2}$ fitting for Rotation Curve of Pseudo-isothermal halo}
\label{chi_square_iso}
\begin{tabular}{lr} 
\hline
 Halo concentration ($c_{iso}$) & $\chi^{2}$ ($r\geqslant 5$ kpc) \\
 \hline\hline
 206 & 22.35  \\
 381 & 21.08  \\
 411 & 21.4 \\
 447 & 19.7 \\ 
 541 & 21.88 \\
 685 & 23.7  \\ 
 \hline
\end{tabular}
\end{table}

\section{Estimation of initial Spin of Pseudo-isothermal Halo} \label{appendix_Iso_halo}
To estimate the initial halo spin for a pseudo-isothermal halo from the spin definition given by \cite{Peebles1969} (Equation (\ref{Peebles_spin3})), we estimate the total energy of a pseudo-isothermal halo using certain assumptions same as used to derive NFW spin in Equation (22) in \cite{Mo.et.al.1998} at $r_{200}$. The total energy of a pseudo-isothermal halo is estimated at the virial radius $r_{200}$ when the halo particle orbits are assumed to be circular about the halo centre of mass.
\begin{equation}
    E_{tot}=- E_{KE}=-\frac{G M^{2}_{dm}(r_{200})}{2 r_{200}} f_{iso}(c_{iso})
\end{equation}
where $c_{iso}= r_{200}/r_{c}$ and
\begin{flalign*}
& f_{iso}(c_{iso}) =   \\
& \frac{ c_{iso} \left( c_{iso} -\tan^{-1}(c_{iso}) -c_{iso}(\tan^{-1}(c_{iso}))^{2} + \int^{c_{iso}}_{0} (\tan^{-1}x)^2 dx \right)}{\left( c_{iso} -\tan^{-1}(c_{iso}) \right)^{2}}
\end{flalign*}
Thus, relating the halo angular momentum and disk angular momentum through Equation (\ref{Jdh}), we can express the halo spin for a pseudo-isothermal halo as
\begin{equation}\label{lambda_iso}
    \lambda_{iso} = \frac{J_{disk} M^{-3/2}_{dm} }{j_{d}\sqrt{2 G r_{200}}} \sqrt{ f_{iso}(c_{iso}) } \texttt{      .}
\end{equation}
Using the estimated $J_{disk}$ from simulations and a core radius $r_{c}=0.25$ kpc, $\lambda_{iso}=0.04$.

\section{Bar strength $A_{2}/A_{0}$} \label{bar_strength}
Bar strength is defined by the Fourier decomposition of the face on the surface density of the stellar disk of a galaxy. The maximum value of the $m=2$ Fourier mode represents the bar strength \citep{Athanassoula.2003}.
\begin{equation}
\frac{A_{2}}{A_{0}}=\frac{\sqrt{ a^{2}_{2} + b^{2}_{2} }}{\Sigma^{N}_{i=1} m_{\star i}}
\end{equation}

\section{Evolution of galaxy disk for 2 Gyrs}
\begin{figure*}
\centering
	\includegraphics[width=0.9\linewidth]{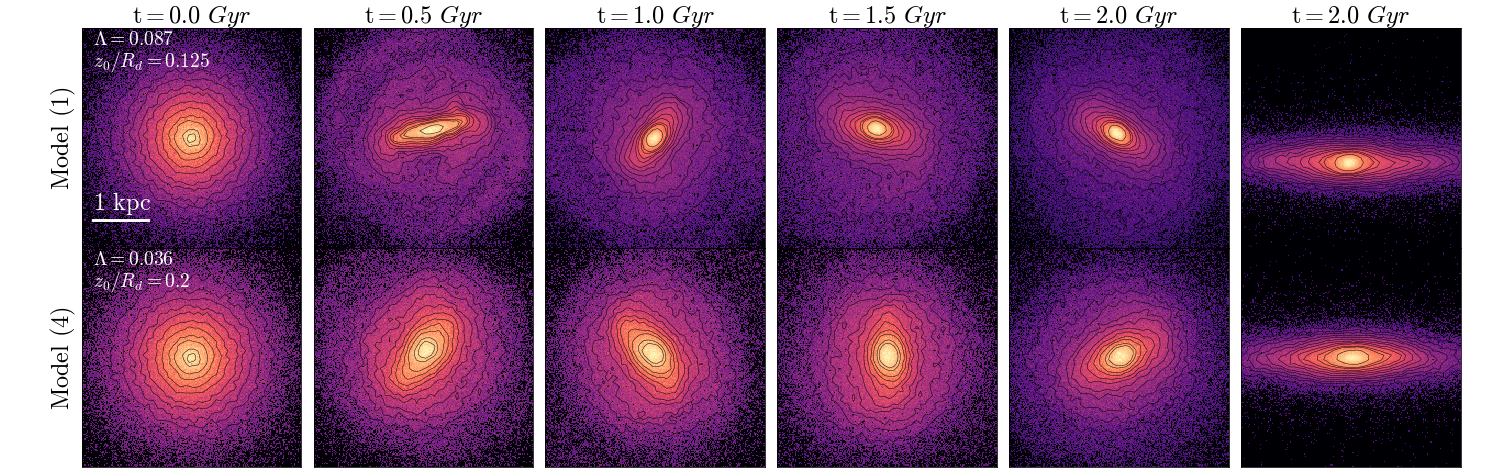}
	\vspace{-1cm}
	\includegraphics[width=0.9\linewidth]{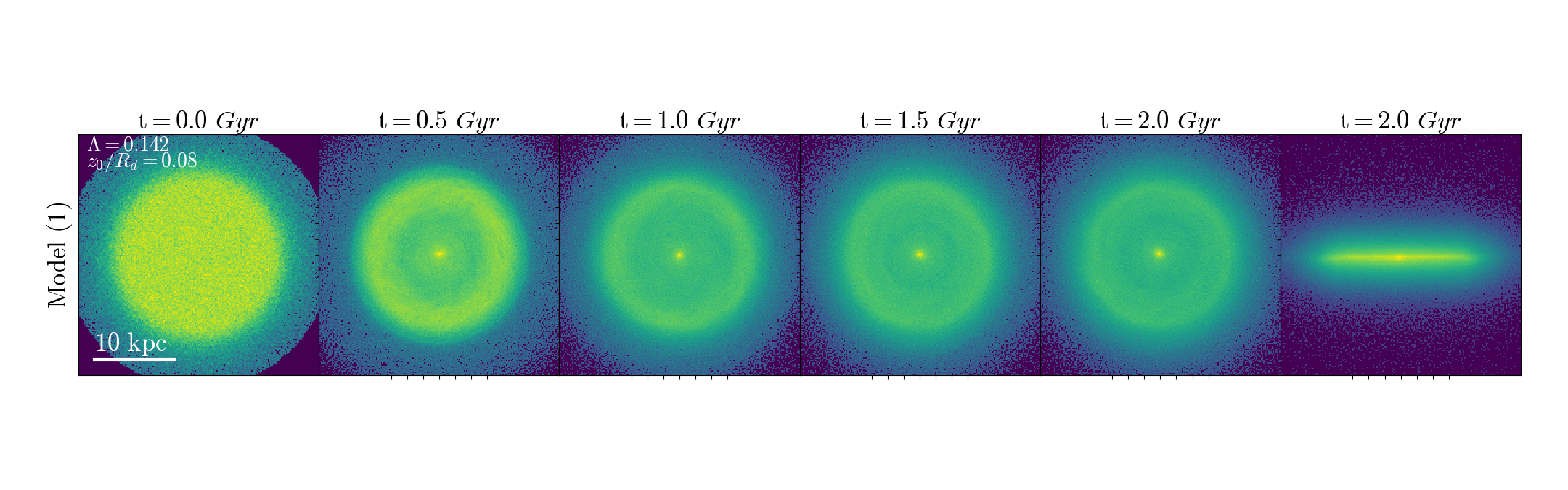}
    \caption{Evolution of galaxy disk for 2 Gyrs for Hernquist halo models with concentration parameter $c=8$ and different halo spin and disk scale height (shown in Log-histogram plots in scales of kpc). The peak ellipticity of the bar in Model (1) is closer to the peak ellipticity of the bar in UGC 5288. Unlike the stellar disk, the gas disk does not show bar formation and the spatial distribution is the same for all the models. The gas disk also accumulates gas at the centre during evolution.}
    \label{evolution_models_c8}
\end{figure*}

\begin{figure*}
\centering
	\includegraphics[width=0.9\linewidth]{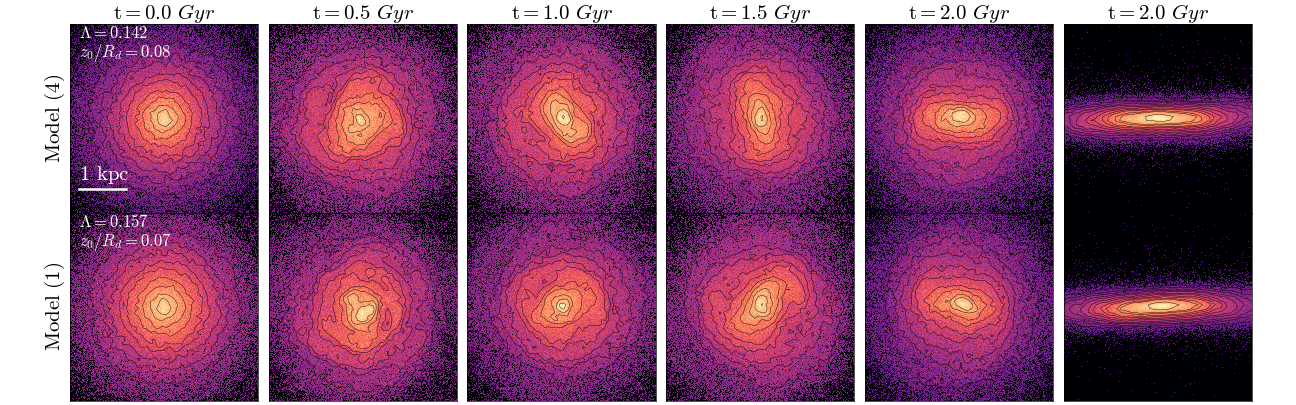}
	\vspace{-1cm}
	\includegraphics[width=0.9\linewidth]{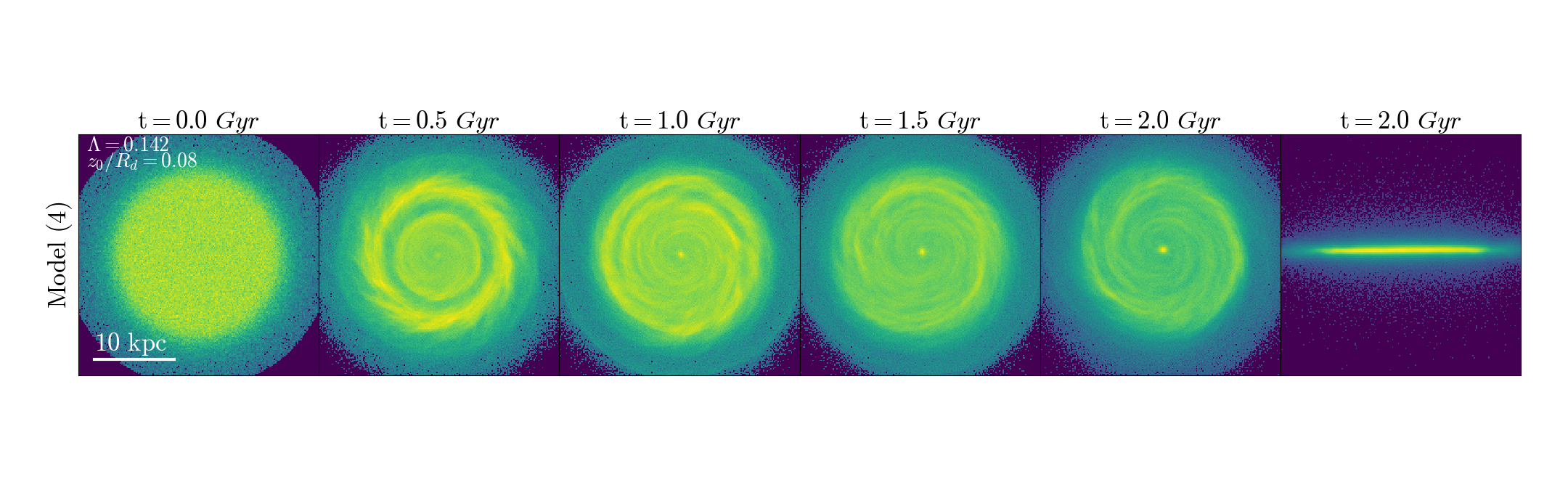}
    \caption{Evolution of galaxy disk for 2 Gyrs for models with Pseudo-isothermal halo core radius $r_{c}=0.23$, and different initial halo spin and disk scale height, shown in Log-histogram plots in scales of kpc. Model (4) starts forming a bar early with a stronger bar at the end of 2 Gyrs. The stellar disks are represented in red and the gas disk in green. Unlike the stellar disk, the gas disk does not show bar formation and the spatial distribution is the same for all the models. The gas disk also accumulates gas at the centre during evolution.}
    \label{evolution_models_ciso}
\end{figure*}

\section{Bar images of TNG50 galaxies}
\begin{figure*}
\centering
	\includegraphics[width=\textwidth]{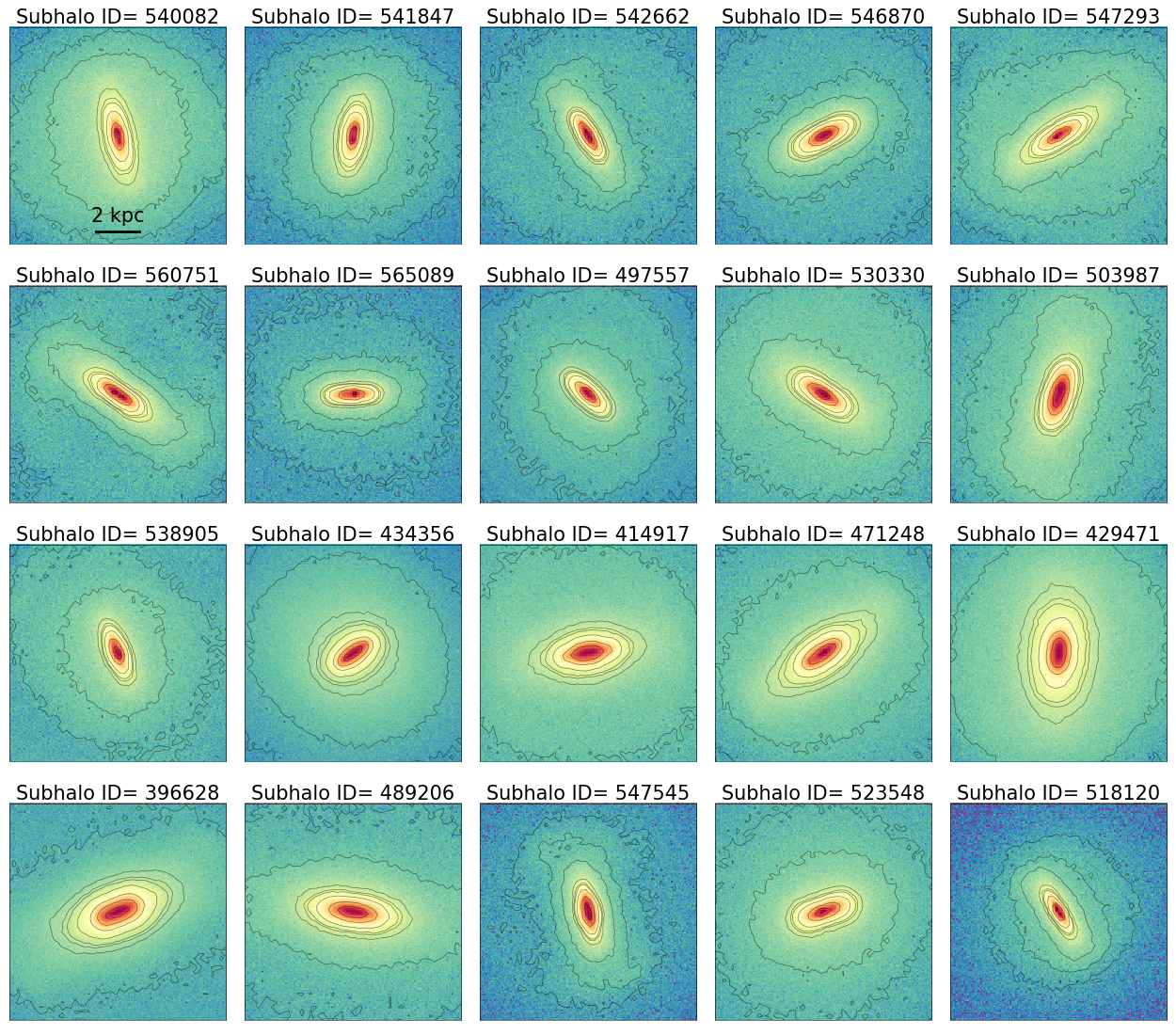}
    \caption{ Here we show 20 TNG50 bar images along with their subhalo IDs that we have used in our analysis. TNG50 have bars having varied morphology, bar strength and bar length for different cases. All the images are in the same scale and the scale length is shown in the first image of the bar.}
    \label{bar_images_tng50}
\end{figure*}
   
\section{Test to check the effect of the presence of the bar in galaxies having similar dark matter mass distribution} \label{appendix:test_CV_bars}
\begin{figure*}
\centering
\includegraphics[width=\columnwidth]{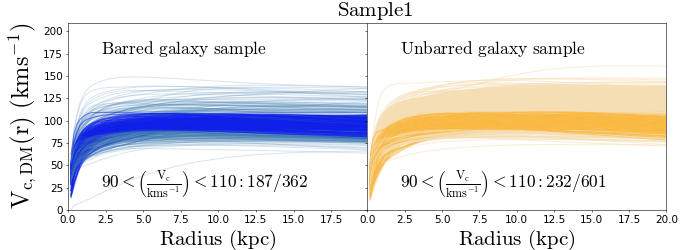}
\includegraphics[width=\columnwidth]{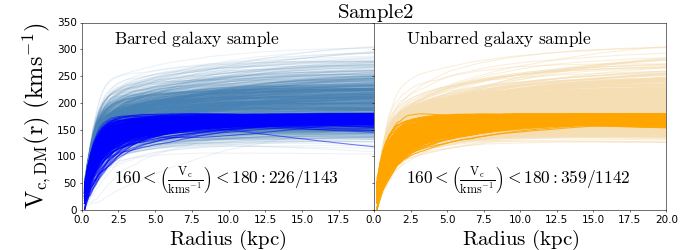} \\
\includegraphics[width=\columnwidth]{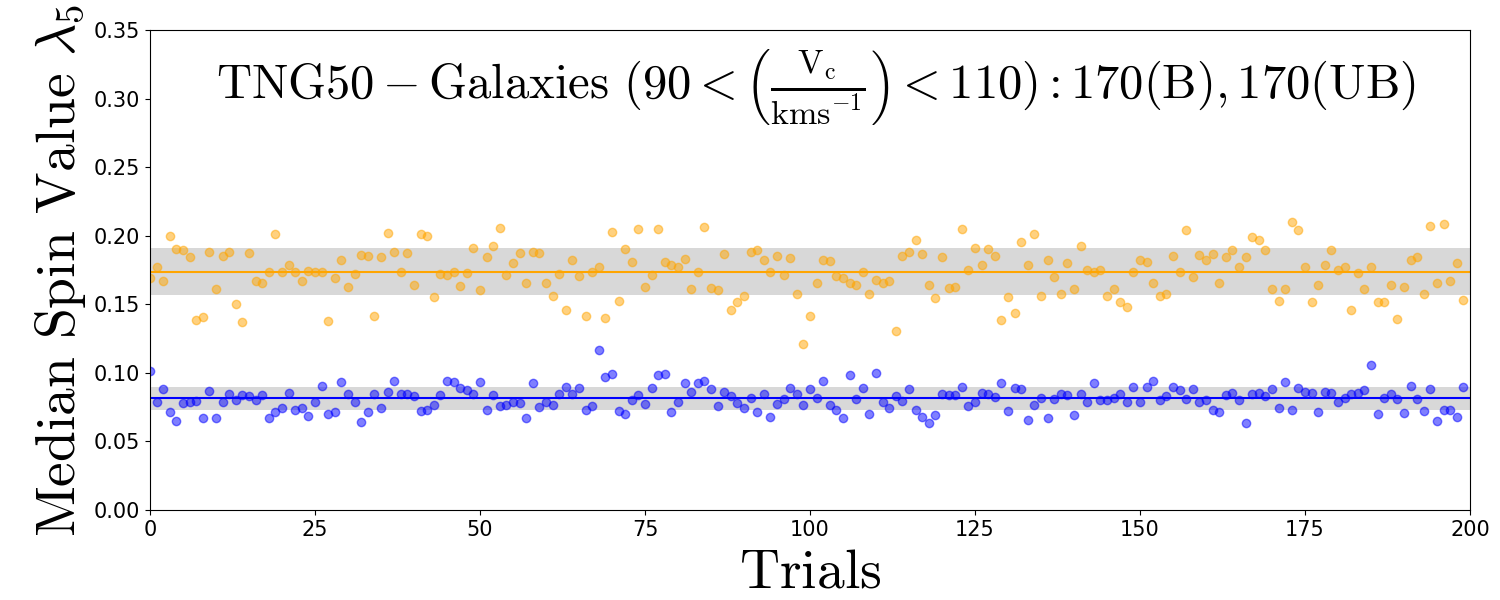}
\includegraphics[width=\columnwidth]{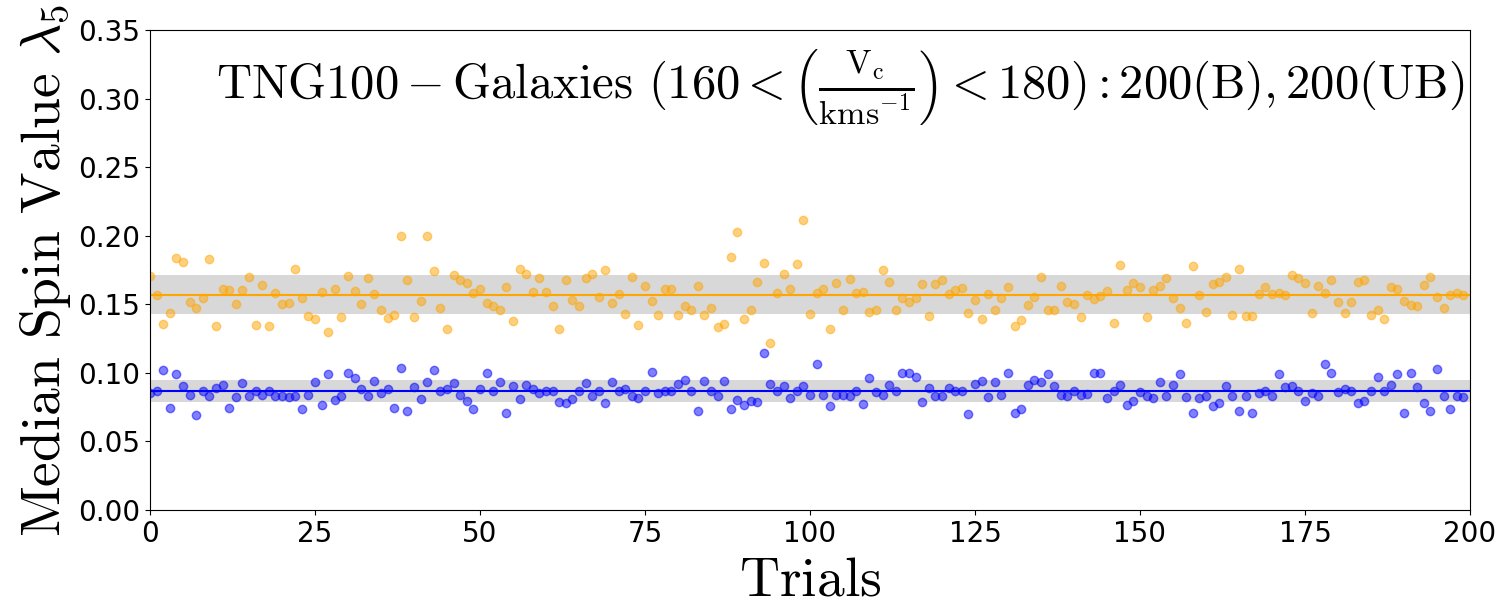} \\
\includegraphics[width=\columnwidth]{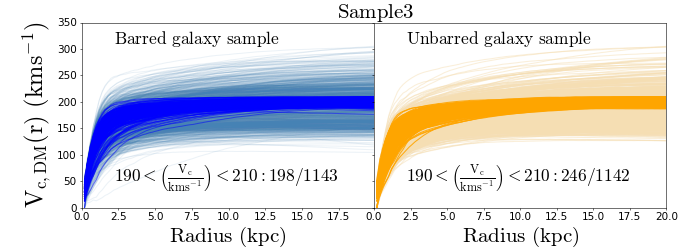} \\
\includegraphics[width=\columnwidth]{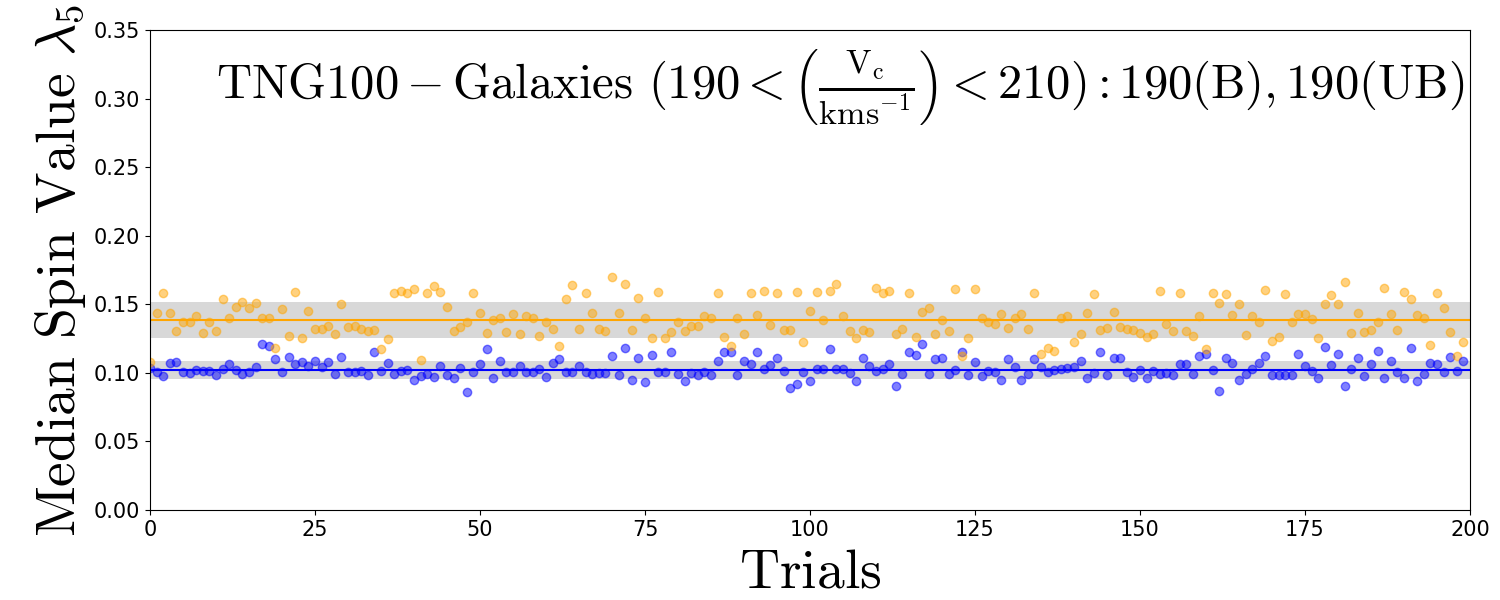}
    \caption{The difference between the median of the halo spin profile for barred and unbarred galaxies in TNG50 with similar dark matter circular velocity, for galaxy stellar mass range $10^9 < M_{\star}/M_{\odot}<10^{10}$. The DM circular velocity of Sample 1 is in the range $90<V_{c, max}/(km/s)<110$, for Sample 2 it is in the range $160<V_{c, max}/(km/s)<180$ and $190<V_{c, max}/(km/s)<210$ for Sample 3 (the darker blue and orange coloured circular velocity curves). The median halo spin for barred galaxies (blue dots) is lower than that for the unbarred galaxies (orange dots) for all three samples with the difference between the median values decreasing for galaxies with higher and higher circular velocities. }
    \label{Sample1}
\end{figure*}

In this section, we investigate the impact of the presence of a bar in the stellar disks of DM halos with a similar type of dark matter mass distribution. We conducted some tests with the TNG50 and TNG100 data sets with barred and unbarred galaxies in different stellar mass ranges, having similar dark matter circular velocity curves. The three samples that we used are as follows. 
\begin{itemize}
    \item {  Sample 1. TNG50 galaxies: $10^9 < M_{\star}/M_{\odot}<10^{10}$ (see top left panels in Figure \ref{Sample1}) and $90<V_{c,max}/(km/s)<110$, containing 187 barred galaxies and 232 unbarred galaxies.}
    \item {Sample 2. TNG100 galaxies: $10^{10} < M_{\star}/M_{\odot}<10^{11}$ (see top right panels in Figure \ref{Sample1}) and $160<V_{c,max}/(km/s)<180$, containing 226 barred galaxies and 359 unbarred galaxies.}
    \item {Sample 3. TNG100 galaxies: $10^{10} < M_{\star}/M_{\odot}<10^{11}$ (see bottom panels in Figure \ref{Sample1}) and $190<V_{c,max}/(km/s)<210$, containing 198 barred galaxies and 246 unbarred galaxies.}
\end{itemize}
We select all the galaxies in the different samples considering the maximum circular velocity $V_{c, max}$ within a radius of 20 kpcs should lie in different ranges of velocities as mentioned above. Once we have selected the barred and unbarred galaxies in the different samples, we estimate the halo spin profile as a function of radius $\lambda(r)$ for all the galaxies. Next, we estimate the median of the distribution of halo spin at $r=5$ kpc, i.e., $\lambda_{5}$ for the barred and unbarred galaxies from the different samples separately. We take an equal number of barred and unbarred galaxies from the samples. For example, we choose 170 barred galaxies out of 187, and 170 unbarred galaxies out of 232 in Sample 1 (see panels for Sample 1 in Figure \ref{Sample1}), and estimate the median of each set of galaxies. This selection is carried out randomly in different trials as shown in the bottom panel of Sample 1 in Figure \ref{Sample1}. Finally, to check if there is any significant difference between the median halo spin of the barred galaxies (blue points) and unbarred galaxies (orange points), we over-plot the median of all the orange points (orange horizontal line) and the blue points (blue horizontal line) and the 1$\sigma$ region around the median (shaded regions). We do a similar analysis for Sample 2 and Sample 3 (see panels corresponding to Sample 2 and 3 in Figure \ref{Sample1}. \\

From the three samples, it is clear that the median spin values are different for the barred and unbarred galaxy groups even though the DM circular velocity curves are very similar for them. This shows that the presence of a bar affects the halo spin irrespective of the effect of the dark matter density distribution. We note that the separation between the median halo spins for the barred and unbarred galaxy groups decreases for higher values of circular velocities (Sample 1 to Sample 3). We are studying this in more detail in a following article (Ansar et al. in prep).\\

\section{Bar properties of TNG50 galaxies}
\label{appendix:bar_properties}

\begin{figure*}
\centering
	\includegraphics[width=0.8\textwidth]{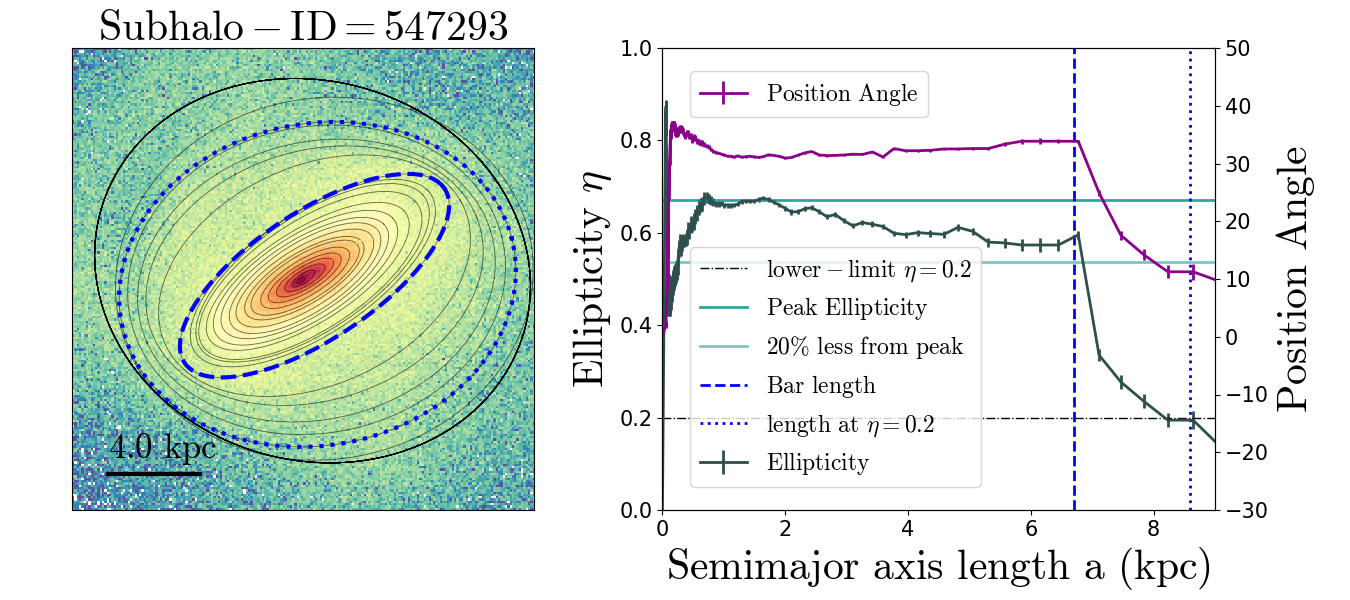}
    \caption{ Estimating bar length from ellipticity and position angle (PA) distribution along the semi-major axis of the bar in the galaxy in subhalo ID 547293. The dashed blue vertical line indicates the bar length estimated from the 20$\%$ decrease in ellipticity in the bar region. It coincides with the fall in both ellipticity and PA at 6.7 kpc. The dotted blue line indicates the outer edge of the bar at 8.6 kpcs where ellipticity $\eta=0.2$. Table (\ref{table:appendix_bar_properties_tng}) shows the bar properties of all the galaxies in S1 and S2.}
    \label{fig:appendix-tng_bar_length}
\end{figure*}

\begin{table*}
	\centering
	\caption{ Bar properties}
	\label{table:appendix_bar_properties_tng}
	\begin{tabular}{lccccr} 
		 \hline
 Subhalo ID &  Bar Strength  & Bar semi-major axis  &  Bar semi-major axis & Ellipticity  &  Sample  \\
           & $\left(\frac{A_{2}}{A_{0}}\right)_{max}$ &  (20$\%$ peak ellipticity) (kpc)  & (at $\eta=0.2$) (kpc)  & (bar region) ($\eta$) &  \\
\hline\hline
540082 &	0.4 &	4.35 &	5.41 &	0.65 & S1, S2 \\
541847 &	0.38 &	3.4 &	5.7  &   0.6 & S1, S2 \\
542662 &	0.56 &	5   & 	5.85 &	0.66 & S1, S2 \\
546870 &	0.507 &	5.6 &	6.8 &  	0.6  & S1, S2 \\
547293 &	0.427 &	6.7 &	8.6 &	0.67 & S1, S2 \\
560751 &	0.55 &	8.5	&   9   &	0.67 & S1, S2 \\    
565089 &	0.56 &	3   &	5.9 &   0.66 & S1, S2 \\
497557 &	0.48 &	2.7	&   3.9 &   0.61 & S1, S2 \\
507784 &	0.47 &	2.75 &	3.4 &	0.415 & S1, S2 \\
513105 &	0.49 &	2.8	& 3.4  &	0.55 & S1, S2 \\
538905 &	0.45 &	3.5 &	4.6	&    0.575 & S1, S2 \\
434356 &	0.38 &	2.5 &	2.75&	0.8 & S1, S2 \\
414917 &	0.406 &	7.15 &	7.5 &	0.48 & S1, S2 \\
402555 &	0.395 &	5.1 &	5.7 &	0.48 & S1, S2 \\
475619 &	0.29 &	7.1 &	7.4 &	0.56 & S1, S2 \\
471248 &	0.41 &	6.8 &	7.4 &	0.52 & S1, S2 \\
429471 &	0.28 &	6.9 &	7.4 &	0.42 & S1, S2 \\
396628 &	0.376 &	9.7 &	10  &	0.52 & S1, S2 \\
379803 &	0.49 &	8.8 &	9.7 &	0.57 & S1, S2  \\
529365 &	 0.27 &	 4.05 &	 4.72 &	 0.48 &	S2 \\ 
530852 &	 0.53 &	 5.9 &	 6.62 &	 0.62 &	 S2 \\ 
530330 &	 0.48 &	 5.3 &	 6.45 &	 0.61 &	 S2 \\ 
531320 &	 0.56 &	 2.7 &	 4.9 &	 0.53 &	 S2 \\ 
534628 &	 0.51 &	 5.9 &	 6.1 &	 0.48 &	 S2 \\ 
540920 &	 0.48 &	 5.1 &	 7.3 &	 0.57 &	 S2 \\ 
551541 &	 0.41 &	 2.1 &	 2.65 &	 0.58 &	 S2 \\ 
547545 &	 0.52 &	 7.0 &	 7.2 &	 0.56 &	 S2 \\ 
483594 &	 0.42 &	 7.1 &	 7.4 &	 0.51 &	 S2 \\ 
489206 &	 0.48 &	 8.7 &	 9.6 &	 0.6 &	 S2 \\ 
503987 &	 0.51 &	 8.2 &	 8.8 &	 0.58 &	 S2 \\ 
514829 &	 0.51 &	 4.65 &	 5.3 &	 0.675 & S2 \\ 
518120 &	 0.5 &	 3.97 &	 4.42 &	 0.673 & S2 \\ 
518682 &	 0.49 &	 1.95 &	 3.4 &	 0.6 &	 S2 \\ 
506720 &	 0.4 &	 5.6 &	 7.2 &	 0.55 &	 S2 \\ 
454171 &	 0.31 &	 5.1 &	 7.5 &	 0.55 &	 S2 \\ 
503437 &	 0.44 &	 2.95 &	 5.5 &	 0.6 &	 S2 \\ 
523548 &	 0.42 &	 2.75 &	 5.2 &	 0.62 &	 S2 \\ 
394621 &	 0.41 &	 5.8 &	 6.3 &	 0.6 &	 S2 \\ 
454171 &	 0.31 &	 5.0 &	 5.5 &	 0.55 &	 S2 \\
517899 &	 0.33 &	 8.0 &	 8.2 &	 0.535 & S2 \\ 
392276 &	 0.32 &	 3.12 &	 3.26 &	 0.4 &	 S2 \\ 
440407 &	 0.51 &	 1.3 &	 2.3 &	 0.7 &	 S2 \\ 
444134 &	 0.52 &	 2.2 &	 5.35 &	 0.6 &	 S2 \\ 
449658 &	 0.44 &	 3.1 &	 4.52 &	 0.565 & S2 \\ 
491426 &     0.44 &  2.0 &   4.53 &  0.6  &  S2 \\
\hline
	\end{tabular}
\end{table*}

\section{Halo spin profile of candidate analogues of UGC5288}
\label{appendix:halo-spin-distribution}
\begin{figure*}
\centering
\includegraphics[width=0.9\textwidth]{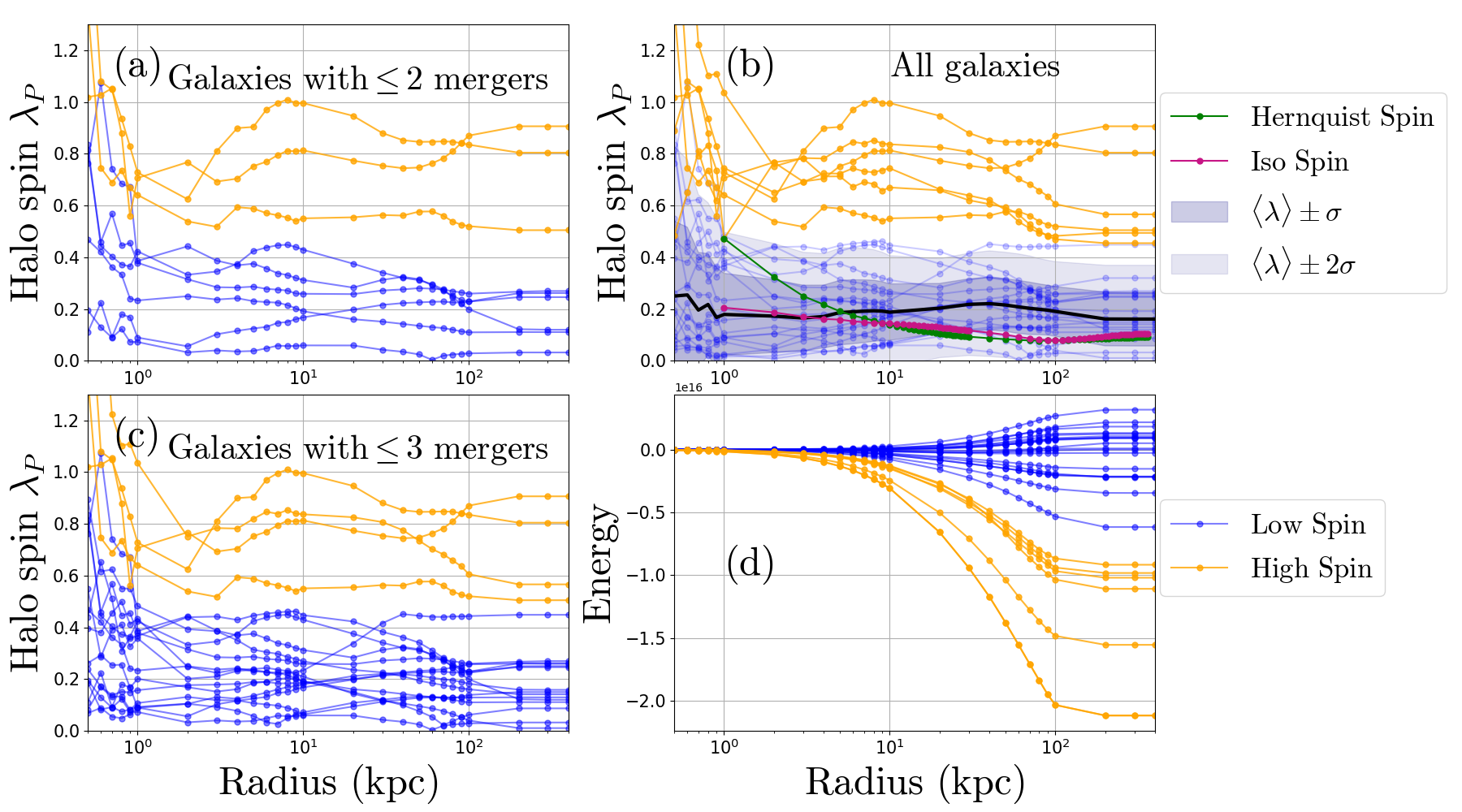}
    \caption{ Halo spin profile of UGC 5288 analogues. In all the panels, the low spin halo is shown in blue and the high spin is shown in orange. Panel (a) and (c) show the halo spin profile of the galaxies that undergo multiple mergers within a redshift range of $0\leq z\leq8$. Nearly 66\% of galaxies that undergo $\leq 2$ mergers have low halo spin, while 79\% of galaxies that undergo $\leq 3$ mergers have low halo spin. Panel (b) consists of all the 24 galaxies in our original sample, along with the median (black solid line) and 1$\sigma$ (darker shade region) and 2$\sigma$ (lighter shade region) regions. The Hernquist spin distribution (green) and pseudo-isothermal spin distribution (dark red) are over-plotted for comparison. The pseudo-isothermal halo spin profile is very similar to the median of the halo spin profiles of the low halo spin galaxies. Panel (d) show the profiles of halo energy at different radii for the 24 galaxies in panel (b). The high halo spin galaxies are in deeper potential wells and hence in denser regions of their local environment. The low-spin halos are in less dense environments similar to the void regions.  }
    \label{fig:appendix-halospin_all_galaxies_energy}
\end{figure*}


\bsp	
\label{lastpage}
\end{document}